\documentclass[a4paper,11pt]{article}
\pdfoutput=1 

\usepackage{jheppub} 
\usepackage[T1]{fontenc} 
\usepackage[italian,english]{babel}
\usepackage{hyperref}
\usepackage{ifpdf}
\usepackage{subfigure}
\usepackage{tikz}
\usepackage[compat=1.1.0]{tikz-feynman}
\usepackage{slashed}
\usetikzlibrary{arrows,shapes}
\usetikzlibrary{trees}
\usetikzlibrary{matrix,arrows} 				
\usetikzlibrary{positioning}				
\usetikzlibrary{calc,through}				
\usetikzlibrary{decorations.pathreplacing}  
\usepackage{pgffor}							

\usetikzlibrary{decorations.pathmorphing}	
\usetikzlibrary{decorations.markings}
\tikzset{
	vector/.style={decorate, decoration={snake}, draw},
	fermion/.style={draw=black, postaction={decorate}}, 
	scalar/.style={dashed,draw=black, postaction={decorate}}}
\tikzstyle{block} = [draw, rectangle, 
minimum height=3em, minimum width=6em]

\usepackage{amssymb}
\usepackage{amsfonts}
\usepackage{epsf}
\usepackage{rotating}
\usepackage{graphicx}
\usepackage{amsmath}
\usepackage{fancyhdr}
\usepackage{lineno}
\usepackage{babel}
\usepackage{graphics}
\usepackage{pstricks}
\usepackage{color}
\usepackage{multirow}
\usepackage{float}
\usepackage{lineno}
\usepackage{mathtools}

\def\sn#1{\textit{Scenario-#1}}
\newcommand{\nn}{\nonumber}
\newcommand{\lsim}{\mathrel{\mathop{\kern 0pt \rlap
			{\raise.2ex\hbox{$<$}}}
		\lower.9ex\hbox{\kern-.190em $\sim$}}}
\newcommand{\gsim}{\mathrel{\mathop{\kern 0pt \rlap
			{\raise.2ex\hbox{$>$}}}
		\lower.9ex\hbox{\kern-.190em $\sim$}}}

\newcommand{\be}{\begin{equation}}
\newcommand{\ee}{\end{equation}}
\newcommand{\bea}{\begin{eqnarray}}
\newcommand{\eea}{\end{eqnarray}}

\newcommand{\gbl}{g_{BL}}
\newcommand{\zbl}{Z_{BL}}

\def\sn#1{\textit{Scenario-#1}}

\def\gev{\ensuremath{\mathrm{\,Ge\kern -0.1em V\,}}}
\def\tev{\ensuremath{\mathrm{\,Te\kern -0.1em V\,}}}

\title{\boldmath  Secluded Dark Matter in Gauged $B-L$ Model}

\author[a]{Priyotosh Bandyopadhyay,}
\author[c,d]{Manimala Mitra,}
\author[c,d]{Rojalin Padhan,}
\author[c,d]{Abhishek Roy}
\author[b]{Michael Spannowsky}

\newcommand{\AddrHBNI}{
	Homi Bhabha National Institute, BARC Training School Complex, Anushakti Nagar, Mumbai 400094, India }
	
\affiliation[a]{Indian Institute of Technology Hyderabad, Kandi,  Sangareddy-50228, Telengana, India}
\affiliation[b]{Institute for Particle Physics Phenomenology, Department of Physics, Durham University, South Road, 
	Durham DH1 3LE, United Kingdom}
\affiliation[c]{Institute of Physics, Sachivalaya Marg, Bhubaneswar, Pin-751005, Odisha}
\affiliation[d]{\AddrHBNI}

\emailAdd{bpriyo@phy.iith.ac.in} 
\emailAdd{manimala@iopb.res.in} 
\emailAdd{rojalin.p@iopb.res.in} 
\emailAdd{abhishek.r@iopb.res.in}
\emailAdd{michael.spannowsky@durham.ac.uk }

\preprint{IITH-PH-0001/22 \\ \hspace*{0pt}\hfill IP/BBSR/2022-13}

\abstract{We consider the gauged $B-L$ model which is extended with a secluded dark sector, comprising of two dark sector particles. In this framework the lightest $\mathcal{Z}_2$-odd particle is the dark matter candidate, having a feeble interaction with all other SM and BSM states. The next-to-lightest $\mathcal{Z}_2$-odd particle in the dark sector is a super-wimp, with large interaction strength with the SM and BSM states. We analyse all the relevant production processes that contribute to the dark matter relic abundance, and broadly classify them in two different scenarios, a) dark matter is primarily produced via the non-thermal production process, b) dark matter is produced mostly from the late decay of the next-to-lightest $\mathcal{Z}_2$-odd particle. We discuss the dependency of the relic abundance of the dark matter on various model parameters. Furthermore, we also analyse the discovery prospect of the BSM Higgs via invisible Higgs decay searches.  }

\begin{document}
\maketitle
\flushbottom
	\section{Introduction}
	
Observational evidence shows that 84$\%$ matter of the universe is in the form of non-baryonic dark matter. However, very little is known about the nature of dark matter (DM) and its origin. The Standard Model (SM) can not explain the observed relic density. One of the most favoured scenarios for DM production has been thermal freeze-out, where a weakly interacting massive particle (WIMP) serves as a DM candidate. The WIMP with electroweak-scale mass, which interacts with other particles via electroweak interaction, can naturally explain the measured DM relic density of $\Omega h^2=0.1199\pm0.0027$ \cite{Ade:2015xua}. Several direct detection experiments so far have searched for a WIMP. However, the lack of conclusive experimental evidence motivates the exploration of alternate dark matter production mechanisms. The production of DM via freeze-in mechanism \cite{Hall:2009bx, Molinaro:2014lfa, Biswas:2015sva, Merle:2015oja, Shakya:2015xnx, Konig:2016dzg, Biswas:2016iyh, Biswas:2016yjr, Barman:2021lot} is one of the most popular production mechanisms. In this framework, DM has a very tiny interaction with the SM and any other particle which are in thermal equilibrium with the plasma and thereby referred to as feebly interacting massive particle (FIMP). Due to significantly suppressed interaction with SM/BSM particles, the FIMP DM never attains thermal equilibrium. Due to a similar suppression in the interaction with the SM particles, FIMP, in general, can not produce any observable signal in the direct detection experiments. See \cite{Hambye:2018dpi} for other alternatives with a light mediator. The DM in the freeze-in scenario is produced from the decay and/or annihilation of SM, and BSM particles which are either in equilibrium with the thermal plasma or also freezing-in along with the DM \cite{Bandyopadhyay:2020qpn}.  

Apart from the DM abundance, the SM fails to explain neutrino masses and mixings. One of the most promising models that explain small neutrino masses is the gauged $B-L$ model, which contains three right-handed neutrinos (RHNs)  \cite{Davidson:1978pm, Mohapatra:1980qe, Wetterich:1981bx, Georgi:1981pg} that generate light neutrino masses via seesaw mechanism \cite{Mohapatra:1979ia, minkowski1977mu}. 
In addition, the model also contains one BSM gauge boson $Z_{BL}$ and a complex scalar field $S$. The scalar field acquires vacuum expectation value and breaks the $B-L$ gauge symmetry. The BSM gauge boson and the heavy neutrinos acquire their masses due to the spontaneous breaking of the $B-L$ gauge symmetry. 
The model can further be extended with a secluded dark sector with a scalar particle $\phi_D$ with non-zero $B-L$ charge and a gauge singlet fermion state $\chi$, where either or both of them can be suitable DM candidates. The dark sector particles are odd under a $\mathcal{Z}_2$ symmetry. The thermal DM for this model has been explored in several works, such as \cite{Bandyopadhyay:2018qcv,Bandyopadhyay:2017bgh}. For a different variation of the $B-L$ gauge model with only a thermal scalar DM, see \cite{Sanchez-Vega:2014rka, Singirala:2017see, Klasen:2016qux, Rodejohann:2015lca, Biswas:2017tce}. For RHN DM in a typical $B-L$ model, see \cite{Basak:2013cga, Okada:2016gsh, Okada:2016tci, Kaneta:2016vkq, Okada:2010wd, Okada:2012sg}. The RHN can also serve as a portal between the SM sector and a secluded dark sector containing DM particles, see \cite{Escudero:2016ksa, Becker:2018rve, Escudero:2016tzx} for the relevant discussion. One of the interesting possibilities is if the fermion state $\chi$ serves as the non-thermal DM candidate, and the scalar particle $\phi_D$, which was in thermal equilibrium in the early universe, has a significant contribution in the production of $\chi$.

In this article, we consider this possibility, where $\chi$ is the DM state, and $\phi_D$ significantly impacts its production. We study the production of DM through a thermal freeze-in mechanism and significant non-thermal freeze-in contribution ~\cite{Molinaro:2014lfa, Garny:2018ali} from the decay of $\phi_D$. The state $\chi$ interacts only with $\phi_D$ and RHN $N$. The DM candidate $\chi$ never thermalises due to very tiny coupling. This work considers that $\phi_D $ is heavier than $\chi$ state. However, it serves as the next-to-lightest $\mathcal{Z}_2$-odd particle~(NLOP). $\phi_D$ thermalises with the SM particles because of large quartic couplings in the scalar potential, sizeable SM-BSM Higgs mixing angle,  and large gauge coupling. DM is primarily produced at high temperatures via a thermal freeze-in mechanism from the decay of bath particle $\phi_D$, which was in equilibrium with the rest of the plasma. $\phi_D$ subsequently decoupled from the thermal bath, and its late decay further produced substantial DM relic density. The abundance of $\phi_D$ at the time of decoupling $\Omega^{FO}_ {\phi_D} h^2$ is governed by the freeze-out mechanism. Depending upon the abundance $\Omega^{FO}_ {\phi_D} h^2$  and its conversion to $\chi$ state through out-of-equilibrium decay, the DM can primarily be produced from the late decay as well, which we refer as non-thermal production of DM. We divide the discussion into two different scenarios and show the importance of thermal and non-thermal freeze-in contributions in determining DM relic abundance. We elaborately discuss the constrain on other model parameters, such as the scalar quartic couplings, SM-BSM Higgs mixing angle and the mass of $\phi_D$ and $\chi$, that appear from the relic density constraint. Other than this, we also explore the discovery prospect of this model at the future High Luminosity run of the LHC~(HL-LHC), mainly focusing on the heavy Higgs searches. Due to non-zero coupling $\lambda_{SD}$ between the state $\phi_D$ and $B-L$ Higgs $S$, the model offers a largely invisible branching ratio of the BSM Higgs state. We analyse the discovery potential of the BSM Higgs $H_2$-$\phi_D$ scalar quartic coupling at the HL-LHC in vector-boson fusion~(VBF) channel. This quartic coupling has a large impact in determining $\phi_D$ abundance, hence the DM relic density.  

The paper is organised as follows. In Section.~\ref{model}, we describe the model. Following this in Section.~\ref{dm}, we discuss the dark matter production in detail and along with the effect of out-of-equilibrium decay of $\phi_{D }$. And later, we discuss the collider prospects on the search of $\phi_D$ at the LHC in Section.~\ref{sec:collider}. We present our conclusion in Section.~\ref{conclusion}. In Appendix.~\ref{appen1},\ \ref{appen2} and \ref{appen3}, we provide the necessary calculation details.

\section{Model}\label{model}

The model is a gauged $ B-L$ model, augmented with a secluded dark sector. In addition to the particles of the gauged $B-L$ model, i.e., the right-handed neutrinos $N$, BSM Higgs field $S$ and BSM gauge boson $Z_{BL}$, the model also contains dark sector particles a complex scalar state $\phi_D$ and a $B-L$ singlet fermion $\chi$. The dark sector particles are odd under $\mathcal{Z}_2$ symmetry, while other particles are even. The $\mathcal{Z}_2$ symmetry ensures the stability of DM. We consider that the BSM sector of the gauged $B-L$ model and the SM Higgs state $h$ act as a portal between  other SM particles and the dark sector~\footnote{In our notation, $h$ represents the SM Higgs doublet field.}. 

The charge assignments of different particles are shown in Table~\ref{tabbml}. Here, the field $S$ represents a complex scalar field, which acquires vacuum expectation value~(vev) $v_{BL}\neq 0$ and breaks $B-L$ gauge symmetry. The state $N$ contributes to the light neutrino mass generation via the seesaw mechanism. Note that the scalar $\phi_D$ is non-trivially charged under $B-L$ gauge symmetry, while the fermion $\chi$ is a singlet under both the $B-L$ and SM gauge group. As we will see in the subsequent sections, this leads to significant differences in the evolution of $\chi$ and $\phi_D$ abundances. The complete Lagrangian of the model has the following form,   
\begin{table}[h]
	\centering
	\begin{tabular}{ |c| c |c |c| c|c |c|c |c|c|c|}
		\hline 
		& $h$ & $N$ &  $L$ & $Q$ & $u_R$ & $d_R$ & $e_R$ & $S$&$\phi_{D}$&$\chi$\\ \hline
		$Y_{B-L}$ & $0$ & $-1$ & $-1$ & $1/3$ &$1/3$ & $1/3$ &$1$& $2$&$1$&$0$ \\
		\hline
		$\mathcal{Z}_2$ & $1$ & $1$ & $1$ & $1$ &$1$ & $1$ &$1$& $1$&$-1$&$-1$ \\
		\hline
	\end{tabular}
	\caption{Charges of all the particles under $B-L$ and $\mathcal{Z}_2$ symmetry.}  \label{tabbml}
\end{table}
       
	\begin{equation}
	\mathcal{L}= \mathcal{L}_{S M}+\mathcal{L}_{D M}+\mathcal{L}_{B-L},
	\end{equation}
	where $\mathcal{L}_{D M}$ is the Lagrangian containing dark sector particles, and $\mathcal{L}_{B-L}$ is the $B-L$ Lagrangian.  The $B-L$ Lagrangian  has the following form, 
	{\small\begin{equation}
	\begin{aligned}
	\mathcal{L}_{B-L}=&\left(D_{\mu} {S}\right)^{\dagger}\left(D^{\mu} {S}\right)-\frac{1}{4} F_{B L \mu \nu} F_{B L}^{\mu \nu}+{i} \bar{N}_{i} \gamma^{\mu} D_{\mu} N_{i}-V_{B-L} \left({h}, {S}\right)\\
	&-\sum_{i=1}^{3} \lambda_{NS}  {S} \bar{N}_{i}^{c} N_{i}-\sum_{i, j=1}^{3} y_{N,i j}^{\prime} \bar{L}_{i} \tilde{h} N_{j}+h . c . \  ,
	\end{aligned}
	\end{equation}}
{\small\begin{equation*}
	V_{B-L}\left(h, S\right)=\mu_{S}^{2} S^{\dagger} {S}+\mu_{h}^{2} {h}^{\dagger} {h}+\lambda_{S}\left({S}^{\dagger} {S}\right)^{2}+\lambda_{h}\left({h}^{\dagger} {h}\right)^{2}+\lambda_{Sh}\left({h}^{\dagger} {h}\right)\left({S}^{\dagger} {S}\right),
	\end{equation*}}\\
The dark sector Lagrangian is given by,
{\small\begin{equation}
	\begin{aligned}
	\mathcal{L}_{D M}=&\bar{\chi}(i\slashed{\partial}-m_{\chi})\chi +(D^{\mu} \phi_{D })^{\dagger}\left(D_{\mu} \phi_{D }\right)-\mu_{D }^{2}\left(\phi_{D }^{\dagger} \phi_{D }\right)-\lambda_{D }\left(\phi_{D }^{\dagger} \phi_{D }\right)^{2}-\lambda_{D h}\left(\phi_{D }^{\dagger} \phi_{D }\right)\left({h}^{\dagger} {h}\right) -\\ &\lambda_{ SD}\left(\phi_{D }^{\dagger} \phi_{D }\right)\left({S}^{\dagger} {S}\right)-(Y_{D\chi}\bar{\chi}\phi_{D }N+h.c.).
	\end{aligned}
\end{equation}}
The interaction strength between $\phi_{D }$ and BSM and SM Higgs bosons($i.e, S$ and $h$) are proportional to $\lambda_{ SD}$ and $\lambda_{ Dh}$. The kinetic energy terms involving $S$, $\phi_D$ and $N$ contains the covariant derivatives which is given by,
\begin{equation}
D_{\mu}X=( \partial_{\mu}+ig_{BL}Y_{B-L}(X)Z_{BL\mu} )X,
\end{equation}
where $X=S,\ N,\ \phi_{D } $ and $Y_{B-L}(X)$ represents $B-L$ charge of the states shown Table~\ref{tabbml}.
\begin{itemize}
	\item {\it SM Higgs and BSM Higgs:} After spontaneous symmetry breaking (SSB), the SM Higgs doublet $h$ and BSM scalar $S$ is given by,
	\begin{equation}
	h=
	\begin{pmatrix}
	0 \\
	\dfrac{v+h_{1}}{\sqrt{2}}
	\end{pmatrix}
	\ \ \ 
	S=
	\begin{pmatrix}
	\dfrac{v_{BL}+h_{2}}{\sqrt{2}}
	\end{pmatrix}.\
	\end{equation}
	Owing to the non-zero $\lambda_{Sh}$, $h_1$ and $h_2$ mixes with each other which leads to the scalar mass matrix  given by,
	\begin{eqnarray}
	\mathcal{M}^2_{scalar} = \left(\begin{array}{cc}
	2\lambda_h v^2 ~~&~~ \lambda_{Sh}\,v_{BL}\,v \\
	~~&~~\\
	\lambda_{Sh}\,v_{BL}\,v ~~&~~ 2 \lambda_S v^2_{BL}
	\end{array}\right) \,\,.
	\label{mass-matrix}
	\end{eqnarray}
	The basis states $h_1$ and $h_2$ can be rotated by suitable angle $\theta$ to the new basis states $H_1$ and $H_2$. The new basis states represents the physical basis states which are given by,
	\begin{eqnarray}
	H_{1}&=&h_{1}\ cos\  \theta - h_{2}\ sin\  \theta,\\
	H_{2}&=&h_{1}\ sin\  \theta + h_{2}\ cos\  \theta,	
	\end{eqnarray}
	where $H_1$ is the SM like Higgs and $H_2$ is the  BSM Higgs. The mixing angle between the two states is defined by,
	\begin{equation}
	tan\ 2\theta =\frac{v v_{BL}\lambda_{ Sh}}{v^{2}\lambda_{h}-v_{BL}^{2}\lambda_{S}}. 
	\end{equation}
	The mass square eigenvalues of $H_1$ and $H_2$ are given by,
	\begin{equation}
	M_{H_1,H_2}^2=\lambda_h v^2+\lambda_{ S}v_{BL}^2\pm \sqrt{(\lambda_{h} v^2-\lambda_{S}v_{BL}^2)^{2}+(\lambda_{ Sh}vv_{BL})^2}. 
	\label{eq:massesscalar}
	\end{equation}
	
	\item {\it Neutrino mass:} The Majorana mass term of RHN's is generated through spontaneous symmetry breaking of $B-L$ symmetry. The mass of RHN's is given by,
	\begin{equation}
	M_{N}=\frac{\lambda_{ NS}\ v_{BL}}{\sqrt{2}}.
	\label{eq:neutrino_mass}
	\end{equation}
	The mass of the SM neutrinos is generated through Type-I seesaw mechanism where the light neutrino mass matrix has the following expression, 
	\begin{equation}
	{m^{\nu}_{ij}}=\frac{y_{N,i k}^{\prime} y_{N,k j}^{\prime} {\langle h \rangle}^2 }{M_{N, k}}.
	\end{equation}
	
	\item {\it Gauge boson mass:} Similar to the  RHN's, the additional neutral gauge boson mass $Z_{\rm BL}$  is generated via spontaneous  breaking of $B-L$ gauge symmetry. The mass of $Z_{\rm BL}$ is related to the symmetry breaking scale $v_{BL}$ as,  
	\begin{equation}
	M_{Z_{\rm BL}}=2g_{BL}v_{BL},
	\end{equation}
	where $g_{BL}$ is the associated $B-L$ gauge coupling constant.
	
	\item {\it Dark sector constituents mass:} The mass square  of  the particle $\phi_{D }$  has the following form,
	\begin{equation}
	m_{\phi_{D}}^2=\mu_{D}^2+\frac{\lambda_{Dh}v^2}{2}+\frac{\lambda_{SD}v_{BL}^2}{2}. 
	\end{equation}
	Note that, both electroweak symmetry breaking vev $v$ and the $B-L$ symmetry breaking vev $v_{BL}$ have impact in determining the mass of $\phi_D$. In this work, we consider  $\lambda_{ SD}$ and $\lambda_{ Dh}$ in between $10^{-1}$ and $10^{-5}$ to have $\phi_{D }$ as the thermal particle. 
	\item
	The DM candidate $\chi$ is singlet under SM and $B-L$ gauge group. It's mass is governed by the bare mass term, $i.e,\ m_{\chi}$.    
\end{itemize}  

\section{Dark Matter \label{dm}  }

The dark sector fields $\chi$ and $\phi_D$ can be DM particles. However, we consider the scenario, where $\chi$ is a non-thermal FIMP DM, and $\phi_D$, the NLOP, is primarily responsible for DM production. In the early universe, the state $\phi_D$ was in thermal equilibrium with the bath particles, and at some later epoch denoted as $T_d$, it decoupled from the rest of the plasma.  
Other than the standard thermal freeze-in contribution via $\phi_D \to \chi N$ process effective up to epoch $T \sim m_{\phi_D} > T_d$, the out-of-equilibrium decay of $\phi_D$ into $\chi$ also contributes significantly in the relic abundance of DM. Below, we explore this possibility in detail.

 \subsection{Super-WIMP $\phi_D$+ FIMP DM $\chi$ \label{phidsw}}

\begin{figure}[H]
	\begin{center}
		\includegraphics[angle=0,height=7cm,width=14cm]{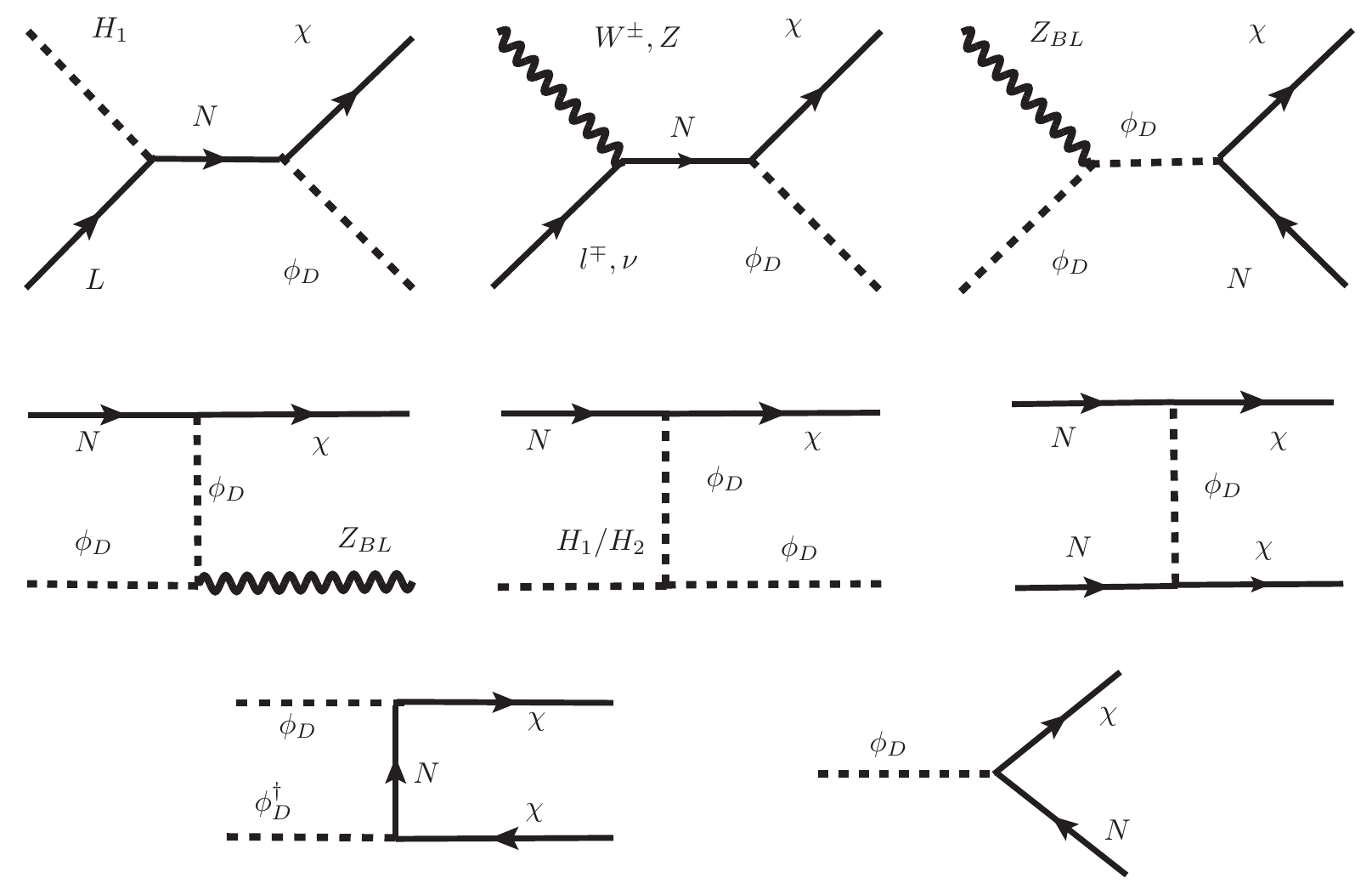}\label{fdCHI}
		\caption{Different contributions for the production of the DM $\chi$ . The $s$ and $t$ channels diagrams give negligible contributions compared to the decay process. }\label{Fig31}
	\end{center}
\end{figure}

This is to note that the particle $\chi$ has only one portal interaction $Y_{D\chi} \bar{\chi} \phi_D N$ with the dark sector field $\phi_D$ and the RHN field $N$, where $Y_{D\chi}$ is the respective coupling. We consider $Y_{D\chi}\sim \mathcal{O}(10^{-10}-10^{-12})$ to be very small, and hence, $\chi$ has feeble interaction with every other particle of this model. Due to tiny interaction, it fails to thermalise with the rest of the plasma. It was produced from the decay and annihilation of the SM and BSM particles in the early epoch. We show Feynman diagram for all possible decay and annihilation processes that contribute to the $\chi$ production in Fig.~\ref{Fig31}.
Among all these processes, the production of $\chi$ from decay processes, however, dominates. The sub-dominant contribution to the production of $\chi$ from the annihilation of SM and BSM states arises due to additional small couplings, heavy propagators, and suppression factors arising from phase space integral. 

\begin{figure}[h!]
	\begin{center}
		\includegraphics[angle=0,height=6cm,width=14cm]{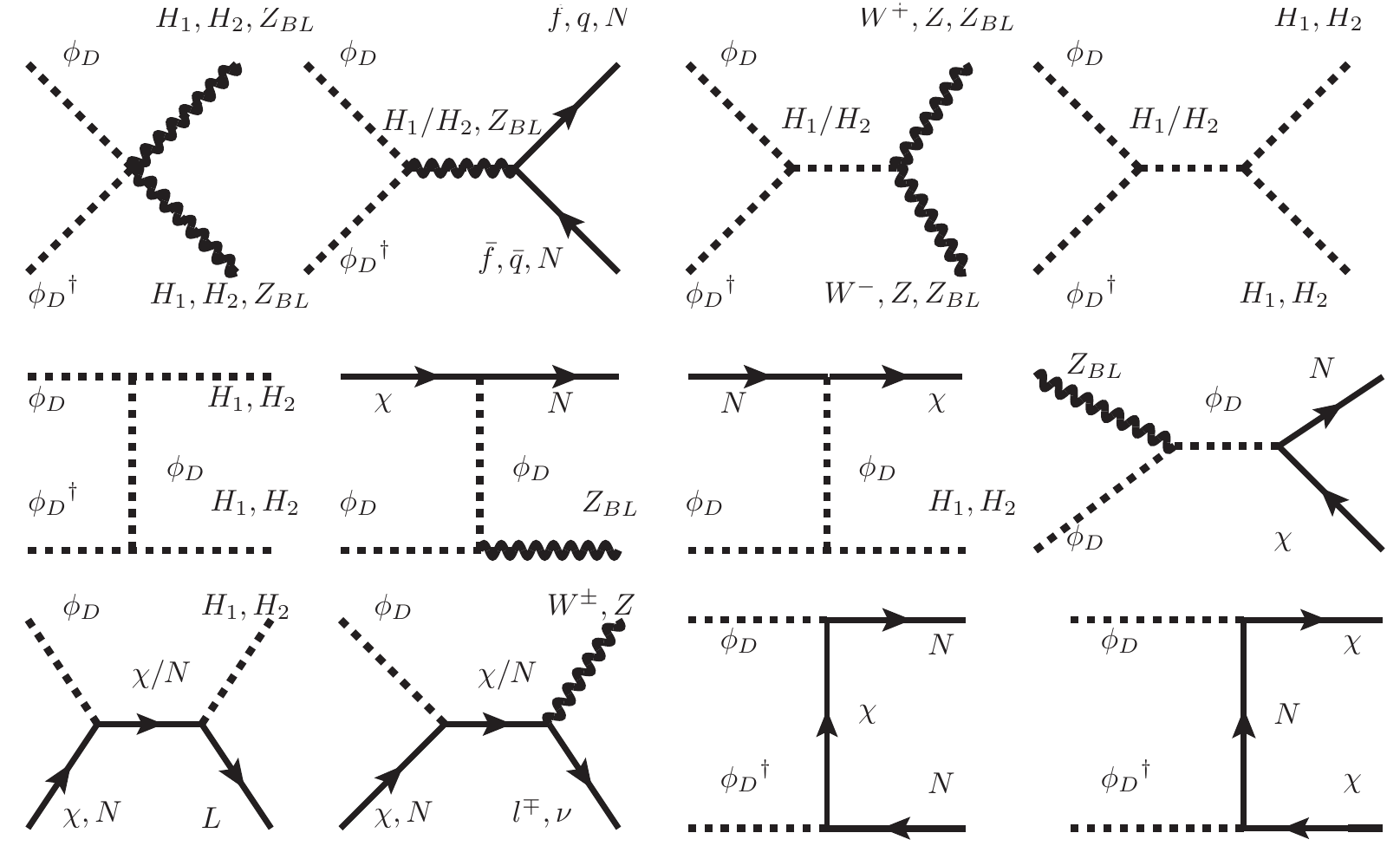}\label{fdPHID}
		\caption{ Annihilation/scattering channels of $\phi_{D }$. }\label{Fig32}
	\end{center}
\end{figure}

This work sticks to the renormalizable interaction between the dark sector, the SM, and $B-L$ particles. Therefore, the production of $\chi$ is insensitive to the reheating temperature, set by reheating/end of inflationary dynamics. The abundance of $\chi$ builds up primarily due to the $\phi_D \to \chi N$ process and increases when $T > m_{\phi_D}$. The production of $\chi$ is most significant when $T \approx m_{\phi_{D }}$. When the temperature falls below $m_{\phi_D}$, i.e., $T < m_{\phi_D}$, Boltzmann suppression of the parent state $\phi_D$ occurs and DM $\chi$ freezes in. This is referred to as thermal freeze-in production, as the parent particle $\phi_D$ during this epoch has been in thermal equilibrium. 
In addition to the standard thermal freeze-in contribution, the abundance of DM $\chi$ can further be enhanced from the out-of-equilibrium decay of $\phi_{D }$. This occurs at a later epoch $T << m_{\phi_D}$ , when $\phi_D$ is in out-of-equilibrium. The production of DM from the late decay of a state which has decoupled from the thermal plasma is referred to as the super-wimp mechanism, which has been discussed in \cite{Covi:1999ty, Feng:2003uy,Molinaro:2014lfa, Garny:2018ali,Feng:2003xh}. In our scenario, the state $\phi_D$ serves as a super-wimp candidate. 
The state $\phi_D$ having non-zero $B-L$ charges and non-zero quartic couplings $\lambda_{SD}$ and $\lambda_{Dh}$ interacts abundantly with the SM and $B-L$ particles. At an earlier epoch, $\phi_D$ hence was in thermal equilibrium with surrounding plasma, maintaining an equilibrium distribution. The non-thermal decay of $\phi_{D }$, which enhances the DM relic density, takes place at a late stage in the thermal history. The dark sector state $\phi_{D }$ tracked equilibrium abundance when the temperature of the universe was greater than its mass. The dark sector states $\phi_{D }$ abundance decreases mainly through annihilation which are efficient up until $\frac{m_{\phi_{D }}}{T}\approx25$. Feynman diagram for $\phi_{D }$ depletion through annihilation to B/SM particles is shown in Fig.{~\ref{Fig32}}. Around this temperature, the interaction rate for the annihilation/scattering of $\phi_{D }$ becomes less than the expansion rate of the universe. Hence, $\phi_{D }$ fails to scatter with surrounding plasma constituents, and it decouples from the cosmic soup. 

To compute the relic density of dark sector constituents, one needs to study the evolution of the number density of its constituents with the temperature of the universe. The evolution of the number density of the dark sector constituents is governed by the Boltzmann equations, which contain all the information of the number changing processes of the dark sector constituents. The Boltzmann equations for the evolution of $\phi_{D }$ and $\chi$ in terms of its co-moving number density $Y_{\phi_{D }/\chi}={n_{\phi_{D }/\chi}}/ s $, where $n_{\phi_{D }/\chi}$ and $s$ are the actual number density of $\phi_{D }$ and $\chi$ and entropy density of the universe are given by,\\
\small\begin{equation}
\begin{aligned}
&\frac{dY_{\phi_{D }}}{dx}=\frac{1}{x^2}\frac{s(m_{\phi_{D }})}{H(m_{\phi_{D }})}\Big[ \sum_{i,j =1}^{2} (\delta_{ij}+\frac{1}{2}|\epsilon_{ij}|){\langle\sigma v\rangle}_{\phi_{D }^{\dagger} \phi_{D } \to H_{i} H_{j}}({Y^{eq}_{\phi_{D }}}^2-{Y^{2}_{\phi_{D }}}) + {\langle\sigma v\rangle}_{\phi_{D }^{\dagger} \phi_{D } \to  W^{+}W^{-}}({Y^{eq}_{\phi_{D }}}^2-{Y^{2}_{\phi_{D }}})+\\& + {\langle\sigma v\rangle}_{\phi_{D }^{\dagger} \phi_{D } \to  ZZ}({Y^{eq}_{\phi_{D }}}^2-{Y^{2}_{\phi_{D }}})+
\sum_{\substack{X= H_{i},Z_{BL}}}{\langle\sigma v\rangle}_{\phi_{D }\chi \to N X}(Y^{eq}_{\phi_{D }}Y^{eq}_{\chi}-Y_{\phi_{D }}Y_{\chi})+\sum_{f =N,t,b}{\langle\sigma v\rangle}_{\phi_{D }^{\dagger} \phi_{D } \to  f\bar{f}}({Y^{eq}_{\phi_{D }}}^2-{Y^{2}_{\phi_{D }}})  \Big]\\&
-\frac{x}{H(m_{\phi_{D }})}\Big[ {\langle \Gamma \rangle}_{\phi_{D }\to \chi N}({Y_{\phi_{D }}}-{Y_{\chi}}
\frac{{Y^{eq}_{\phi_{D }}}}{{Y^{eq}_{\chi}}}) + 
\theta(x-x_{ew}) {\langle \Gamma \rangle}_{\phi_{D }\to \chi \nu}({Y_{\phi_{D }}}-{Y_{\chi}}
\frac{{Y^{eq}_{\phi_{D }}}}{{Y^{eq}_{\chi}}}) \Big],
\end{aligned}\label{beqnPhiD}
\end{equation}

\small\begin{equation}
\begin{aligned}
&\frac{dY_{\chi}}{dx}=\frac{1}{x^2}\frac{s(m_{\phi_{D }})}{H(m_{\phi_{D }})}\Big[\sum_{i=1}^{2}{\langle\sigma v\rangle}_{L H_{i} \to \phi_{D }\chi}({Y^{eq}_{\phi_{D }}}{Y^{eq}_{\chi}}-{Y_{\phi_{D }}}{Y_{\chi}})+{\langle\sigma v\rangle}_{W^{\pm} l^{\mp} \to \phi_{D }\chi}({Y^{eq}_{\phi_{D }}}{Y^{eq}_{\chi}}-{Y_{\phi_{D }}}{Y_{\chi}})+ \\&
+{\langle\sigma v\rangle}_{Z \nu \to \phi_{D }\chi}({Y^{eq}_{\phi_{D }}}{Y^{eq}_{\chi}}-{Y_{\phi_{D }}}{Y_{\chi}})+{\langle\sigma v\rangle}_{\phi_{D }^{\dagger} \phi_{D } \to \chi \chi}({Y^{2}_{\phi_{D }}}-{Y_{\chi}}^{2}\frac{{Y^{eq}_{\phi_{D }}}^{2}}{{Y^{eq}_{\chi}}^{2}} )+
{\langle\sigma v\rangle}_{N N \to \chi \chi}({Y^{eq}_{N}}^2-{Y_{\chi}}^{2}\frac{{Y^{eq}_{N}}^2}{{Y^{eq}_{\chi}}^{2}} ) \Big]+\\&
\frac{x}{H(m_{\phi_{D }})}\Big[  {\langle \Gamma \rangle}_{\phi_{D }\to \chi N}({Y_{\phi_{D }}}-{Y_N}{Y_{\chi}}
\frac{{Y^{eq}_{\phi_{D }}}}{{Y^{eq}_{N}}{Y^{eq}_{\chi}}})+\theta(x-x_{ew}){\langle \Gamma \rangle}_{\phi_{D }\to \chi \nu}({Y_{\phi_{D }}}-{Y_{\chi}}
\frac{{Y^{eq}_{\phi_{D }}}}{{Y^{eq}_{\chi}}}) \Big].
\end{aligned}\label{beqnChi}
\end{equation}
The entropy density and Hubble parameter in terms of $m_{\phi_{D }}$ are 
\begin{equation}
s(m_{\phi_{D }}) =\frac{2 \pi^{2}}{45} g_{*}^{s}m_{\phi_{D }}^3, \ \ \  H(m_{\phi_{D }}) = \frac{\pi}{\sqrt{90}}\frac{\sqrt{g_{*}}}{M_{pl}^{r}} m_{\phi_{D }}^{3},
\end{equation}
where $M_{pl}^{r} = 2.44 \times 10^{18}$ is the reduced Plank Mass. $g_{*}$ and $g_{*}^{s}$
are the effective degree of freedom related to the energy and entropy density of the universe, respectively at temperature $T = \frac{m_{\phi_{D }}}{x}$.
$Y_{i}^{eq}$ is the equilibrium number density of species $i$ in comoving volume and is given by,
\begin{equation}
Y_{i}^{eq} =  \frac{n_i^{eq}}{s}  =\frac{45}{4 \pi^{4}}\frac{g_{i}}{g_{*}^{s}}\Big(\frac{m_{i}}{m_{\phi_{D }}}x\Big)^2 K_{2}\Big(\frac{m_{i}}{m_{\phi_{D }}}x\Big), 
\end{equation}
with $m_i$ and $g_i$ are the mass and the internal degree of freedom for particle $i$, and $K_{2}$ the order-2 modified Bessel function of the second kind.
The thermal average width $\langle \Gamma_i \rangle $ of the species $i$ is given by,
\begin{equation}
\langle \Gamma_{i \to j k} \rangle = \Gamma_{i \to j k} \frac{K_1 \Big(\frac{ m_i}{m_{\phi_{D }}} x\Big) }{K_2 \Big(\frac{ m_i}{m_{\phi_{D }}} x\Big)}.
\end{equation}
The thermal average cross-section is given by \cite{Gondolo:1990dk},
\begin{equation}
\langle \sigma_{i j \to k l} v \rangle =\frac{x}{128 \pi^{2} m_{\phi_{D }}}\frac{1}{m_{i}^2 m_{j}^2 K_2 \Big(\frac{ m_i}{m_{\phi_{D }}} x\Big) K_2 \Big(\frac{ m_j}{m_{\phi_{D }}} x\Big) } \int_{\big(m_i + m_j\big)^2}^{\infty}ds \frac{p_{ij} p_{kl} K_1\Big(\frac{\sqrt{s}}{m_{\phi_{D }}} x\Big)}{\sqrt{s}}\int \bar{|M|}^{2}d\Omega,
\end{equation}
where $s$ is the centre of mass energy, and $p_{ij}(p_{kl})$ are initial(final) centre of mass momentum.
Finally, the relic density of the DM state $\chi$ is given by,
\begin{equation}
\Omega_{\chi} h^{2}  = \Omega^{TFI}_{\chi} h^{2} + \frac{m_{\chi}}{m_{\phi_{D }}} \Omega^{FO}_{\phi_{D }} h^{2},
\label{eq:a1}
\end{equation} 
where $  \Omega^{TFI}_{\chi} h^{2}$ is the relic density obtained from the thermal freeze-in mechanism and $\Omega^{FO}_{\phi_{D }} h^{2}$ is the abundance of $\phi_D$ at the decoupling epoch $T_d$. In the above, the second term represents the super-wimp contribution to the DM relic abundance, which occurs due to the late decay of $\phi_D$.
The analytical expression for $  \Omega^{TFI}_{\chi} h^{2}$ for the production of DM $\chi$ through the decay of $\phi_{D }$ is given by,
\begin{equation}
\Omega^{TFI}_{\chi} h^{2}\simeq\frac{1.09\times10^{27}}{g_{*}^{s}\sqrt{g_{*}}}m_{\chi}\frac{g_{\phi_{D }}\Gamma_{\phi_{D } \to \chi N}}{m_{\phi_{D }}^{2}}
\end{equation}  
where $g_{\phi_{D }}$ is the internal degree of freedom for $\phi_{D }$. For the analysis, we consider the following mass spectra, $M_N=50$ GeV, $M_{Z_{BL}}=7$ TeV, $M_{H_2}=500$ GeV, $m_{\phi_{D }}= 100$ GeV (unless mentioned otherwise), and the gauge coupling $g_{BL}=0.9$. The choice of $M_{Z_{BL}}$ and $g_{BL}$ is consistent with the constraint from CMS and ATLAS searches~\cite{CMS:2021ctt,ATLAS:2019erb}. The right-hand side of Eq.~\ref{beqnPhiD} takes into account all possible number changing processes of $\phi_{D }$ to B/SM states as well as its decay. 
\begin{itemize}
\item
This is to note, the depletion rate of $\phi_{D }$ via $Z_{BL}$ mediated annihilation processes, $i.e. ,$ $\phi_{D }^{\dagger} \phi_{D } \to Z_{BL} \to N N, \bar{f} f, H_2 H_2 $ is suppressed due to large $Z_{BL}$ mass. Such processes decoupled from the cosmic soup much earlier compared to the annihilation  of $\phi_{D }$ through contact interactions and processes mediated via B/SM Higgs($H_1,H_2$), $i.e. ,$ $\phi_{D }^{\dagger} \phi_{D } \to H_1/H_2 \to N N, \bar{f} f, H_2 H_2,  etc $. 
\item
The annihilation of $\phi_{D }$ through contact interactions and s-channel mediated process via B/SM Higgs keeps the $\phi_{D }$ in the thermal bath for a longer time. When the respective interaction rate becomes less than the universe's expansion rate, $\phi_D$ decouples from the thermal bath. 
\item
The depletion rate of $\phi_{D }$ via processes that are dependent on the dark sector Yukawa coupling $Y_{D\chi}$, such as $\chi \phi_{D } \to N Z_{BL}, N \phi_{D } \to \chi Z_{BL}, etc $ are highly suppressed because of small coupling strength $Y_{D\chi}$ and negligible abundance of $\chi$ at an early epoch.

\item The decay of $\phi_{D}$ through $\phi_{D} \to \chi \nu$ process happens because of the active and sterile neutrino mixing which takes place after electroweak symmetry breaking (EWSB). The Heaviside step function, $\theta(x-x_{ew})$ ensures that decrease in the number density of $\phi_{D}$ via $\phi_{D} \to \chi \nu$ happens only after EWSB. 
\end{itemize}
In Eq.~\ref{beqnChi}, the right-hand side contains all relevant processes to study the evolution of $\chi$. As discussed earlier, production of $\chi$ is dominated by the decay process, i.e., $\phi_{D } \to \chi N$ compared to the annihilation of the bath particles. Based on the primary production mechanism of $\chi$, there are two different scenarios.
\begin{itemize}
	\item
	\sn{I}: The DM is primarily produced via the thermal freeze-in mechanism. This corresponds to the case where $\phi_{D }$ stays in the thermal bath for a more extended period of time owing to more considerable coupling strength with the bath particles. This tends to reduce its number density significantly. Thus its late decay gives negligible contribution to $\chi$ number density. Therefore, the correct relic density of $\chi$ is mostly obtained from the thermal freeze-in mechanism. This is illustrated in the left panel of Fig.~\ref{Fig33}.
	\begin{figure}[h]
		\begin{center}
		
			\includegraphics[angle=0,height=6.2cm,width=7.0cm]{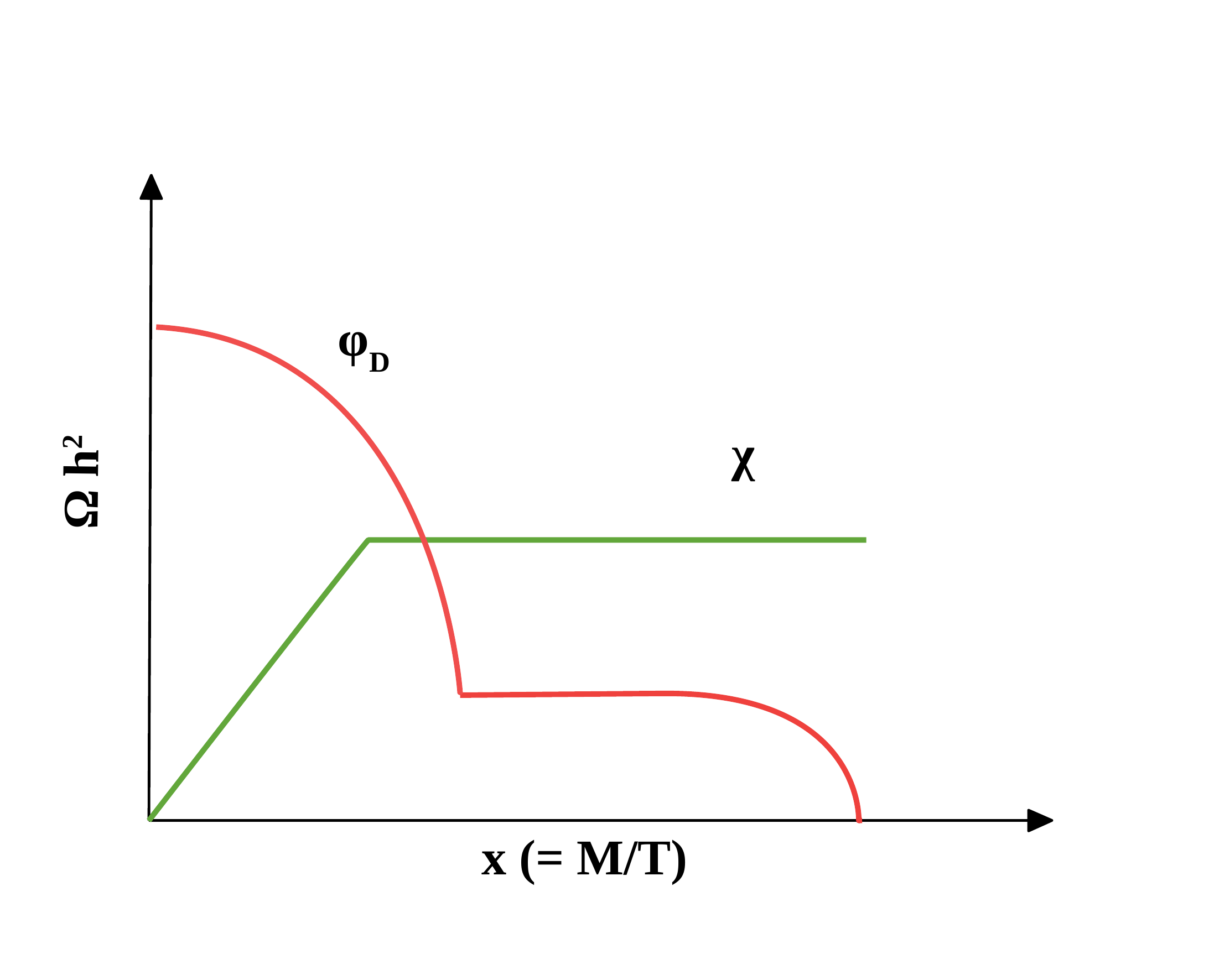}\label{mcSWIMP}
			\includegraphics[angle=0,height=6.2cm,width=7.0cm]{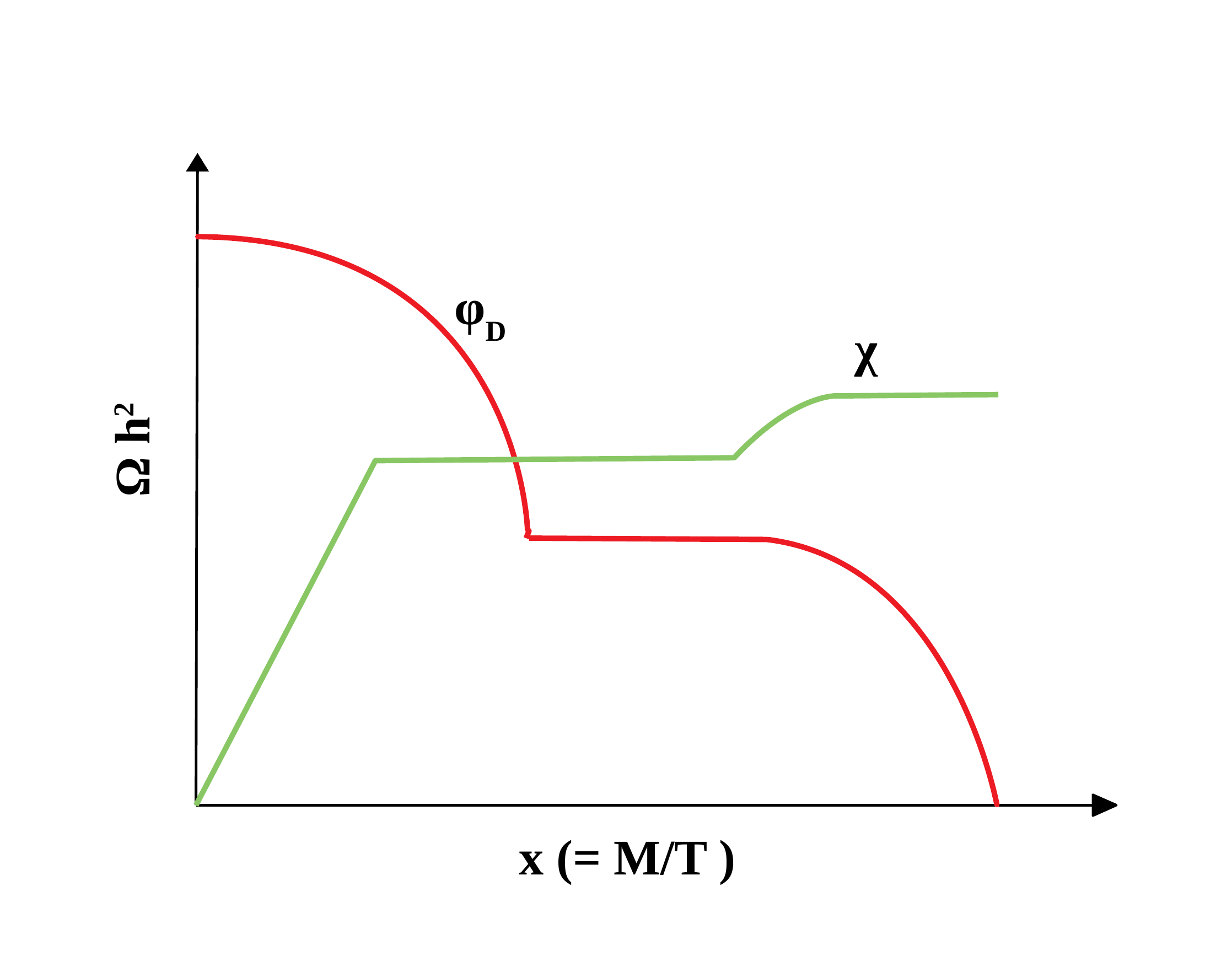}\label{mcFIMP}
			\caption{Left panel: schematic diagram representing freeze-in dominated scenario. Right panel: schematic diagram representing super-wimp dominated scenario. }\label{Fig33}
		\end{center}
	\end{figure}
	\item
	\sn{II}: The DM $\chi$ is primarily produced from the decay of $\phi_{D }$ after a freeze out, referred to as super-wimp mechanism. 
	In this scenario, the number density of $\chi$ increases gradually through the thermal freeze-in mechanism, and at a later epoch, its number density increases significantly from the late decay of $\phi_{D }$. This is to note that this scenario is possible to realise if $\phi_{D }$ decouples much earlier from the thermal bath due to suppressed interaction which leads to a large $\phi_{D }$ abundance at the time of decoupling. $\phi_D$  later decays completely to $\chi$, significantly increasing the DM number density. We illustrate this schematically in the right panel of Fig.~\ref{Fig33}.
	

\end{itemize}

\begin{figure}[h]
    	\begin{center}
    		\includegraphics[angle=0,height=6.5cm,width=8.0cm]{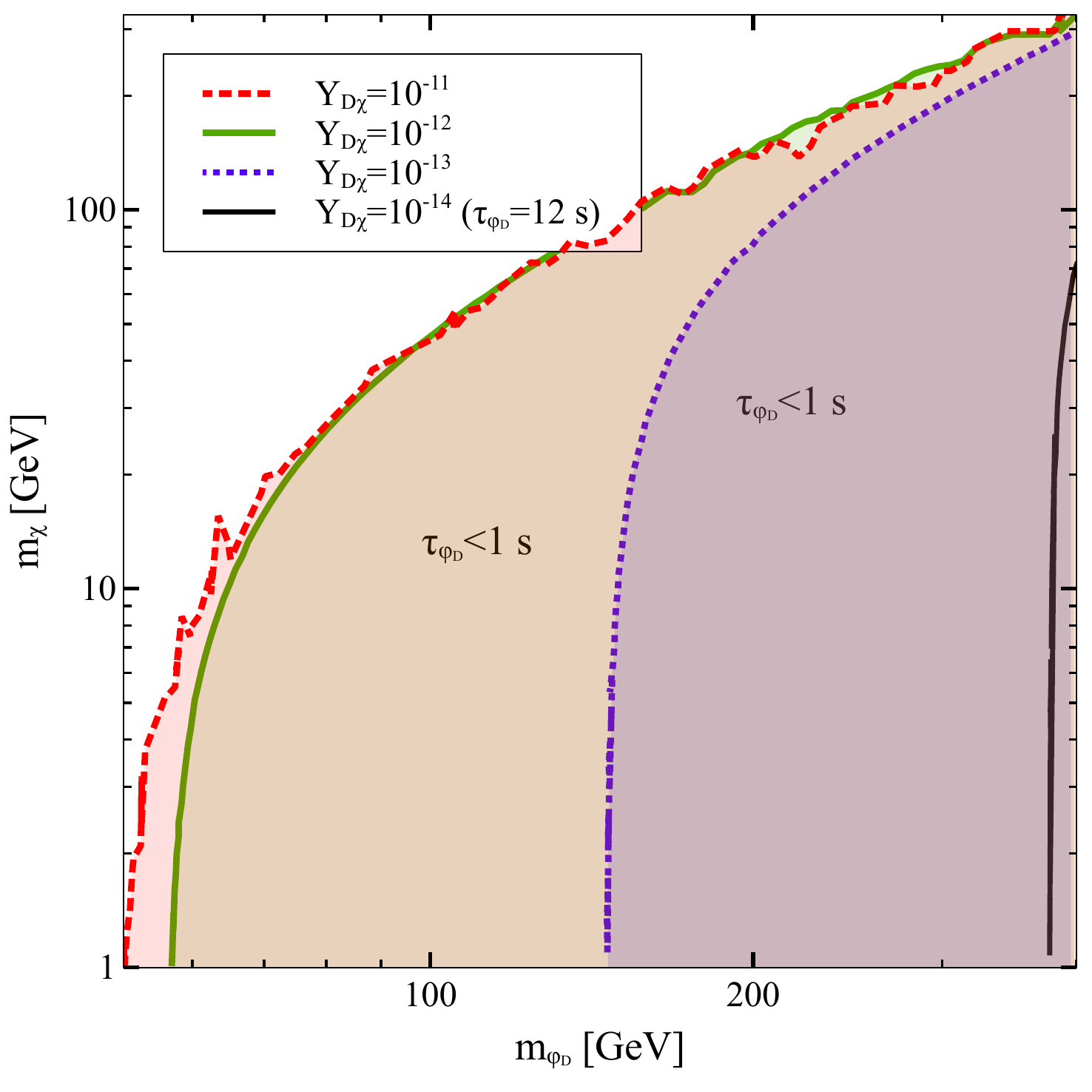}\label{bbn}
    		\caption{Lifetime contours of $\phi_{D }$ in $m_{\chi}$ and $m_{\phi_{D }}$
    			plane for fixed values of the coupling $Y_{D \chi}$. For each of the  chosen $Y_{D \chi}$, the shaded region is allowed from BBN, where lifetime of $\phi_D$ is less than 1 second.}\label{Fig35}
    	\end{center}
    \end{figure}

      
Before focusing on the main study of this work, we want to bring attention to the readers that $\phi_{D }$ must decay before the big bang nucleosynthesis (BBN). The decay of $\phi_{D }$ adds relativistic species to the thermal bath, which may alter the standard BBN scenario, and hence will be severely constrained. To avoid such conflict, we demand that lifetime of $\phi_{D }$ must be less than $ 1\ sec$ which puts a lower bound on dark sector Yukawa coupling $Y_{D\chi}$, via which $\phi_D$ decays. In Fig.~\ref{Fig35}, we show the lifetime contour of the $\phi_D$ state for each of the given Yukawa coupling $Y_{D\chi}$, where we vary the mass $m_{\chi}$ of the DM and the mass of the parent particle $m_{\phi_{D }}$. For each of the chosen $Y_{D\chi}$ values, the shaded region is allowed from BBN. We note that one of the products of the $\phi_{D }$ decay is $N$, where $N$ further decays to the SM particles and adds relativistic species to the thermal bath. The decay length of RHN depends on the active-sterile neutrino mixing, which depends on the light neutrino masses and PMNS mixing angles. We provide the expressions for the decay width of $\phi_D$ and $N$ in the appendix. For $M_{N} > m_{W^{\pm}}, m_{Z} $, $N$ decays dominantly through 2 body processes, such as, $N \to l^{\pm} W^{\mp}$ and $N \to Z \nu$. For $M_{N} < m_{W^{\pm}}, m_{Z} $ which we consider in this study, RHN decays to the three SM fermions through off-shell $W$, and $Z$ gauge bosons. For our choice of mass parameter $M_N$, we have checked that the decay of $N$ to SM states takes place instantaneously and is not constrained from BBN. 

One can additionally set  a upper bound on dark sector Yukawa coupling $Y_{D\chi}$ by demanding that 
$\phi_{D }$ decays to $\chi$ after $\phi_{D }$ freezes-out (at $T \sim m_{\phi_{D }}/25$). This requirement implies,
\begin{eqnarray}
\Gamma_{\phi_{D } \to \chi N} \le H (m_{\phi_{D}} )  \xRightarrow{m_{\phi_{D }} = 100 GeV  } Y_{D\chi} \le 10^{-8}
\label{eq:eqsecond}
\end{eqnarray}  
In this work we consider such values of $Y_{D\chi}$, so that both the BBN and the above mentioned constrained are satisfied. The entropy of the universe increases after the decay of $\phi_{D }$. The increase in entropy  can be approximated \cite{Cheng:2020gut} as,
\begin{eqnarray}
\frac{\Delta s}{s}\approx \frac{{n_{\phi_{D}}(m_{\phi_{D }}-m_{\chi})}}{sT}\approx \frac{{Y_{\phi_{D}}(m_{\phi_{D }}-m_{\chi})}}{T}
\end{eqnarray}
Due  to the small value of $Y_{D\chi}$, $\phi_{D }$ decays in the late epoch of the universe when the universe
temperature is around MeV. This leads to negligible amount of increase in entropy density in the universe, otherwise such  increase in entropy can dilute the DM produced in an early epoch.  \\
\begin{figure}[]
	\begin{center}
		\mbox{\subfigure[]{\includegraphics[angle=0,height=5.5cm,width=7.5cm]{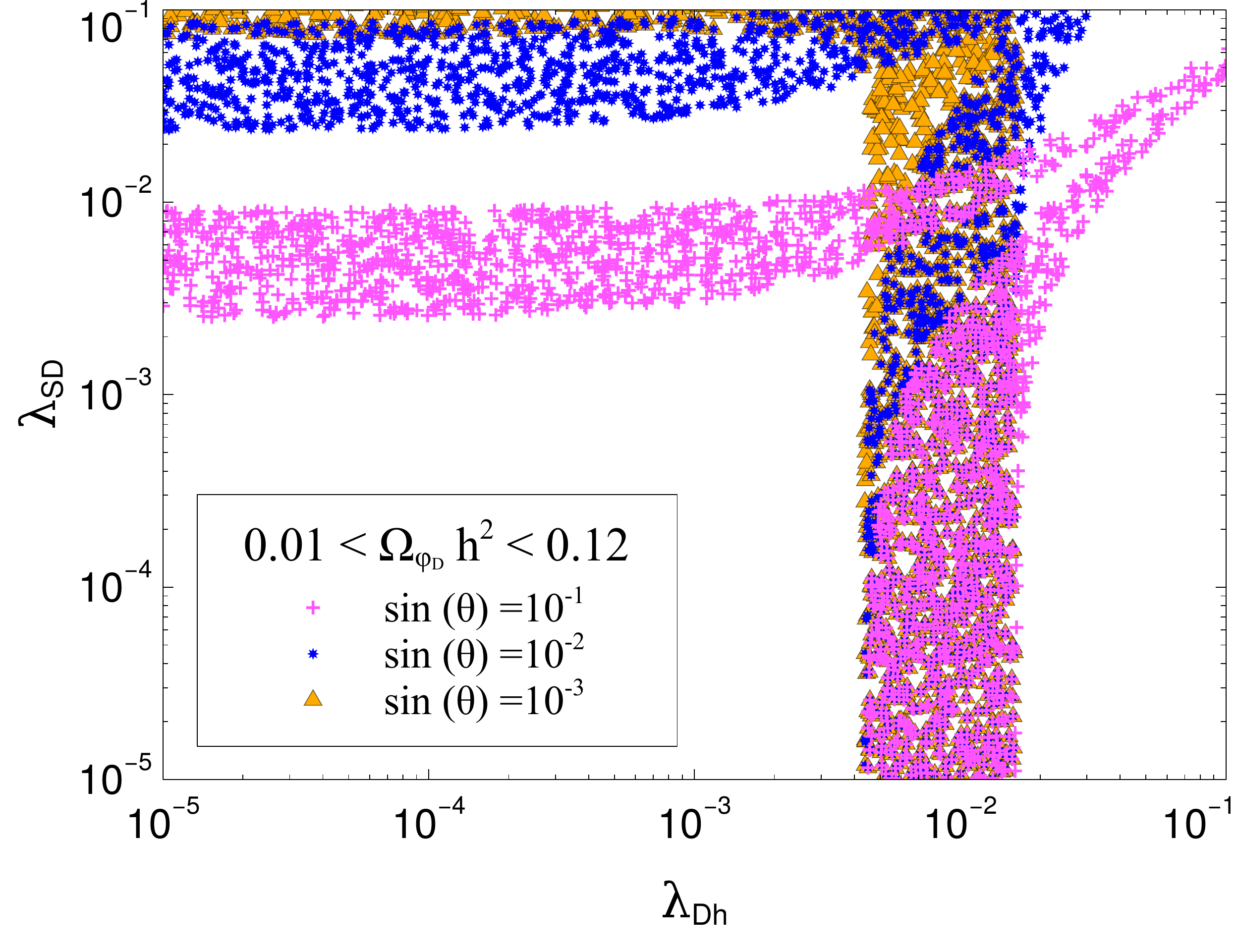}\label{ydvsmp1}}
			\subfigure[]{\includegraphics[angle=0,height=5.5cm,width=7.5cm]{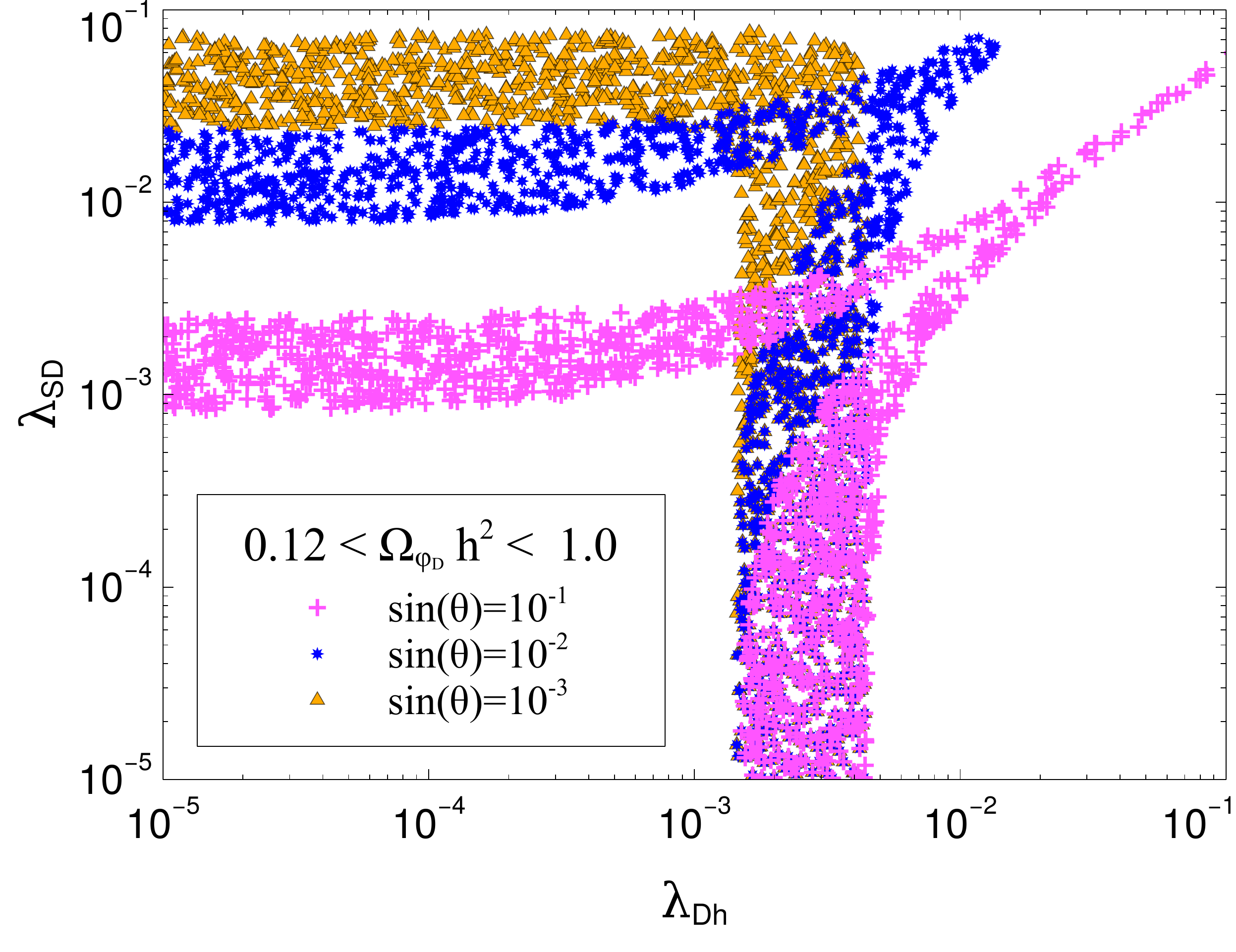}\label{ydvsmp2}}}
		\subfigure[]{\includegraphics[angle=0,height=5.5cm,width=7.5cm]{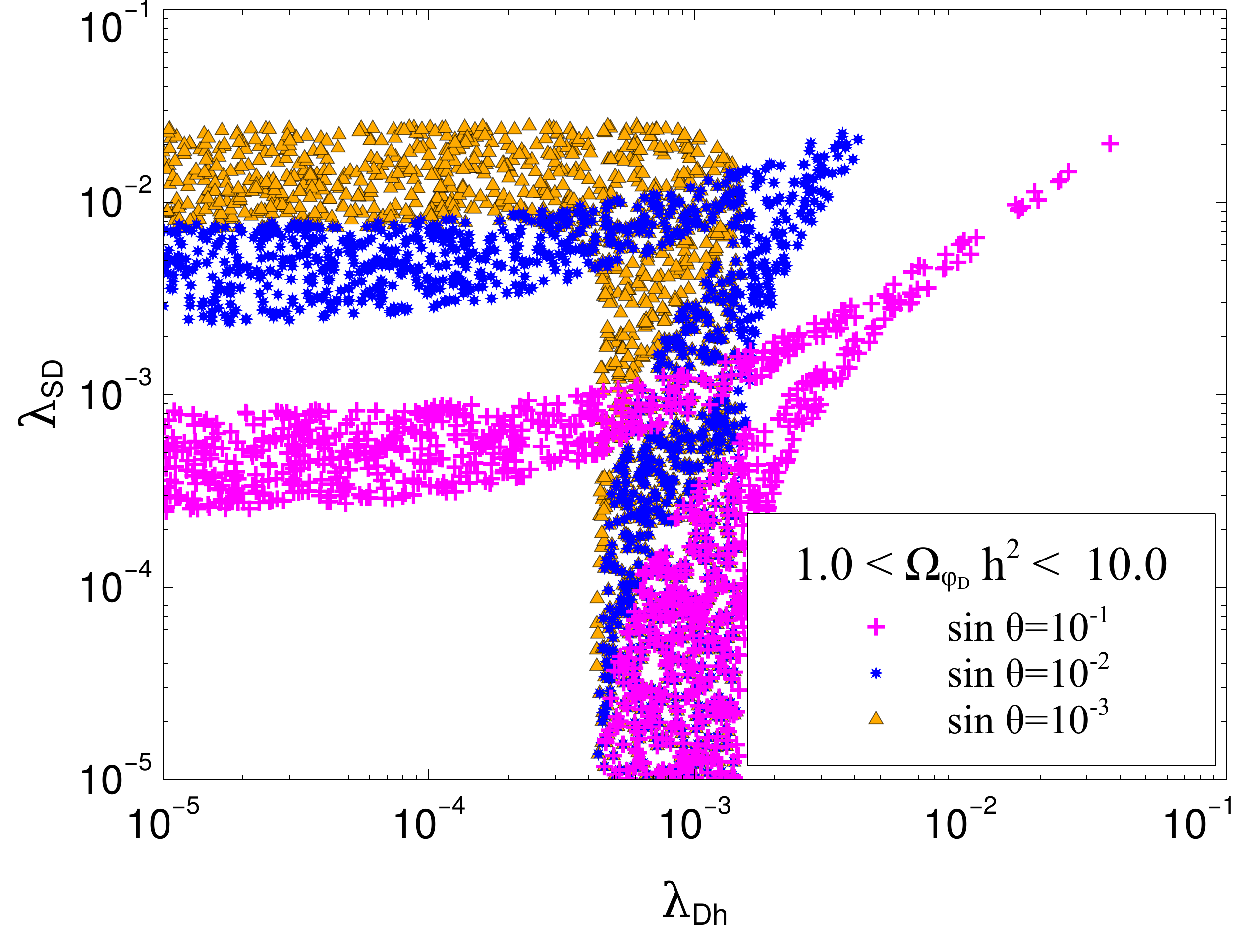}\label{ydvsmp3}}
		\caption{ We show the scatter plot in $\lambda_{ Dh}$-$\lambda_{ SD}$ plane for the different values of $\sin\theta$ and demanding  $\Omega_{\phi_{D }}^{FO}h^{2}$ in the range 0.01 to 0.12 in Fig.~\ref{ydvsmp1}. Similarly, demanding $\Omega_{\phi_{D }}^{FO}h^{2}$ in the range 0.12 to 1.0 shown in Fig.~\ref{ydvsmp2} and for $\Omega_{\phi_{D }}^{FO}h^{2}$ greater than 1.0 shown in Fig.~\ref{ydvsmp3}.}\label{Fig36}
	\end{center}
\end{figure}
\subsection{$\phi_D$ abundance} 
\begin{figure}[h]
	\begin{center}
		\mbox{\subfigure[]{\includegraphics[angle=0,height=5.5cm,width=7.5cm]{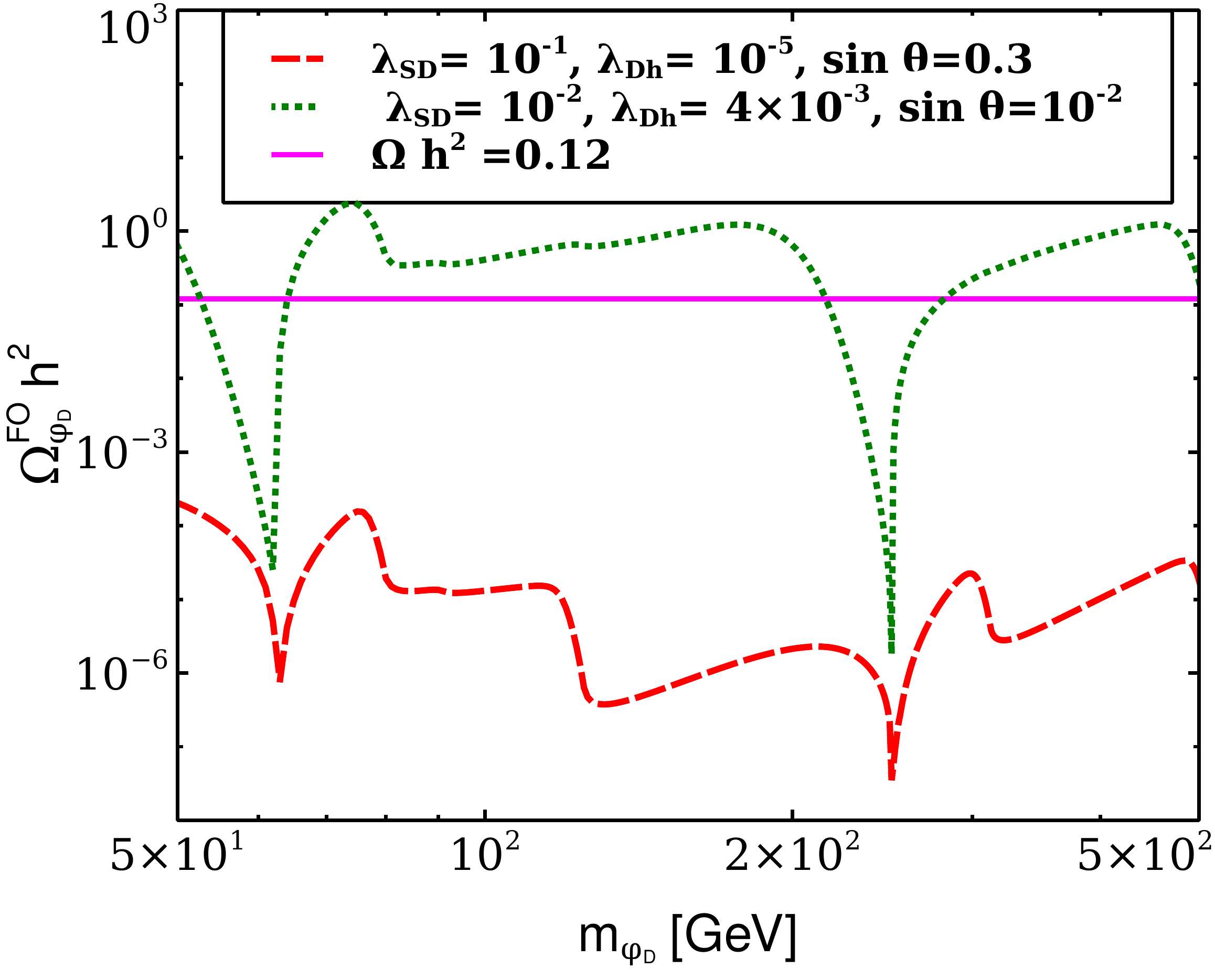}\label{phirelic}}
			\subfigure[]{\includegraphics[angle=0,height=5.5cm,width=7.5cm]{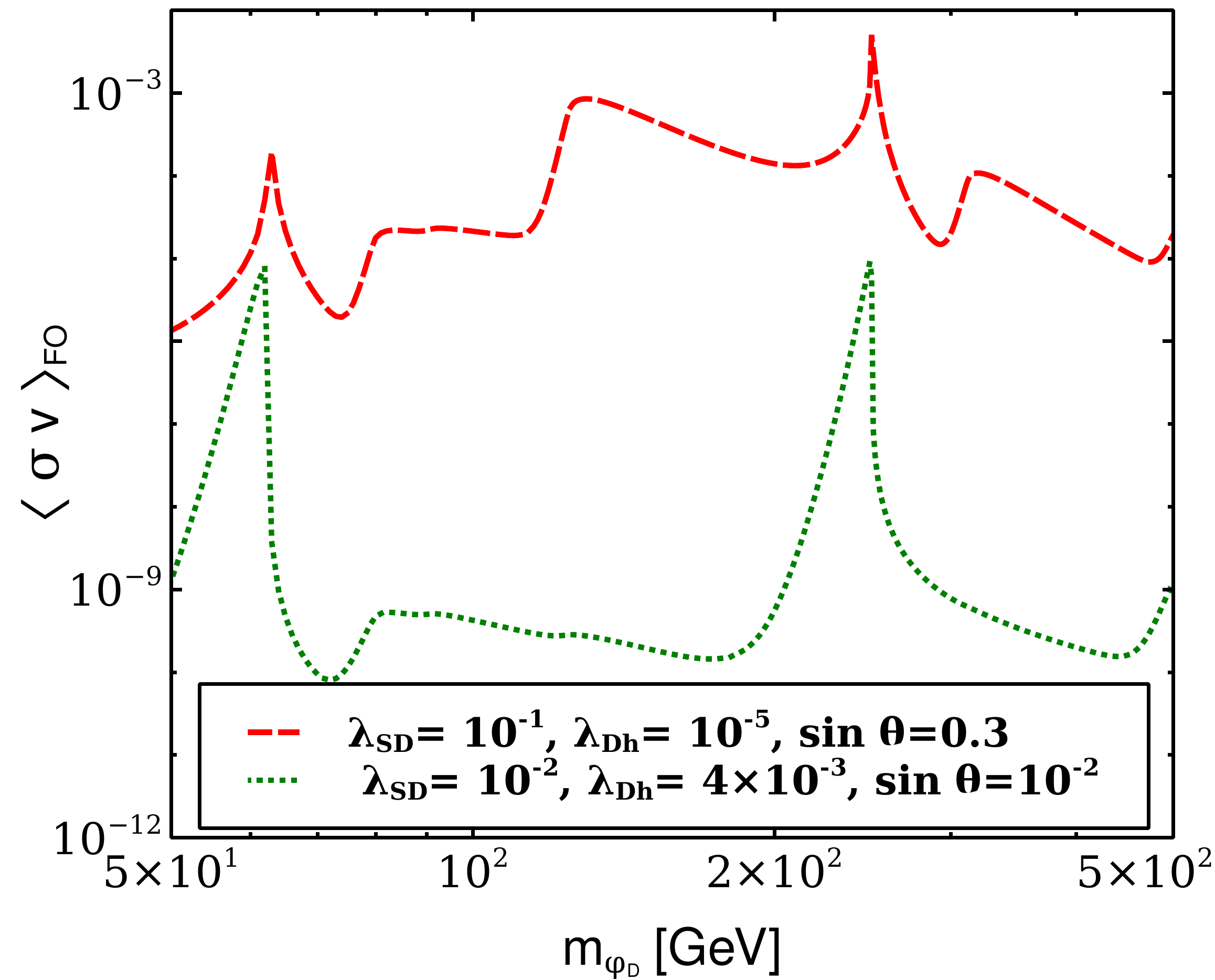}\label{phixsec}}}
			\subfigure[]{\includegraphics[angle=0,height=5.5cm,width=7.5cm]{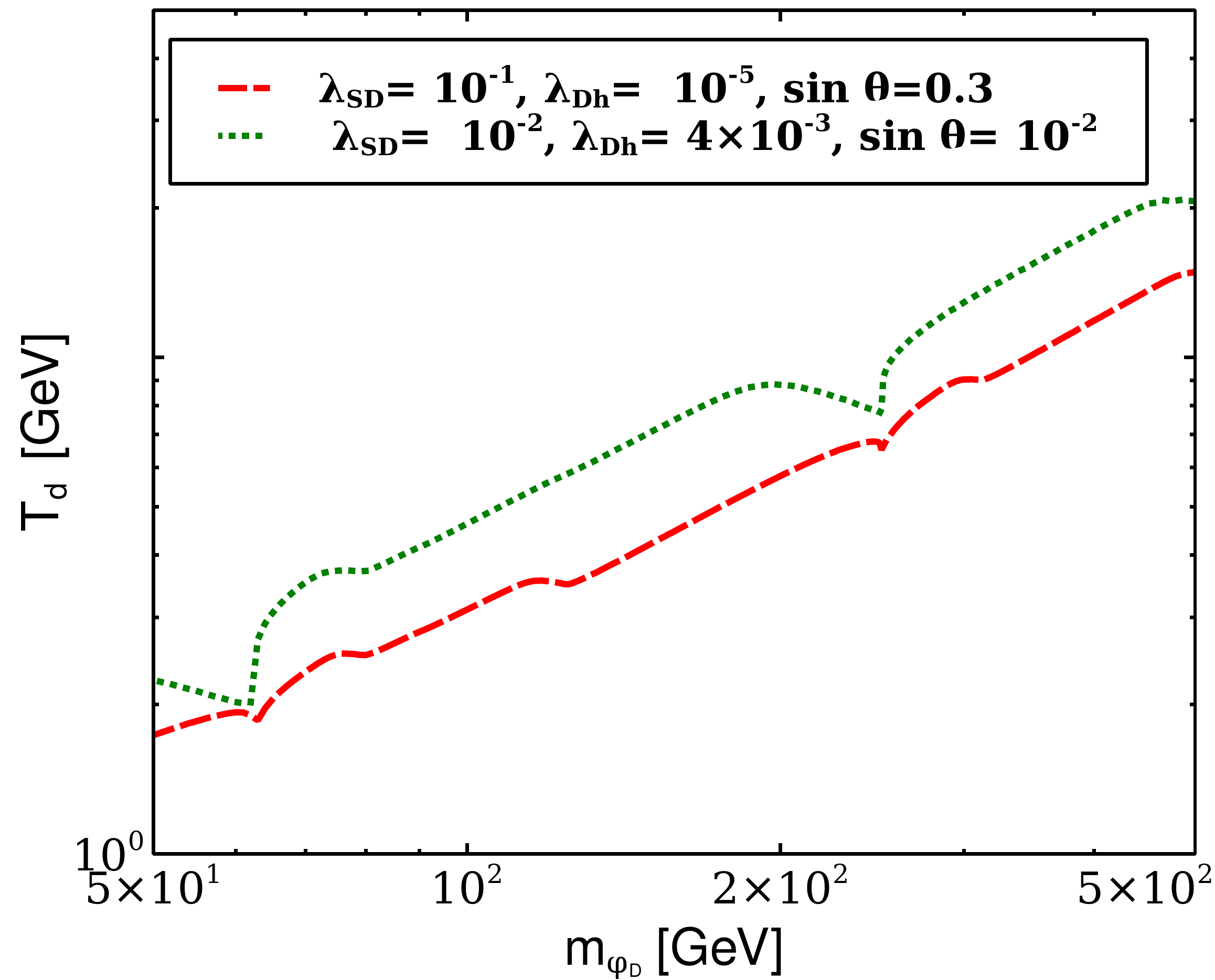}\label{phitf}}
		\caption{Fig.~\ref{phirelic}, and Fig.~\ref{phixsec} represent the variation of $\Omega_{\phi_D}^{FO} h^2$, and $\langle \sigma v \rangle_{FO}$ with mass of $\phi_D$,   and Fig.~\ref{phitf} represent the variation of the freeze-out temperature of  $\phi_D$ w.r.t its mass. 
\label{Fig37} }
	\end{center}
\end{figure}

Since a significantly large number of $\chi$ is produced from the late decay of $\phi_D$; therefore the abundance of $\phi_D$ at the time of decoupling plays a vital role in determining the correct DM relic density. A large abundance of $\phi_D$ will contribute significantly in the $\chi$ production via $\phi_D \to \chi N$ late decay. As discussed before, $\phi_D$ was in equilibrium with the SM particles because of the large scalar quartic couplings $\lambda_{SD}, \lambda_{Dh}$ and SM-BSM Higgs mixing angle $\sin\theta$. The most dominant annihilation channels for $\phi_D$, $ \phi^{\dagger}_D \phi_D \to W^+ W^-, ZZ, NN$ mediated via SM-like Higgs state $H_1$, for which we provide the analytic expressions of the cross-section in the appendix, and show the variation of thermal average cross-section $\langle \sigma v \rangle_{FO} $ at $T_d$ in Fig.~\ref{phixsec}. The $H_2$ mediated contribution is relatively smaller due to heavy propagator suppression, except the region of the $H_2$ resonance. In Fig.~\ref{Fig36}, we show the scatter plots in the $\lambda_{ SD}$ and $\lambda_{ Dh}$ plane for the three different values of mixing angle $sin\, \theta$, for which $\Omega^{FO}_{\phi_{D }} h^{2}$ varies in between a) $0.01<\Omega^{FO}_{\phi_{D }} h^{2}<0.12$ (for the top left panel plot), b) $0.12<\Omega^{FO}_{\phi_{D }} h^{2}<1$ (for the top right panel plot), and c) $1.0<\Omega^{FO}_{\phi_{D }} h^{2}<10.0$ (for the bottom plot). The $\phi_D$ abundance at $T_{d}$ is mostly governed by the couplings $\lambda_{SD}$, $\lambda_{Dh}$ and $\sin \theta$. Comparing the { horizontal and vertical bands between different panels}, for a fixed value of $\sin \theta$, decreasing  $\lambda_{ SD}$ and $\lambda_{ Dh}$ will lead to a higher $\Omega_{\phi_{D }}^{FO}h^{2}$. This typically occurs, as with the decrease in the relevant quartic coupling, the interaction rate of $\phi_D$ decreases, resulting in an early freeze-out of $\phi_D$, which subsequently gives a larger $\phi_D$ abundance. On the other hand, in each of these three panels, for a fixed value of $\lambda_{Dh}$, as we decrease $\sin \theta$ from 0.1 to 0.01 and further, a larger $\lambda_{SD}$ coupling is required to compensate the decrease in the interaction strength and to maintain $\Omega^{FO}_{\phi_{D }} h^{2}$   in the given range. The cone-shaped region in each of the plots represents a cancellation in the $\phi_{D }^{\dagger}\phi_{D}H_1$ vertex that we will discuss later. Due to a relative suppression in the vertex factor, the interaction rate decreases, thereby leading to a higher value of $\Omega^{FO}_{\phi_{D }} h^{2}$. To maintain $\Omega^{FO}_{\phi_{D }} h^{2}$ in the given range, hence a larger value of couplings $\lambda_{SD}$ and $\lambda_{Dh}$, and a large thermal averaged cross-section $\langle \sigma v \rangle$ are required.  
To further explore the effect of an early and late decoupling of $\phi_D$ on DM number density, we consider two benchmark points which are as follows,
\begin{enumerate}
\item $\lambda_{ SD}=10^{-1},\  \sin\theta=0.3,\  \lambda_{ Dh}=10^{-5} $.
\item $\lambda_{ SD}=10^{-2},\  \sin \theta=10^{-2},\  \lambda_{ Dh}=4 \times10^{-3} $.
\end{enumerate}

In Fig.~\ref{phirelic}, we show the variation of $\phi_D$ abundance at $T_d$ with the change in $\phi_{D }$ mass for these two above mentioned benchmark points. The red line in Fig.~\ref{phirelic} corresponds to our first benchmark point, where $\phi_{D }$ stays in thermal equilibrium for a longer period due to a large interaction rate with the bath particles {which happen because of a large $\sin \theta$ and $\lambda_{ SD}$}. In Fig.~\ref{phixsec}, we show the variation of thermal average cross-section $\langle \sigma v\rangle_{FO}$ at $T_{d}$ with the mass of $\phi_{D }$. As we can see there are few sudden increases in $\langle \sigma v\rangle_{FO}$ w.r.t $m_{\phi_D}$. When the mass of $\phi_{D }$ becomes half of the mass of SM-like Higgs $H_1$, $s$ channel resonance mediated via $H_1$ takes place, and $\langle \sigma v\rangle_{FO}$ increases significantly. After which, it decreases with the increase in mass of $\phi_{D }$. For $m_{\phi_{D }}\approx 80.4\, \textrm{GeV}$, thermal average cross-section $\langle \sigma v\rangle_{FO}$ further increases. This occurs as the channel $\phi_{D }^{\dagger}\phi_{D }\to W^{+}W^{-}$ opens up. Similar increase in $\langle \sigma v\rangle_{FO}$ occurs when $m_{\phi_{D }}\approx m_{H_1}$, it is when $\phi_{D }^{\dagger}\phi_{D } \to H_1 H_1$ opens up. Thereafter $\langle \sigma v\rangle_{FO}$ decreases with the increase in mass of $\phi_{D }$ except around $m_{\phi_{D }}\approx 250 \, \textrm{GeV}$. It is where $s$ channel resonance mediated via $H_2$ occurs which enhances $\langle \sigma v\rangle_{FO}$ significantly. For thermal dark sector particle $\phi_D$, the abundance $\Omega_{\phi_D}^{FO}h^{2}$ is inversely proportional to $\langle \sigma v\rangle_{FO}$, which is evident from the figure. In each of these three panels, the green line corresponds to the second benchmark point, for which interaction of $\phi_{D }$ is suppressed owing to a smaller coupling $ \lambda_{ SD}$ and mixing angle $\sin \theta$. The variation of $\Omega_{\phi_{D }}^{FO} h^2$ and $\langle \sigma v\rangle_{FO}$ with the change in $\phi_{D }$ mass is similar to the first benchmark point. It is important to note that due to suppressed interaction, $\phi_{D }$ decouples from the thermal bath much earlier, leading to a large relic density of $\phi_{D }$. This is also reflected in Fig.~\ref{phitf}, which shows the variation of the freeze-out temperature of $\phi_{D }$ with its mass for these two benchmark points. As it is evident, the freeze-out temperature is relatively smaller for the first benchmark point, and hence freeze-out of $\phi_D$ occurs at a later epoch. The  sudden dip in the freeze-out temperature around $\phi_{D } \sim 60 $ GeV and $250 $ GeV occur because of $s$-channel resonance mediated via $H_1$ and $H_2$. As the $\langle \sigma v \rangle_{FO}$ increases significantly in this region, this enables $\phi_{D }$ to remain in a thermal bath for a long time. The first benchmark point corresponds to {\it Scenario-I}, and the second benchmark point corresponds to {\it Scenario-II}, discussed earlier. 

\begin{table}[h!]
	\begin{center}
		\vskip 0.05cm
		\begin{tabular} {|c|c|}
		
			\hline
			Vertex & Vertex Factor\\
			\hline
			 $\phi^\dagger_{D}\,\phi_{D}\,{\zbl}_{\mu}$
			&  $\lambda_{Z_{BL}} =\gbl (p_2-p_1)^\mu$\\
			\hline
		    $\phi_{D}^\dagger\,\phi_{D}\,H_1$ & $\lambda_{H_1}=\,(\lambda_{Dh} v
			\cos \theta - \lambda_{SD} v_{BL} \sin \theta)$\\
			\hline
			 $\phi_{D}^\dagger\,\phi_{D}\,H_2$ & $\lambda_{H_2}=-\,(\lambda_{Dh} v
			\sin \theta + \lambda_{SD}v_{BL} \cos \theta)$ \\
			\hline
		
		\end{tabular}
	\end{center}
	\caption{Couplings of $\phi_{D}$ with $\zbl$, $H_1$ and $H_2$.}
	\label{vertex_table}
\end{table}
\begin{table}[h]	
	\begin{center}
		\vskip 0.05cm
		\begin{tabular} {|c|c|}
			\hline
			Vertex & Vertex Factor\\
			\hline
			$H_{1}VV (V=\ W,Z)$ & $\,2 m_{V}^2 cos \theta / v$ \\
			\hline
			$H_{2}VV (V=\ W,Z)$ & $\,2 m_{V}^2 sin \theta / v$ \\
			\hline
			$H_{1}ff (f=\ t,b)$ & $\,2 m_{f} cos \theta / v$ \\	
			\hline
			$H_{2}ff (f=\ t,b)$ & $\,2 m_{f} sin \theta / v$ \\
			\hline
			$H_{1}NN $ & $\,y_{N} sin \theta / \sqrt{2}$ \\	
			\hline
			$H_{2}NN $ & $\,y_{N} cos \theta / \sqrt{2}$ \\	
			\hline
		\end{tabular}
	\end{center}
\caption{Couplings of SM and BSM Higgs with SM fermions, RHN and gauge bosons.}
\label{vertex_table_Higgs}
\end{table}
\begin{figure}[]
	\begin{center}
		\mbox{\subfigure[]{\includegraphics[angle=0,height=5.50cm,width=7.50cm]{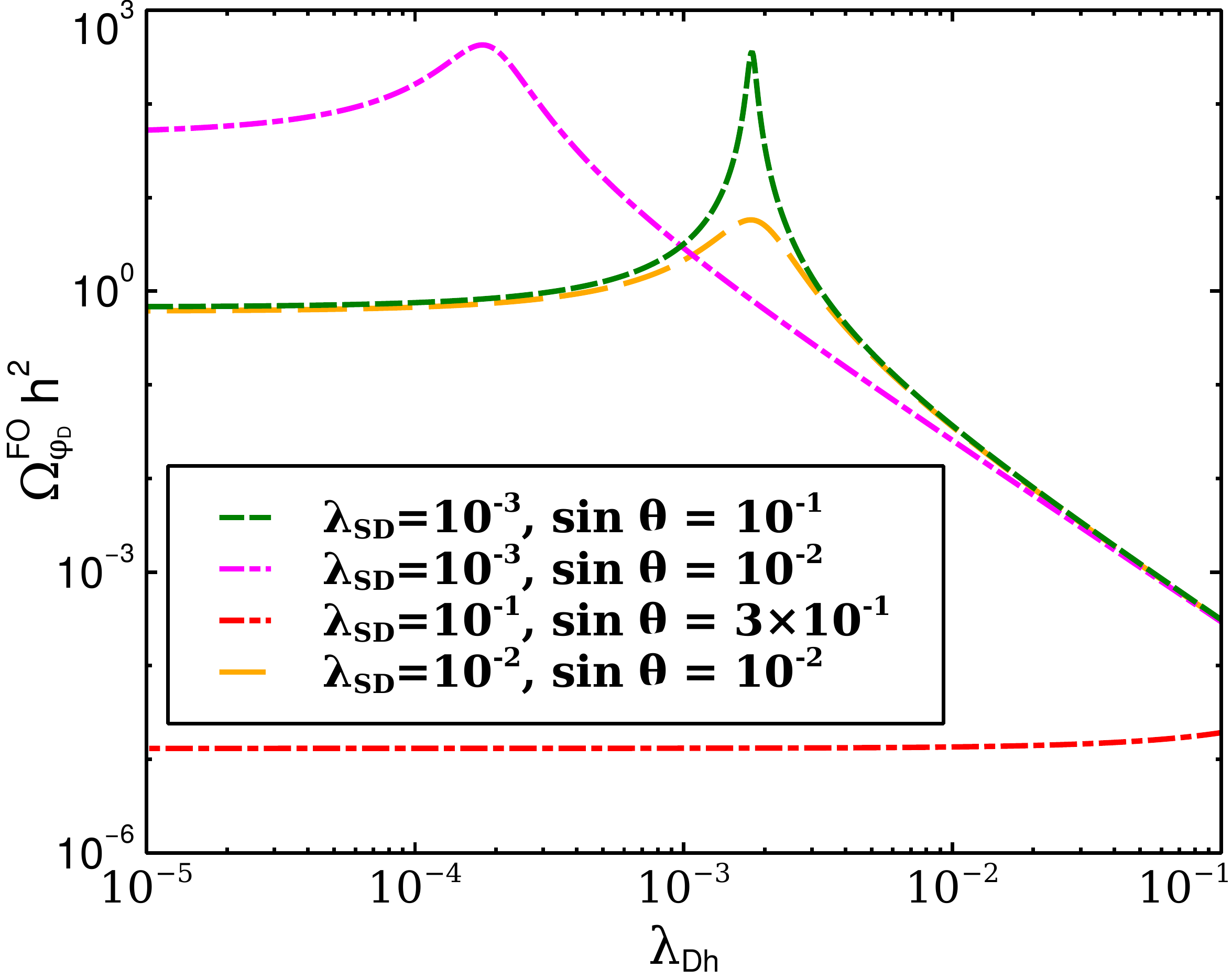}\label{pf1}}
			\subfigure[]{\includegraphics[angle=0,height=5.50cm,width=7.50cm]{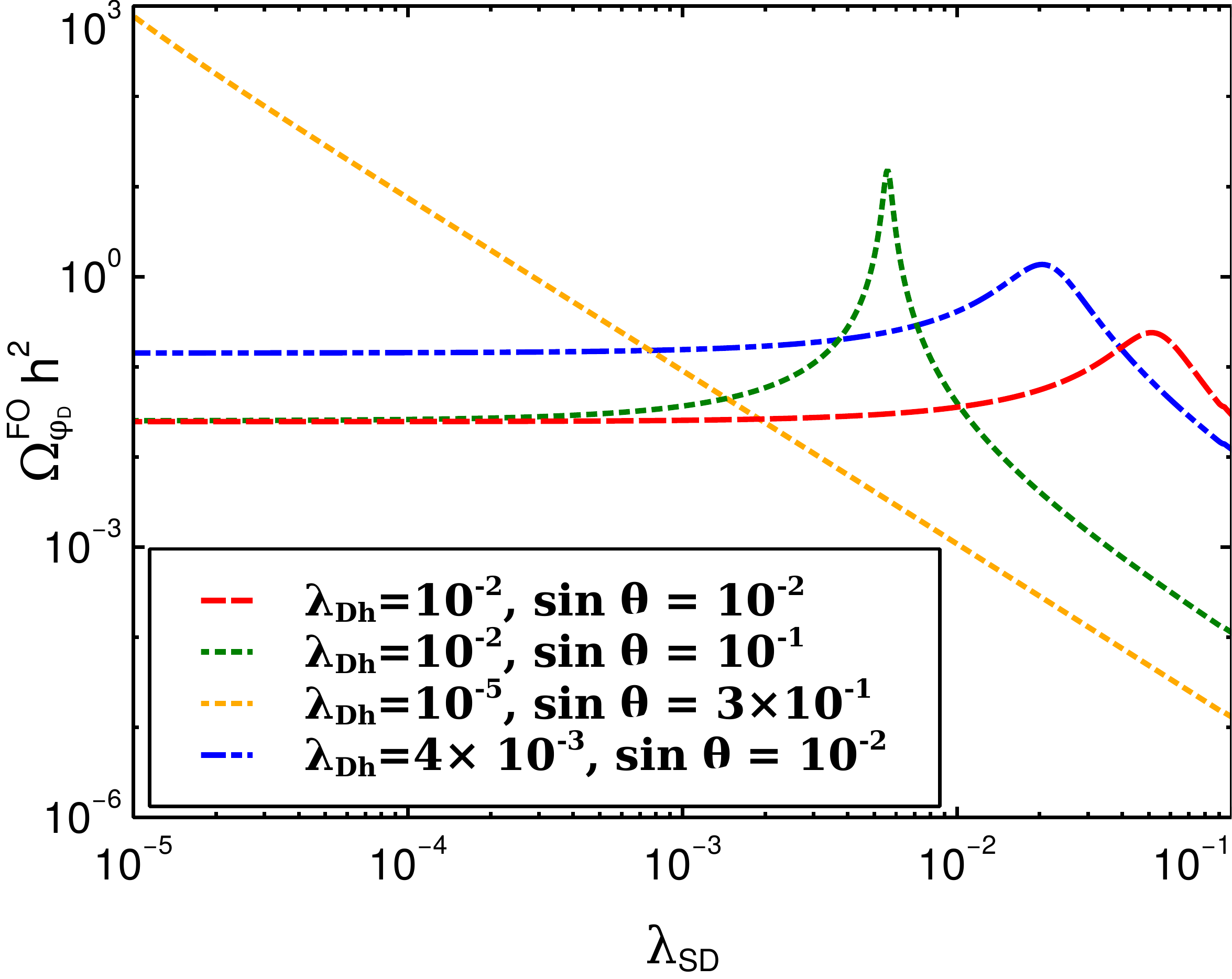}\label{pf2}}}
		\subfigure[]{\includegraphics[angle=0,height=5.50cm,width=7.50cm]{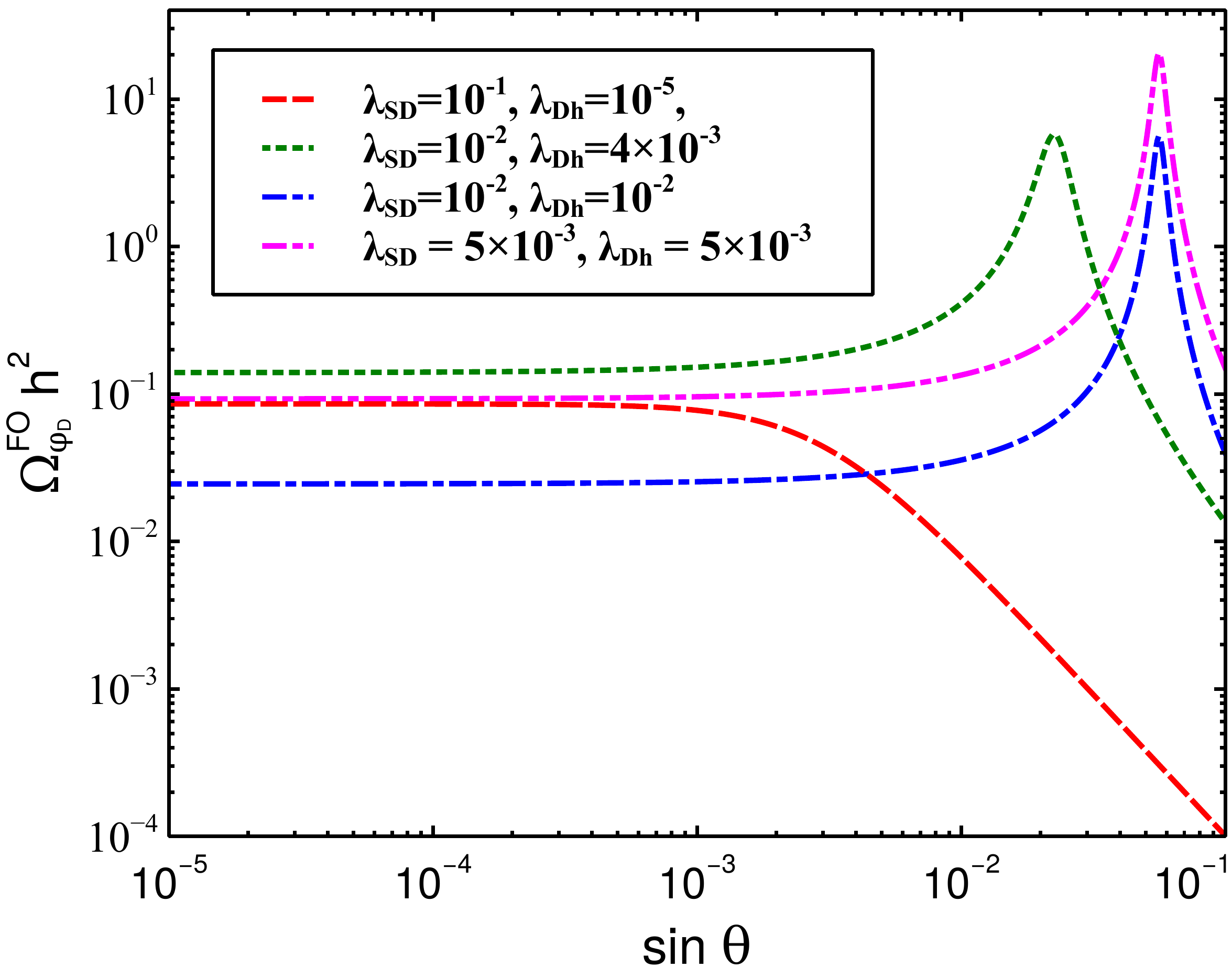}\label{pf3}}
		\caption{Fig.~\ref{pf1} and Fig.~\ref{pf2} show the variation of the relic abundance of $\phi_D$ with  the couplings $\lambda_{DH}$ and $\lambda_{SD}$, respectively. Fig.~\ref{pf3} shows the variation of relic abundance  of $\phi_{D }$ with $\sin \theta$. 
			\label{Fig38} }
	\end{center}
\end{figure}

In Fig.~\ref{pf1}, \ref{pf2} and \ref{pf3}, we show the variation of  $\phi_D$ abundance at $T_{d}$ with the parameters $\lambda_{ SD}$, $\lambda_{ Dh}$ and $\sin \theta$. We re-emphasize, the dominant annihilation mode for $\phi_{D }$ are $\phi_{D }^{\dagger}\phi_{D }\to W^{+}W^{-},ZZ,NN$ which are mediated via  $H_1$ and $H_2$ \footnote{For $\phi_{D }^{\dagger}\phi_{D }\to W^{+}W^{-},ZZ$, $H_2$ mediated process is suppressed due to small $\sin \theta$ and heavy mass of $H_2$. However, for $\phi_{D }^{\dagger}\phi_{D }\to NN$, $H_1$ mediated process is suppressed due to $\sin \theta$. }.
The other process, such as  $\phi_{D }^{\dagger}\phi_{D }\to \bar{f}f$ contribute negligibly in $\phi_{D }$ annihilation, as they are suppressed by the small mass of the final state fermions. In Fig.~\ref{pf1}, we show the variation of $\Omega^{FO}_{\phi_{D }}h^{2}$ with  $\lambda_{ Dh}$ while keeping $\lambda_{SD}$ and $\sin \theta$ fixed to few sets of values. As we can see  from the green line which corresponds to $\lambda_{ SD}=10^{-3}$ and $\sin\theta = 10^{-1}$, $\Omega^{FO}_{\phi_{D }}h^{2}$ remains independent of $\lambda_{ Dh}$ in between  $10^{-5}$ to $10^{-4}$.  This occurs as the effective vertex factor involving $\phi_{D }^{\dagger}\phi_{D}H_1$ for such a small value of $\lambda_{Dh}$ is governed by $\lambda_{ SD}$ and $\sin \theta$ rather than $\lambda_{ Dh}$. This can be understood from the expression of the vertex factors, which we provide in Table.~\ref{vertex_table}. We also provide the vertex factors of different $H_{1,2}$ interactions with the SM fermions and gauge bosons  in Table.~\ref{vertex_table_Higgs}.  In between  $10^{-5} < \lambda_{ Dh}<10^{-4}$,  dominant annihilation modes are $\phi_{D }^{\dagger}\phi_{D } \to W^{+}W^{-}, ZZ$ mediated by SM Higgs boson $H_1$. As $\lambda_{ Dh}$ increases, effective vertex factor involving $\phi_{D }\phi_{D}^{\dagger}H_1$ decreases due to a relative cancellation between different terms in the respective expression (see Table.~\ref{vertex_table}),   leading to a suppressed  annihilation rate. This results in an increase in the $\Omega^{FO}_{\phi_{D }}h^{2}$ as $\lambda_{ Dh}$ increases from $\lambda_{Dh}\approx10^{-4}$ to $\lambda_{Dh}\approx 2\times10^{-3}$. This is to note that in this region annihilation rate $\phi_{D }^{\dagger}\phi_{D }\to NN$ contributes more than  $\phi_{D }^{\dagger}\phi_{D }\to W^{+}W^{-},ZZ$.  This occurs as the process $\phi_{D }^{\dagger}\phi_{D }\to N N$ mediated by  $H_2$ dominates as $\lambda_{H_2}$ becomes larger than $\lambda_{H_1}$ and also the process is not suppressed by scalar mixing angle. As  $\lambda_{ Dh}$ further increases,  $\lambda_{H_1}$ increases which  makes  the annihilation rate of $\phi_{D }$ large. This results in a decrease  in $\Omega^{FO}_{\phi_{D }}h^{2}$ as $\lambda_{Dh}$ in between  $2\times 10^{-3} < \lambda_{Dh}< 10^{-1}$. The magenta and the yellow line follow the  same characteristics as the green line. The difference in the abundance of $\phi_{D }$ for different  lines arises due to  different choices of $\lambda_{ SD}$ and $\sin \theta$.  For a very large $\lambda_{ Dh}$,   the green, yellow and magenta lines merge as  $\lambda_{H_1}$ is governed by the  $\lambda_{ Dh}$ only. The red line corresponds to a much larger value of $\lambda_{ SD}=10^{-1}$ and $\sin \theta = 0.3$. For this chosen parameter, $\lambda_{H_1}$ is  governed by $\lambda_{SD}$ only. This makes  the relic abundance of $\phi_{D}$ at $T_d$ independent of $\lambda_{Dh}$.  

In Fig.~\ref{pf2}, we show the variation of $\Omega^{FO}_{\phi_{D }}h^{2}$ w.r.t the coupling $\lambda_{SD}$. For the chosen parameters corresponding to the red, green and blue lines, $\Omega^{FO}_{\phi_{D }}h^{2}$  is independent of $\lambda_{SD}$ for minimal coupling. In this scenario, $\lambda_{H_1}$ is governed by $\lambda_{Dh}$ only. As for both the green and red lines, the chosen value of $\lambda_{Dh}$ is the same and $\cos \theta \approx 1$, therefore both of these lines give similar contributions to the $\Omega^{FO}_{\phi_{D }}h^{2}$ for small $\lambda_{ SD}$. The blue line corresponds to a relatively smaller value of $\lambda_{Dh}$ compared to the green and red line, because of which  $\Omega^{FO}_{\phi_{D }}h^{2}$ is larger for blue line due to low interaction rate, in the region where  $\lambda_{ SD}$ is very small. As $\lambda_{ SD}$ increases, the second term in the $\lambda_{H_1}$ becomes larger. This leads to cancellation in the respective vertex due to a difference in the sign between the first and second terms. Similar to Fig.~\ref{pf1}, the suppression in the effective vertex leads to an increase in $\Omega^{FO}_{\phi_{D }}h^{2}$. In the cancellation region, the most dominant annihilation channel is $\phi_{D }^{\dagger}\phi_{D }\to NN$. For the yellow line, $\lambda_{ Dh}$ is considered to be negligible because of which $\lambda_{H_1}$ is governed by its second term only. Thus, as $\lambda_{ SD}$ increases, the annihilation rate of $\phi_{D }$ also increases which result in a decrease in $\Omega^{FO}_{\phi_{D }}h^{2}$. In Fig.~\ref{pf3}, we show the variation of $\Omega^{FO}_{\phi_{D }}h^{2}$ w.r.t $\sin \theta$. Similar to Fig.~\ref{pf1} and Fig.~\ref{pf2}, for very small $\sin \theta$, few of the lines represent similar values of $\phi_D$ abundance, as the interaction rate is totally governed by $\lambda_{ Dh} \cos \theta$ combination in the $\lambda_{H_1}$. The cancellation in $\lambda_{H_1}$ takes place only at a larger value of $\sin \theta$, for which $\Omega^{FO}_{\phi_{D }}h^{2}$ increases significantly. For the chosen parameter corresponding to the red line, $\lambda_{ SD}$ is much larger than $\lambda_{Dh}$. Therefore, in this region, the annihilation rate is governed by the second term of $\lambda_{H_1}$, because of which $\Omega^{FO}_{\phi_{D }}h^{2}$ decreases for larger values of $\sin \theta$ due to an increase in the annihilation rate.    

\begin{figure}[]
	\begin{center}
		\mbox{\subfigure[]{\includegraphics[angle=0,height=7.0cm,width=7.5cm]{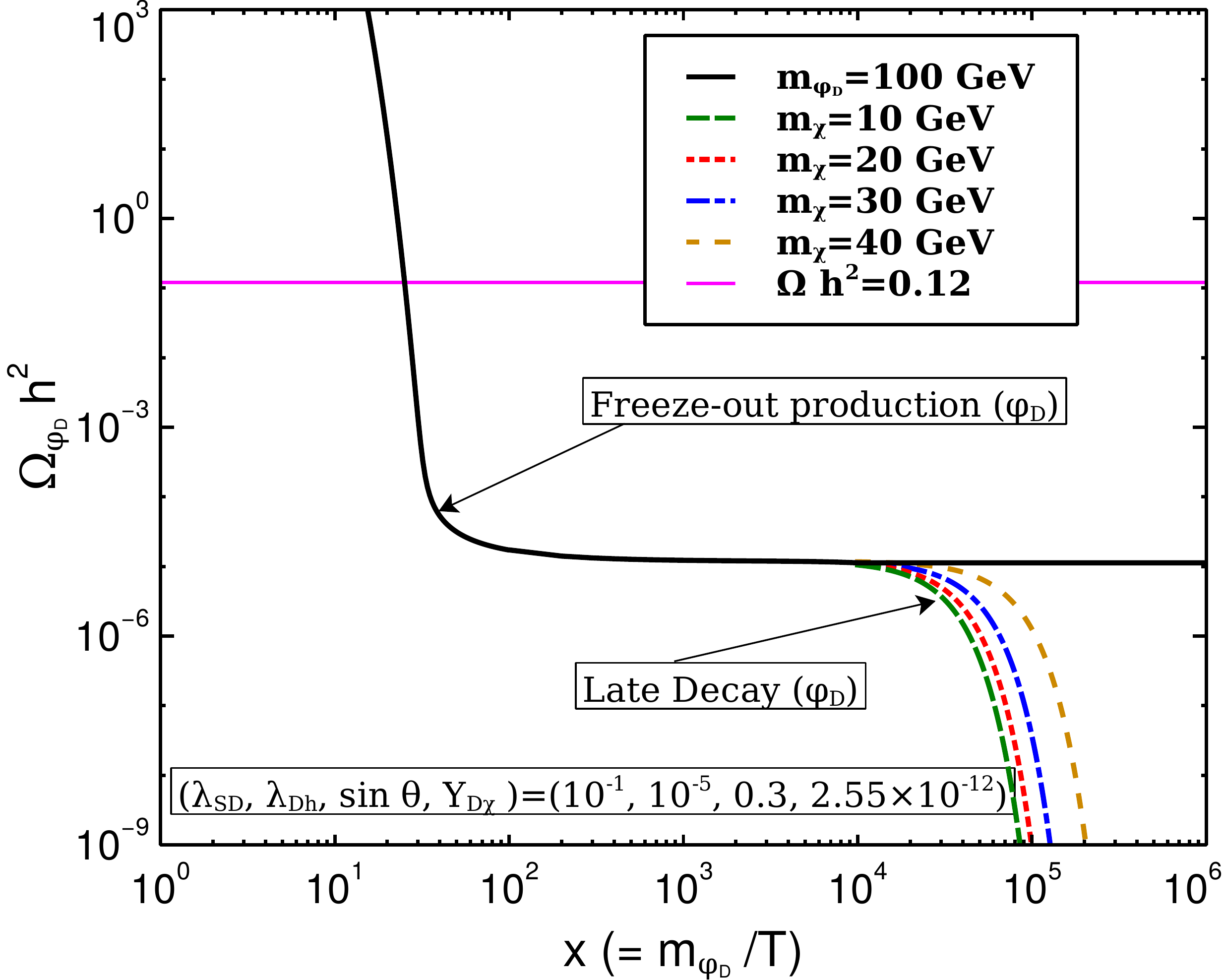}\label{chi11}}
			\subfigure[]{\includegraphics[angle=0,height=7.0cm,width=7.5cm]{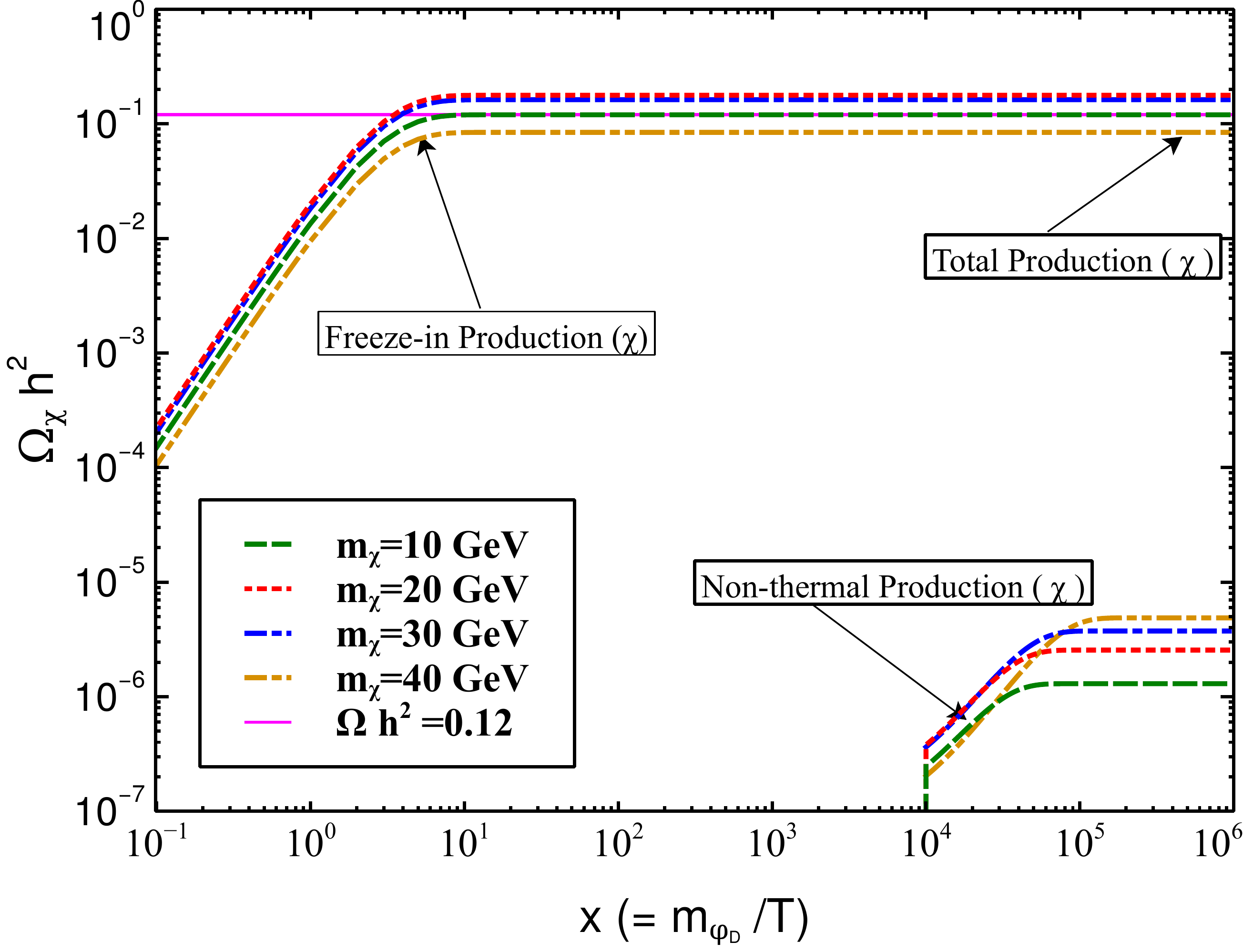}\label{chi12}}}
		\caption{Left panel:Fig~\ref{chi11} shows the variation of $\Omega_{\phi_D} h^2$ w.r.t $x$. Right panel: Fig~\ref{chi12} shows the variation of $\Omega_{\chi} h^2$ w.r.t $x$. This corresponds to the freeze-in dominated scenario.  }\label{Fig41}
	\end{center}
\end{figure}
\subsection{$\phi_D$ and $\chi$ abundance in {\it Scenario-I} and {\it Scenario-II} } 
Fig.~\ref{Fig41} corresponds to {\it Scenario-I}, for which primary contribution to the DM relic density arises from thermal freeze-in production of $\chi$. For the evaluation of $\Omega^{FO}_{\phi_{D }}h^{2}$ and $\Omega_{\chi}h^{2}$, we consider benchmark point 1. This is evident from Fig.~\ref{chi11} that $\phi_{D }$ stays in the thermal bath for a significantly longer time owing to a large $\lambda_{ SD}$ and $\sin \theta$. Due to this, the abundance of $\phi_{D }$ is reduced significantly before it freezes out, and therefore, its contribution to the production of $\chi$ via late decay is negligible. For this figure, we consider a $Y_{D \chi}$ which is in agreement with the BBN constraint as well as Eq.~\ref{eq:eqsecond}.  
The lifetime of $\phi_{D }$  increases with the increase in $\chi$ mass, thereby leading to the differences that can be seen from yellow, blue, green and red lines in the plot. \\ 
In Fig.~\ref{chi12}, we show the thermal freeze-in production of $\chi$, production of $\chi$ from the out of equilibrium decay of $\phi_D$, and the relic abundance of $\chi$ including both the contributions. At a very early epoch, the abundance of $\chi$ was vanishingly small due to suppressed interaction of $\chi$ with bath particles owing to small coupling strength $Y_{D\chi} \sim 10^{-12}$. For our chosen parameters, the production of $\chi$ is, however, primarily governed by the decay of $\phi_{D }$ to $\chi N$ state. The thermal freeze-in production of $\chi$ ceases as soon as the temperature of the thermal bath becomes less than the mass of $\phi_{D }$. In evaluating the relic abundance of $\chi$, we have neglected the inverse decay process $\chi N \to \phi_D$, as due to a very small abundance of $\chi$, inverse decay is entirely negligible. Due to this, the freeze-in temperature of $\chi$ in our analysis is independent of abundance of $\chi$ but rather depends on $m_{\phi_{D}}$. The abundance of $\chi$ can further be enhanced through the out of equilibrium decay of $\phi_{D }$. However, in this scenario, non-thermal production of $\chi$ from the late decay of $\phi_{D }$ is tiny, as has been shown in Fig.~\ref{Fig41}. 
Therefore, the total production of $\chi$, in this case, is determined by the thermal freeze-in mechanism. It is also important to highlight that the production of $\chi$ through the late decay of $\phi_{D }$ increases as the mass of $\chi$ increases, as can be understood from Eq.~\ref{eq:a1}.  
And it is also evident from the lower right side of Fig. \ref{chi12}, where we show the non-thermal contribution to relic abundance of $\Omega_{\chi} h^2 $ for different $\chi$ masses. \\ 
\begin{figure}[]
\begin{center}
\mbox{\subfigure[]{\includegraphics[angle=0,height=6.5cm,width=7.0cm]{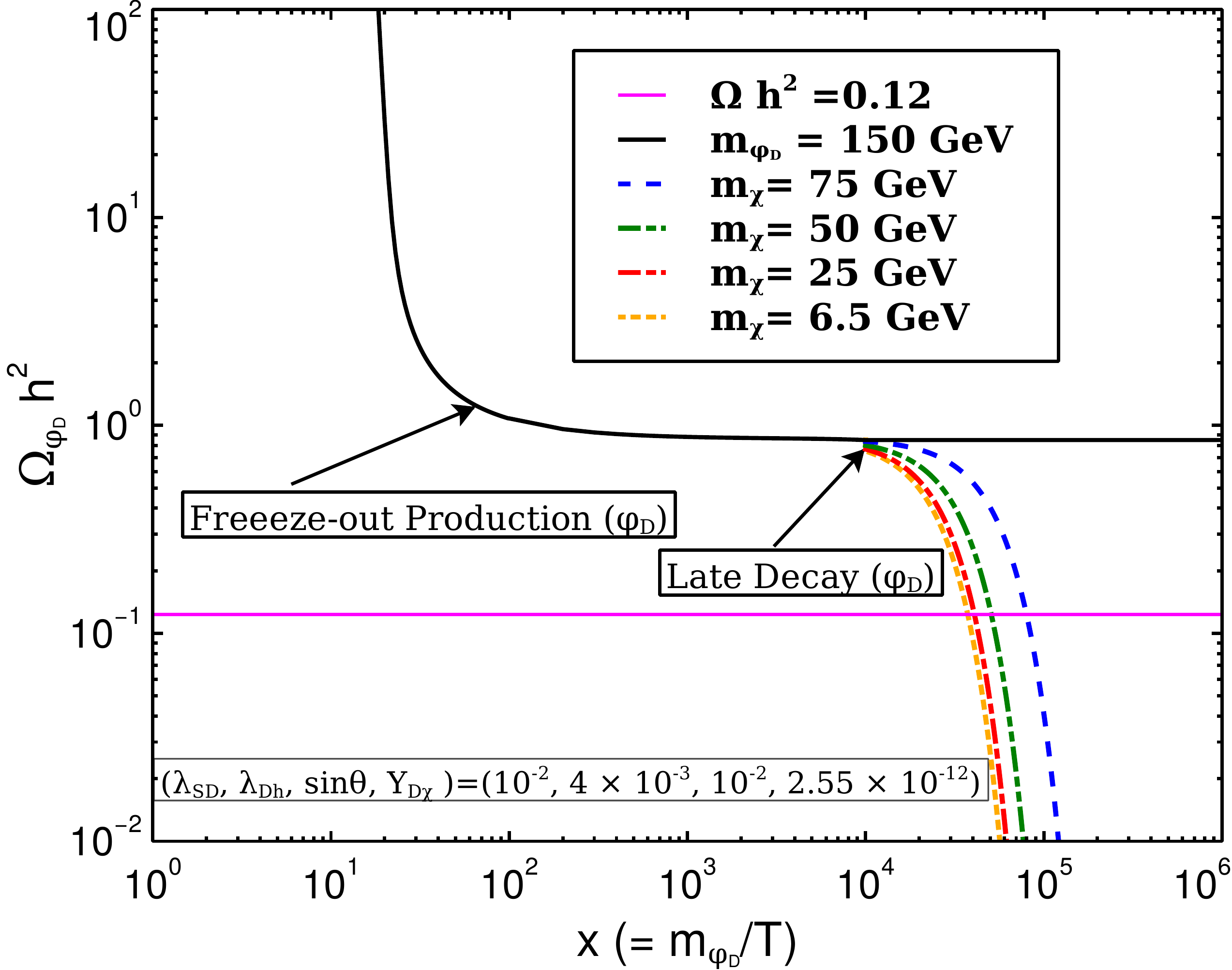}\label{chi21}}
\subfigure[]{\includegraphics[angle=0,height=6.5cm,width=7.0cm]{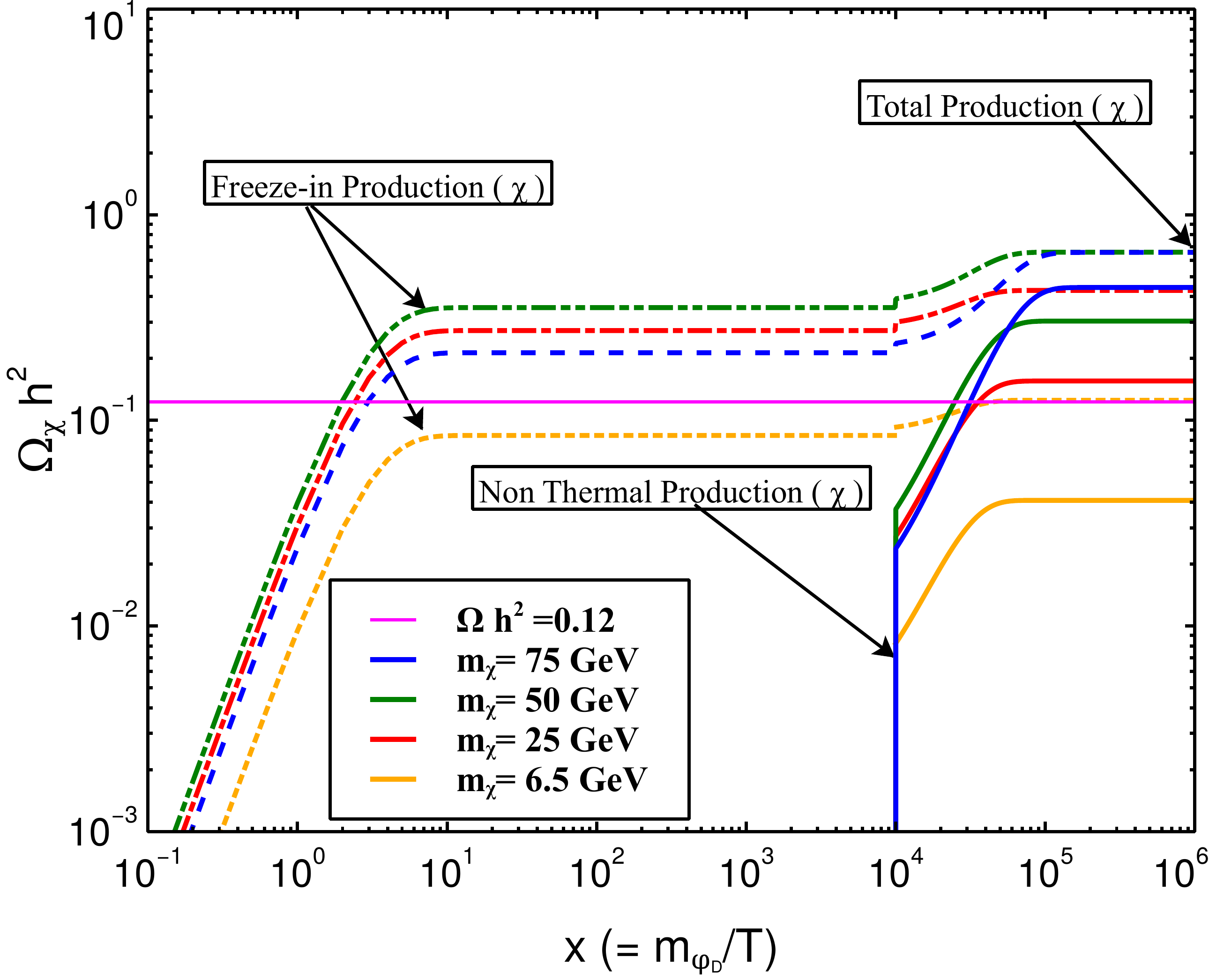}\label{chi22}}}
\caption{Left panel: Fig~\ref{chi21} shows the variation of $\Omega_{\phi_D} h^2$ w.r.t $x$. Right panel: Fig~\ref{chi22} shows the variation of $\Omega_{\chi} h^2$ w.r.t $x$. This corresponds to the super-wimp dominated scenario.}\label{Fig42}
\end{center}
\end{figure}
Contrary to the previous scenario {\it Scenario-I}, Fig.~\ref{Fig42} captures all the details about {\it Scenario-II}, where late decay of $\phi_{D }$ contributes significantly in the production of $\chi$. As it is evident from Fig.~\ref{chi21} that $\phi_{D }$ decouples from the thermal bath much earlier due to suppressed interaction with the bath particles. The abundance of $\phi_{D }$ at $T_d$ is therefore much larger than the $\phi_D$ abundance for {\it Scenario-I}, and its out-of-equilibrium decay can contribute significantly to the $\chi$ abundance. This has been shown in Fig.~\ref{chi22}, where we show the evolution of $\chi$ abundance. At high temperatures, the thermal freeze-in mechanism governs the production of $\chi$. Similar to the previous scenario, at this very early epoch, the dominant freeze-in production mode of $\chi$ is from $\phi_{D }$ decay. It is important to highlight that as the mass of $\chi$ increases from $m_{\chi}=50$ GeV to $m_{\chi}=75$ GeV, the thermal freeze-in contribution decreases instead of increase. It is due to the fact that phase space suppression for $\phi_{D }\to \chi N$ increases as mass of $\chi$ increases from $m_{\chi}=50$ GeV to $m_{\chi}=75$ GeV . At a later epoch, out of equilibrium decay of $\phi_{D }$ starts to contribute to the production of $\chi$. As one can see, the out of equilibrium decay of $\phi_{D }$ alone can overproduce the $\chi$. For our choice of parameters, the DM relic abundance is satisfied if $m_{\chi}=6.5$ GeV.

\begin{figure}[]
	\begin{center}
		\mbox{\subfigure[]{\includegraphics[angle=0,height=6.5cm,width=7.5cm]{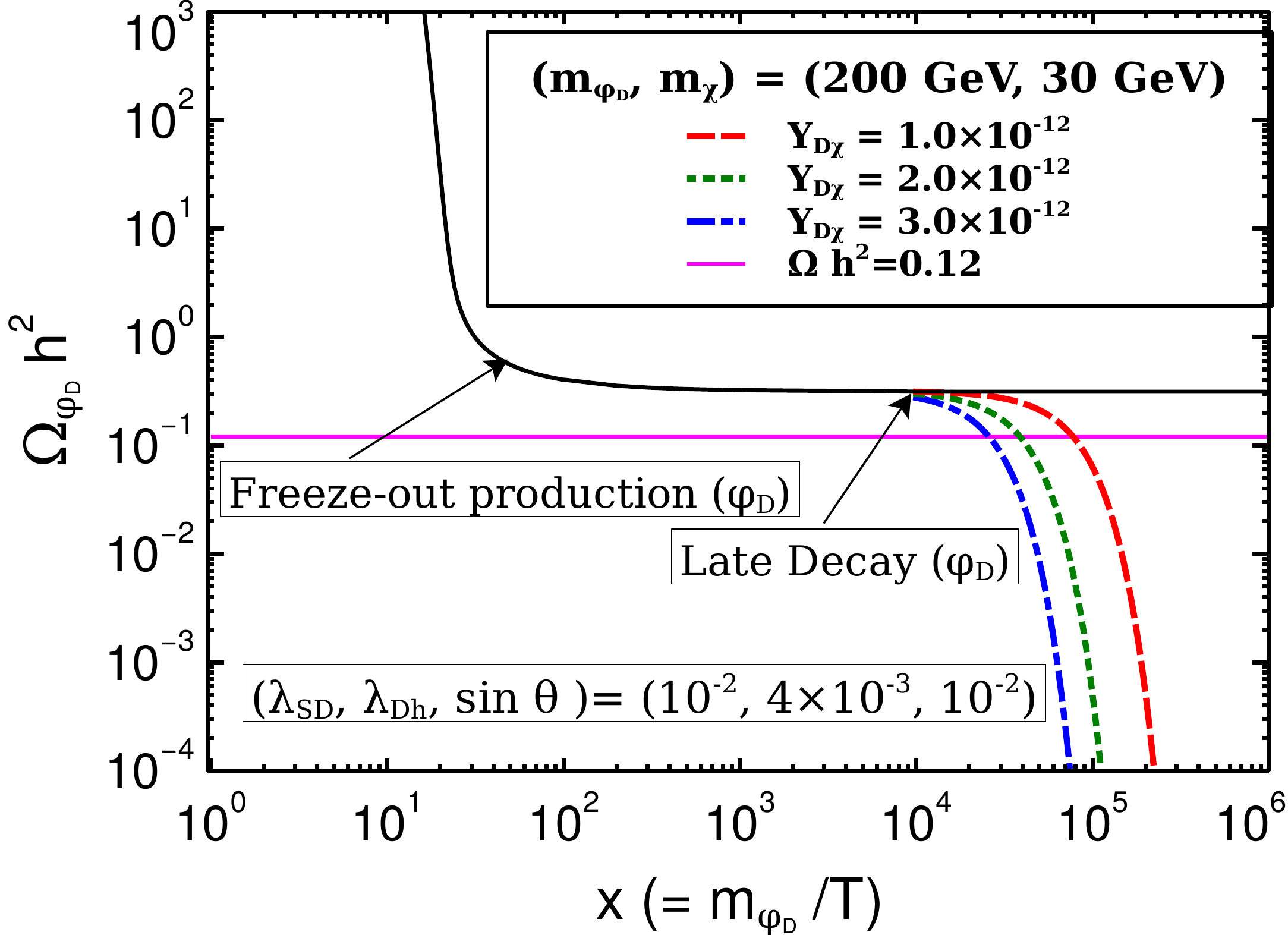}\label{chi31}}
			\subfigure[]{\includegraphics[angle=0,height=6.5cm,width=7.5cm]{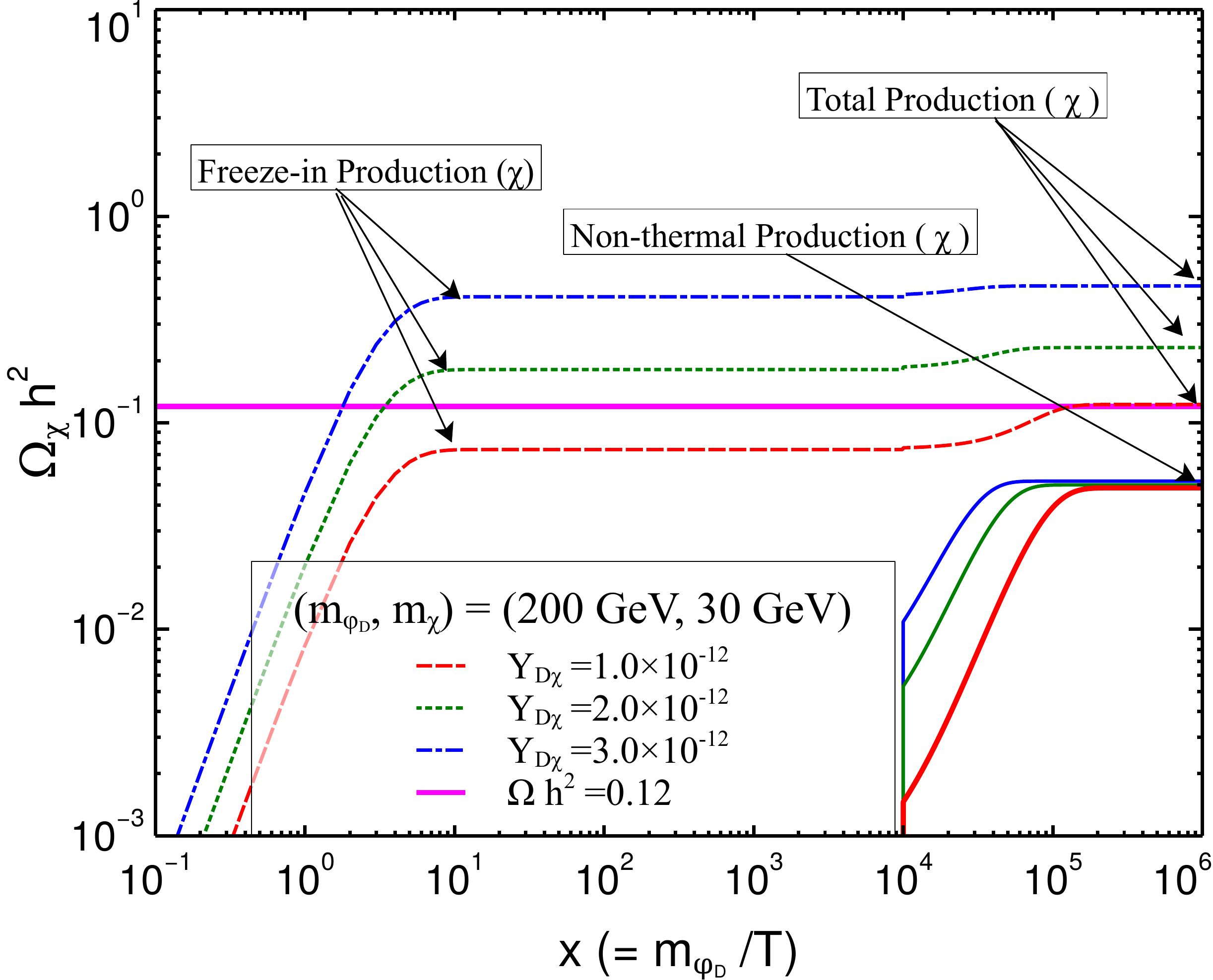}\label{chi32}}}
		\caption{Left panel: Fig~\ref{chi31} shows the variation of $\Omega_{\phi_D} h^2$ w.r.t $x$. Right panel:Fig~\ref{chi32} shows the variation of $\Omega_{\chi} h^2$ w.r.t $x$. This corresponds to a mixed scenario, where both the freeze-in and super-wimp contributions are significant. }\label{Fig43}
	\end{center}
\end{figure}

In Fig.~\ref{Fig43}, we show the effect of dark sector coupling $Y_{D\chi}$ on the production of $\phi_D$ and $\chi$. As one sees from Fig.~\ref{chi31} that change in $Y_{D\chi}$ only changes the lifetime of $\phi_{D }$. Owing to a small value of $Y_{D\chi}$, it does not have any effect in the freeze-out processes of $\phi_{D }$. In Fig.~\ref{chi32}, we show that the thermal freeze-in production of $\chi$ increases as $Y_{D\chi}$ increases. Since $\phi_D$ abundance is independent of $Y_{D\chi}$ coupling, hence, production of $\chi$ from late decay of $\phi_{D }$ is also independent of dark sector coupling $Y_{D\chi}$. For the parameter choice, both the thermal freeze-in and non-thermal contributions are significant in the production of DM $\chi$.
\begin{figure}[]
	\begin{center}
		\mbox{\subfigure[]{\includegraphics[angle=0,height=6.5cm,width=7.5cm]{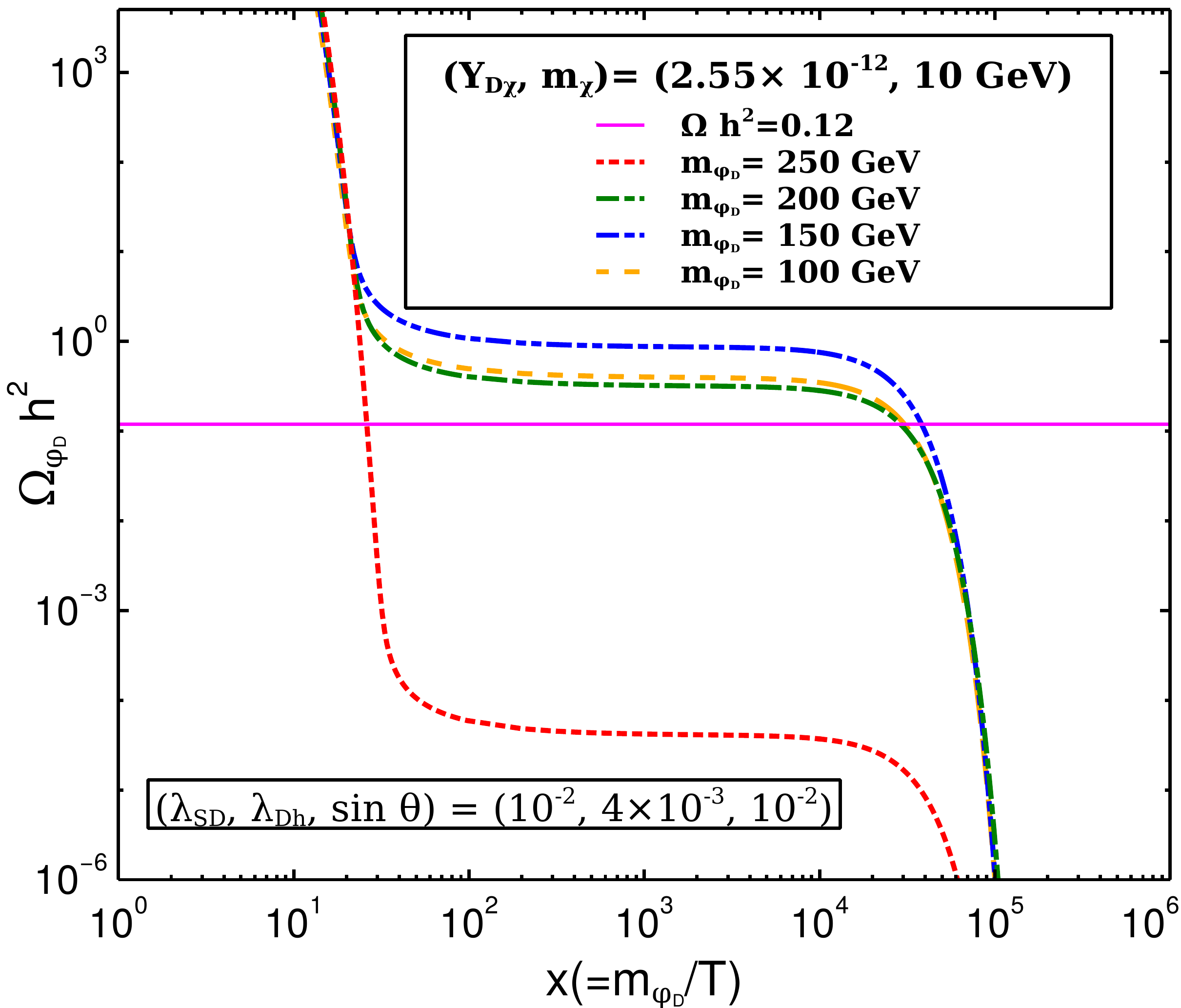}\label{chi41}}
			\subfigure[]{\includegraphics[angle=0,height=6.5cm,width=7.5cm]{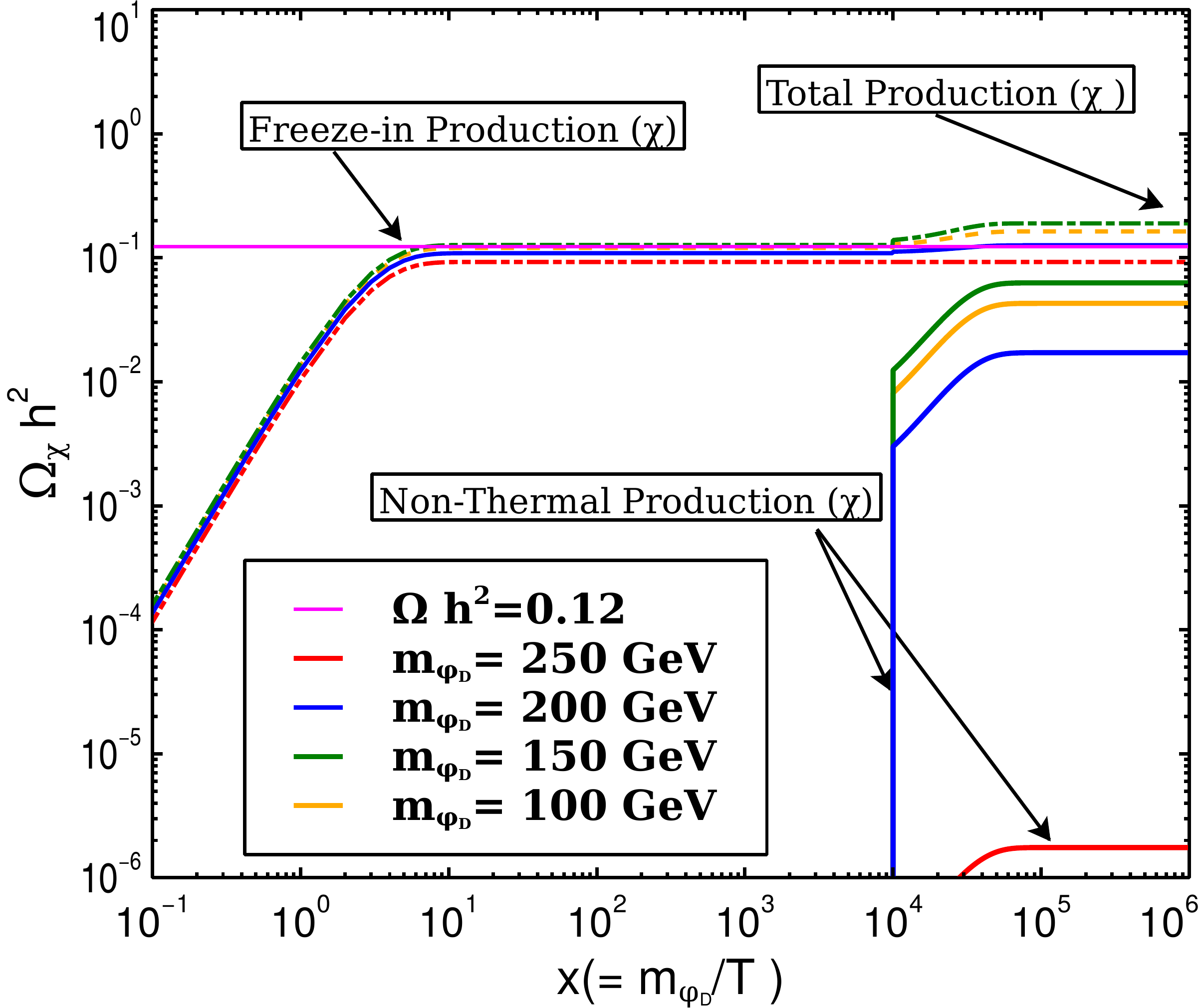}\label{chi42}}}
		\caption{Left panel: Fig~\ref{chi41} shows the variation of $\Omega_{\phi_D} h^2$ w.r.t $m_{\phi_{D }}$. Right panel:Fig~\ref{chi42} shows the variation of $\Omega_{\chi} h^2$ w.r.t $x$.}\label{Fig44}
	\end{center}
\end{figure}
 Fig.~\ref{Fig44} shows the variation of  $\Omega^{FO}_{\phi_D} h^2$ w.r.t $x $ for different choices of $m_{\phi_{D }}$. As one can see from the left panel (Fig.~\ref{chi41}) that $\phi_{D }$ yield increases as we vary  $m_{\phi_{D }} = 100$ GeV to $m_{\phi_{D }} = 150$ GeV. However, for even larger $m_{\phi_D}$ values, such as, 200 and 250 GeV,  the  $\Omega^{FO}_{\phi_D} h^2$ decreases  as annihilation processes of $\phi_{D }$ approaches s-channel resonances mediated  via $H_2$. This in turn effects  the production of $\chi$ from the  late decay of $\phi_{D }$. As one can see from Fig.~\ref{chi42}, the thermal freeze-in production of $\chi$ at high temperature increases as the mass of $\phi_{D }$ decreases. This occurs because $T_{FI}\sim m_{\phi_D}$, and a lower $T_{FI}$ leads to higher production.  At a  later epoch, non-thermal production of $\chi$ from the late decay of $\phi_{D }$ starts to contribute. The  non-thermal production of $\chi$ increases as mass of $\phi_{D }$ increases from $100$ to  $150$ GeV,  but later decreases with the increase in the mass of $\phi_{D }$. 

\begin{figure}[]
	\begin{center}
		\mbox{\subfigure[]{\includegraphics[angle=0,height=6.5cm,width=7.5cm]{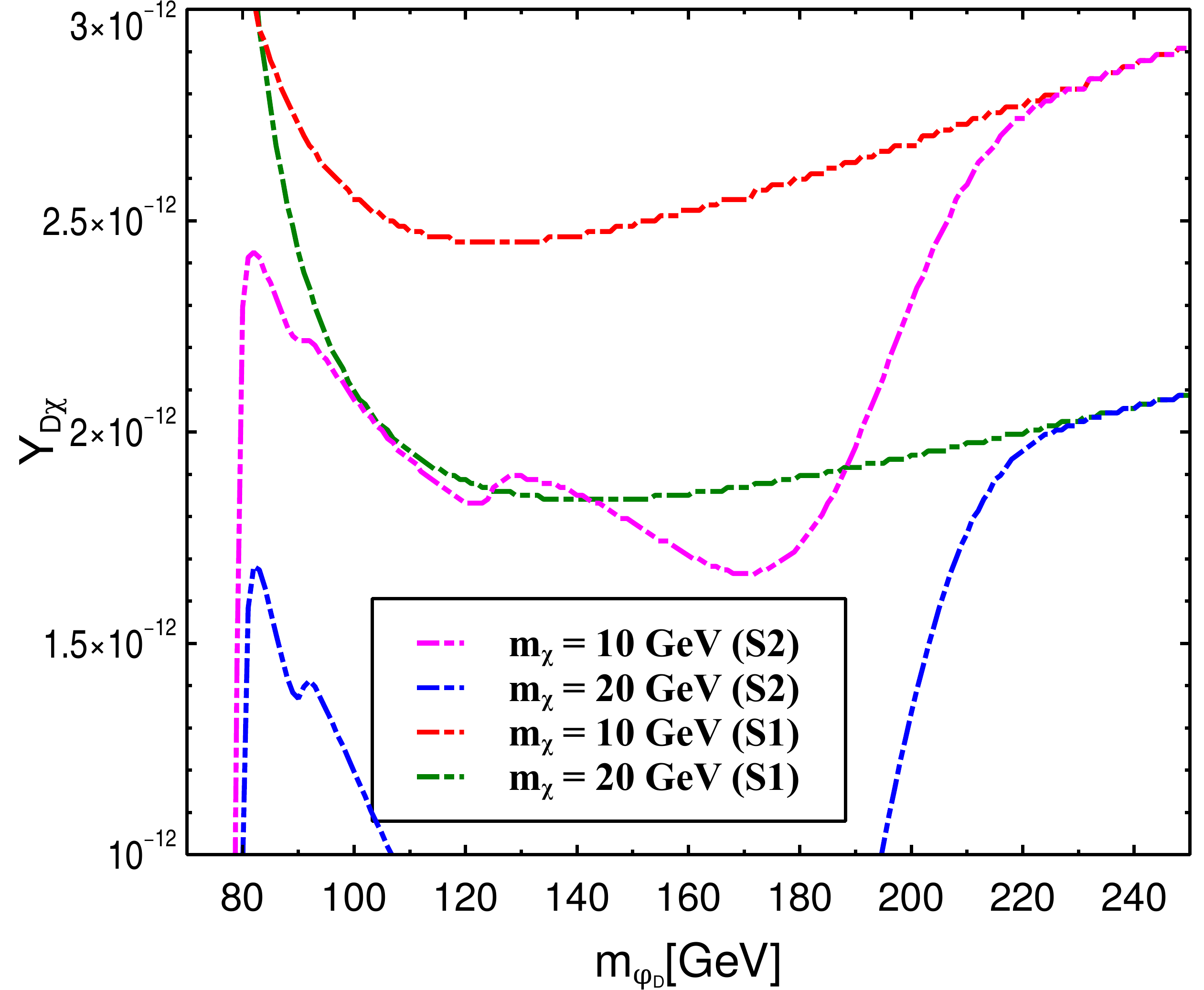}\label{a31}} 
			\subfigure[]{\includegraphics[angle=0,height=6.5cm,width=7.5cm]{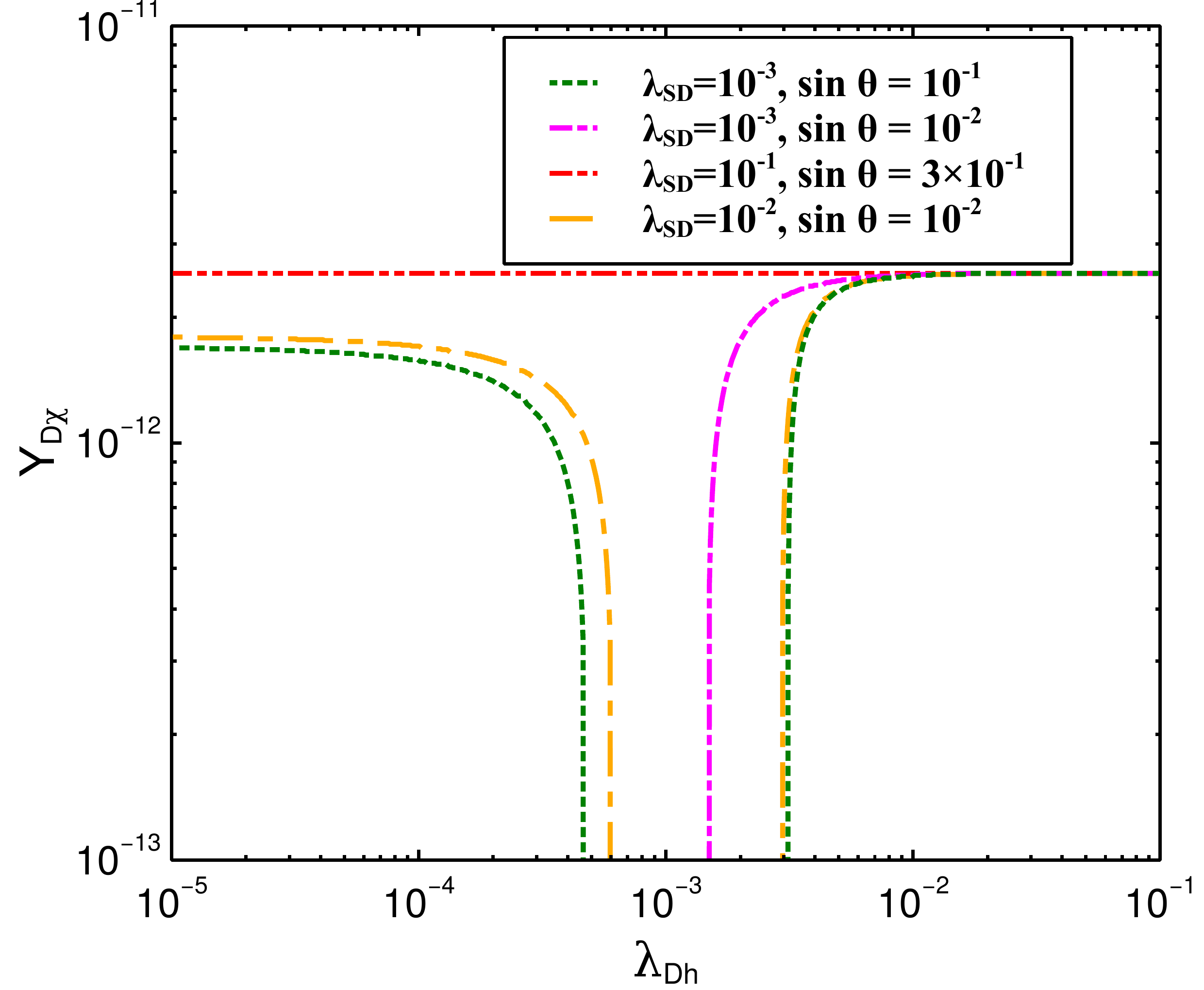}\label{a32}}}
		\mbox{\subfigure[]{\includegraphics[angle=0,height=6.5cm,width=7.5cm]{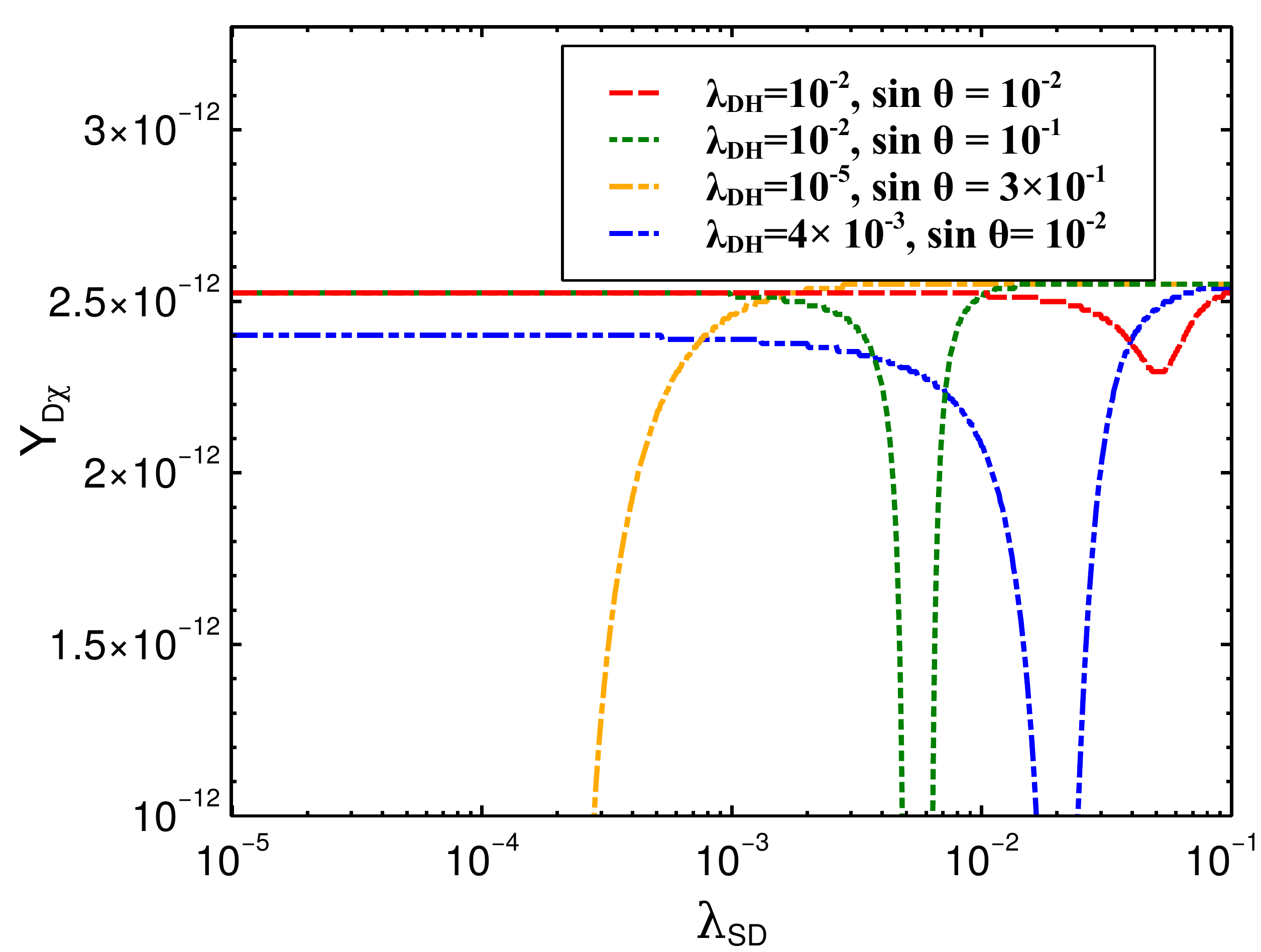}\label{a33}}
			\subfigure[]{\includegraphics[angle=0,height=6.5cm,width=7.5cm]{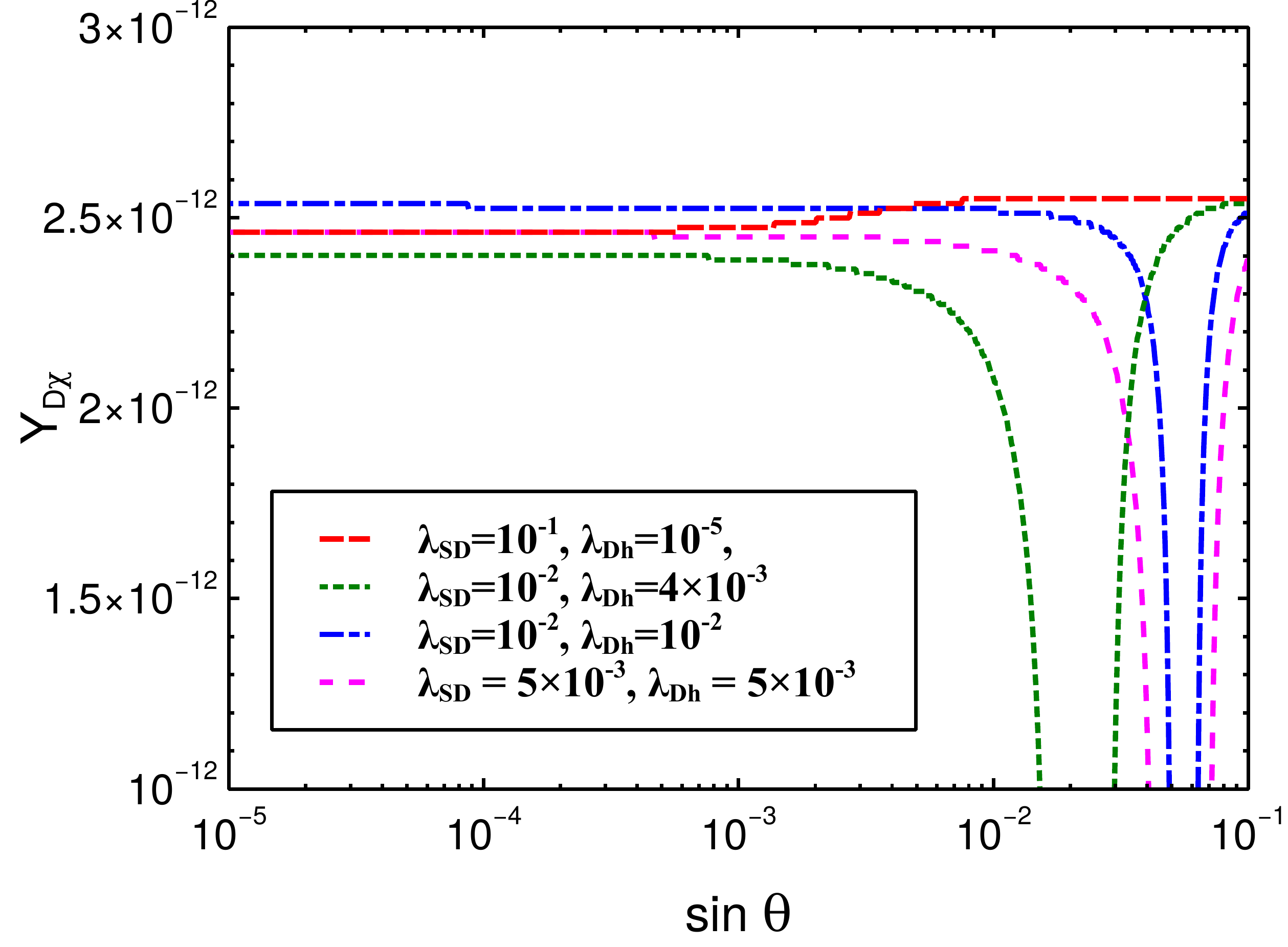}\label{a34}}}
		\caption{Fig.~\ref{a31} shows the contour of $\Omega h^2_{\chi}=0.12$ in the coupling $Y_{D\chi}$ and mass of $\phi_D$ ($m_{\phi_D}$) plane. Fig.~\ref{a32} shows the same in $Y_{D\chi}$ and $\lambda_{ Dh}$ plane,  Fig.~\ref{a33} and Fig.~\ref{a34} show the same in the   $Y_{D\chi}$-$\lambda_{ SD}$, and $Y_{D\chi}$- $\sin \theta$ plane. The parameter chosen for this plot are as follows, $m_S = 500$ GeV, $ M_N = 50$ GeV, $ m_{\chi} = 10 $ GeV, $m_{\phi_{D }} =100$ GeV, $  g_{BL}=0.9,\  m_{Z_{BL}}= 7$\ TeV }\label{Fig46}
	\end{center}
\end{figure}
\subsection{ Dependence of  $Y_{D\chi}$ on model parameters } 
In Fig.~\ref{Fig46}, we show the dependence of dark sector Yukawa coupling $Y_{D\chi}$ which governs the abundance of $\chi$ via thermal freeze-in production on parameters which determine the abundance of $\phi_{D }$ at the time of decoupling.
In Fig.~\ref{a31}, we show dependence of $Y_{D\chi}$ on $m_{\phi_{D}}$ for our two scenarios.
\begin{itemize}
	\item The thermal freeze-in dominated scenario, i.e., {\it Scenario-I} is represented by the red and green lines in the figure. In this scenario, $\phi_{D }$ has a negligible abundance after freeze-out from the thermal bath, which is evident from the red line of Fig.~\ref{phirelic}.
	To show the dependence of $Y_{D\chi}$ on $m_{\phi_{D }}$, we have considered two different masses of $\chi$ which are $10$ and $20$ GeV. The thermal freeze-in production ceases when the temperature of the thermal bath becomes less than $m_{\phi_{D }}$. The freeze-in temperature drops as we consider a lighter $\phi_{D }$ state. For the lower mass of $\phi_{D }$, production of $\chi$, therefore, takes place for a longer time which in turn increase the abundance of $\chi$ significantly. To satisfy the correct relic density, we can decrease the production of $\chi$ by decreasing the dark sector Yukawa coupling $Y_{D\chi}$. This behaviour is opposite for a large mass of $\phi_{D }$. The freeze-in production of $\chi$ suffers Boltzmann suppression at a much higher temperature compared to the case of the low mass of $\phi_{D}$. To compensate this effect, the production rate needs to be increased, which is done by increasing the magnitude of $Y_{D\chi}$. 
	It is important to highlight that as mass of $\phi_{D }$ decreases from $m_{\phi_{D }}= 120\ $GeV to $m_{\phi_{D }}=80$ GeV, the magnitude of $Y_{D\chi}$ increases instead of decrease. It is due to fact that the phase space suppression for $\phi_{D }\to \chi N$ process increases as mass of $\phi_{D }$ decreases from $m_{\phi_{D }}= 120\ $GeV to $m_{\phi_{D }}= 80$ GeV. The behaviour of these two curves, even though similar, however the required value of $Y_{D\chi}$ is more prominent for the lower mass of $\chi$ compared to the higher mass of $\chi$. 	
	\item The variation of the required $Y_{D\chi}$ for {\it Scenario-II} which satisfies the DM relic abundance is shown by the pink and blue lines in Fig.\ref{a31}. $\Omega^{FO}_{\phi_{D }} h^{2}$ for {\it Scenario-II} is shown via the green coloured line in Fig.~\ref{phirelic}, which indicates $\Omega^{FO}_{\phi_{D }} h^{2}\ge0.12$ except the region near $s$-channel resonance around $m_{\phi_D} =M_{H_1}/2 \sim 62.5$ GeV and $m_{\phi_D} =M_{H_2}/2 \sim 250$ GeV. The large abundance of $\phi_D$ enhances the relic density of $\chi$. The late decay of $\phi_D$ producing $\chi$ is independent of the Yukawa coupling $Y_{D\chi}$; however, the thermal contribution depends on $Y_{D\chi}$. Hence, depending upon the abundance of $\phi_{D }$, the dark sector Yukawa coupling $Y_{D\chi}$ needs to be tuned accordingly to satisfy the relic density constraint. It is important to highlight that the out-of-equilibrium decay of $\phi_{D }$
	in the production of $\chi$ can be so significant that to satisfy the correct relic abundance of $\chi$, the thermal freeze-in production of $\chi$ is required to be small, which is possible to achieve for a small $Y_{D\chi}$. However, note that the coupling $Y_{D\chi}$ can not be made arbitrarily small, as the BBN imposes a strong lower bound on $Y_{D\chi}$. \\
	The pink and blue lines in Fig.~\ref{a31} along which the DM relic abundance $\Omega_{\chi}h^{2} = 0.12$ show the variation of the required $Y_{D\chi}$ w.r.t $\phi_D$ mass for this scenario. The pink line clearly shows that a smaller value of $Y_{D\chi}$ is required in order to satisfy the correct relic density for {\it Scenario-II} when compared to the red line, which corresponds to the thermal freeze-in dominated scenario of {\it Scenario-I}. Also, note that the pink and red lines merge for the mass of $\phi_{D }$ greater than $200$ GeV. For $m_{\phi_D}\sim 250$ GeV, due to the $s$-channel resonance, the abundance of $\phi_{D }$ decreases significantly (see Fig.~\ref{phirelic}). Therefore, a large thermal freeze-in contribution is required to satisfy the correct relic abundance, which in turn demands a larger value of the Yukawa coupling $Y_{D\chi}$. The blue line traces the pink line in part of the parameter space, with the notable difference that the out of equilibrium decay of $\phi_{D }$ producing $\chi$ is more dominant in between 110-190 GeV. The relic abundance of $\chi$ in this region becomes larger than the observed relic density due to the large contribution from the late decay of $\phi_D$. Hence, for no value of $Y_{D\chi}$, the DM relic density constraint $\Omega_{\chi}h^{2} = 0.12$ is satisfied. 
\end{itemize}
In Fig.~\ref{a32}, we show dependence of $Y_{D\chi}$ on $\lambda_{ Dh}$. The observations are listed as follows:

\begin{itemize}
	\item  The red line in this figure represents {\it Scenario-I}, i.e., the thermal freeze-in dominated scenario. In Fig.~\ref{pf1}, the red line shows the variation of $\Omega^{FO}_{\phi_{D }} h^{2}$ with $\lambda_{ Dh}$ for $\lambda_{ SD} =10^{-1}$ and $sin\  \theta = 0.3$. $\Omega^{FO}_{\phi_{D }} h^{2}$ is significantly small for all values of $\lambda_{ Dh}$. Therefore, out of equilibrium decay of $\phi_{D }$ can not produce significant number of $\chi$. Due to this, the correct relic density of $\chi$ is obtained only through thermal freeze-in production which depends on $Y_{D\chi}$, and  not on the coupling $\lambda_{Dh}$. Therefore, the required coupling $Y_{D\chi}$  is independent of $\lambda_{ Dh}$. 
	
	\item  The  variation of $Y_{D\chi}$ w.r.t the  variation of $\lambda_{Dh}$ in  Fig.~\ref{a32} can be understood from Fig.~\ref{pf1}. For fixed value of $m_{\phi_{D }}$ and $m_{\chi}$, production of $\chi$ through out of equilibrium decay of $\phi_{D }$ is proportional to $\Omega^{FO}_{\phi_{D }} h^{2}$. As we can see from the green and yellow lines in Fig.~\ref{pf1} that $\Omega^{FO}_{\phi_{D }} h^{2}> 0.12$ but remain constant in the range $10^{-5}<\lambda_{ Dh}< 10^{-4}$. For $\lambda_{Dh}> 10^{-4}$,  $\Omega^{FO}_{\phi_{D }} h^{2}$ increases significantly as $\lambda_{ Dh}$ increases.  This sudden jump occurs  due to the cancellation in $\lambda_{H_1}$, described earlier. $\Omega^{FO}_{\phi_{D }} h^{2}$ again  falls  much below 0.12 as $\lambda_{ Dh}$ increases further. For a large $\phi_D$ abundance, the out-of-equilibrium contribution from $\phi_D \to \chi N$ will be substantial. In  Fig.~\ref{a32}, for $\lambda_{ Dh}>3\times10^{-3}$, due to very suppressed $\phi_D$ abundance, out of equilibrium decay contribution is small, and the thermal freeze-in contribution alone satisfies the DM abundance. This is represented via the red, yellow and green lines which  merge near $\lambda_{ Dh} \sim 3\times10^{-3}$. For small value of $\lambda_{ Dh} < 10^{-4}$,  both the out of equilibrium decay of $\phi_{D }$ as well as  thermal freeze-in production contribute substantially to the relic abundance of $\chi$. In this region, due to the presence of a finite out-of-equilibrium decay contribution, the required value of  $Y_{D\chi}$ to satisfy correct DM relic density is typically less  when compared to the only thermal freeze-in dominated scenario, represented via the red line. In between $5 \times10^{-4}<\lambda_{Dh}<2\times10^{-3}$, the $\phi_D$ abundance is very large due to cancellation in $\lambda_{H_1}$ leading to a large out-of-equilibrium decay contribution which results in $\Omega_{\chi}h^2>0.12$. Hence, this region is disallowed.
	\item  The magenta line in this figure corresponds to $\lambda_{SD}=10^{-3}$ and $\sin \theta=10^{-2}$. The behaviour can again be understood by referring to the magenta line in Fig.~\ref{pf1}.  $\Omega^{FO}_{\phi_{D }} h^{2}$ is much larger than 1.0 for  $\lambda_{ Dh} < 2\times 10^{-3}$. This leads to the overproduction of  $\chi$  via out-of-equilibrium decay of $\phi_{D }$. Hence, the correct relic density of $\chi$ in this case is only obtained for $\lambda_{ Dh}$ larger  than $2\times 10^{-3}$.  
\end{itemize}

In  Fig.~\ref{a33}, we show dependency of $Y_{D\chi}$ on  $\lambda_{ SD}$. The observations are listed as follows:

\begin{itemize}
	\item  For the yellow curve in Fig.~\ref{a33}, the correct relic density of $\chi$ is possible to obtain for $\lambda_{ SD}>3\times 10^{-4}$. For $\lambda_{ SD}<3\times 10^{-4}$, abundance of $\phi_{D }$ is significantly large (see the yellow curve of  Fig.~\ref{pf2}), leading to the overproduction of $\chi$ via  out of equilibrium decay of $\phi_{D }$. 
	\item  We can see in Fig.~\ref{pf2} that red, green and blue lines have similar features. The cancellation in $\lambda_{H_1}$ takes place for different values of $\lambda_{ SD}$. As we can see the funnel shaped region in Fig.\ref{a33} that $Y_{D\chi}$ decreases significantly in the region where cancellation in $\lambda_{H_1}$ is effective.  
\end{itemize}
In  Fig.~\ref{a34}, we show dependence of $Y_{D\chi}$ on  $\sin\theta$. The observations are listed as follows:
\begin{itemize}
	\item  The  nature of the red, green and blue lines can be understood from  Fig.~\ref{pf3}. The required value of $Y_{D\chi}$ to satisfy correct DM  relic density decreases significantly when $\phi_{D }$ abundance gets enhanced. For each of these three lines, in the funnel shaped region, the $\phi_D$ abundance becomes  very large due to the cancellation in $\lambda_{H_1}$. This larger $\phi_D$ abundance leads to  $\Omega h^2_{\chi}>0.12$, which is ruled out. 
	\item In Fig.\ref{a34}, the red line represents a scenario, where both the thermal freeze-in production and out-of-equilibrium production of $\chi$ can contribute. For $\sin \theta > 0.1$, due to a smaller $\phi_D$ abundance, mostly thermal freeze-in contribution dominate. Hence a larger value of $Y_{D\chi}$ is required to satisfy the DM relic abundance. In the blue, pink, and green lines, the effect of cancellation in the $\phi_{D }^{\dagger}\phi_{D }H_1$ vertex is clearly visible. For each of these lines, in the funnel shaped region, $\phi_D$ abundance is very large, leading to an overproduction of $\chi$. For small $\sin \theta$, both the thermal freeze-in and out-of-equilibrium decay  can contribute significantly (see Fig.~\ref{pf3}).
\end{itemize} 

\section{Collider Prospects}\label{sec:collider}
This section focuses on the search for the  BSM Higgs $H_2$ via its invisible decay, i.e., $H_2\to\phi_D^\dagger\phi_D$ at the LHC. The partial decay width for this decay mode is determined by the couplings $\lambda_{SD}$. The other production mode of $\phi_D$ from $Z_{BL}$ is suppressed due to a very heavy $Z_{BL}$. The possible decay mode of $\phi_D$ is $\phi_D \to \chi N $ which is controlled by the coupling $Y_{D\chi}$. As mentioned before we assume $Y_{D\chi}=\mathcal{O} (10^{-12})$ to realize the freeze-in production of the DM $\chi$. As a result of this tiny coupling, $\phi_D$ escapes the detector without leaving any visible footprint. However, its production can be confirmed by the observed imbalance in the transverse momentum. For collider analysis, we consider the mass of $\phi_D$ to be $m_{\phi_D}=100$ GeV. Other parameters are set to $M_{Z_{BL}}=7$ TeV, $g_{BL}=0.9$, and $M_N=50$ GeV, which we also consider for the DM study. 
We first discuss existing constraints on the model parameters from the collider experiment and then project the future sensitivity to probe the coupling  $\lambda_{SD}$ at the HL-LHC.

\subsection{LHC Constraints}\label{sec:const}

We first discuss the different constraints applicable on the SM-BSM Higgs mixing angle $\theta$, the mass of the BSM Higgs and the quartic coupling $\lambda_{SD}$. \\
\\
{\bf {Measurement of Higgs signal strength and coupling constant modifiers:}}
The signal strength of SM Higgs decaying into two SM states $a,b$, such as $WW^*, ZZ^*, \tau \tau, b\bar{b}, \mu^+\mu^-$ is,
\begin{equation}
\mu_{H_1\to a b}=\frac{\sigma(H_1)}{\sigma(H_1)_{\text{SM}}}\frac{\text{BR}(H_1\to a b)}{\text{BR}(H_1\to a b)_{\text{SM}}}.
\end{equation} The global signal strength of $H_1$ using $139 \ \text{fb}^{-1}$ data at LHC is measured as $\mu=1.06\pm0.07$~\cite{ATLAS:2020qdt}. Due to the presence of the BSM Higgs, which mix with the SM like Higgs states, the standard couplings of the SM Higgs with $W^+W^-, ZZ, \tau \tau$ and others will be modified. We adopt the constant coupling modifiers - $\kappa$ framework, where $\kappa$'s are defined as
\begin{eqnarray}
\kappa_x=\frac{\lambda_{xxh}}{\lambda^{SM}_{xxh}}=\cos \theta, 
\end{eqnarray}
where $\lambda_{xxh}$ is the couplings of SM-like Higgs field $H_1$ with two SM fields in the model considered, and $\lambda^{SM}_{xxh}$ is the respective coupling in the SM. We consider the ATLAS search \cite{ATLAS:2020qdt}, and translate the measurements of each measured $\kappa$'s to the upper limit on the SM and BSM Higgs mixing angle $\sin \theta$. The results are shown in Table.~\ref{tab:kappa}. In our collider analysis for HL-LHC, we adopt a conservative approach and consider relatively smaller values of $\sin \theta=0.3$, which agrees with the LHC constraints. 
\\
	\begin{table}[t]
		\centering
		\begin{tabular}{|l|l|l|l|l|l|l|l|}\hline
			Parameter&$\kappa_Z$& $\kappa_W$ &$\kappa_t$ &$\kappa_b$ &$\kappa_{ta}$& $\kappa_g$&$\kappa_\gamma$   \\ \hline	   
			$ \sin\theta\  \text{at} \ 95 \% \text{CL}$&0.46&0.45&0.65&0.69&0.65&0.61 &0.42\\ \hline          
		\end{tabular}
		\caption{Upper limit on $\sin\theta$ obtained from Higgs boson coupling modifiers, $\kappa_{Z/W/t/b/\tau/g/\gamma}$ $ $ at $\ 95 \% \ \text{CL}$~\cite{ATLAS:2020qdt}. }\label{tab:kappa}
	\end{table}	
{\bf{SM Higgs decaying to long-lived particle~(LLP):}}  
The theory under consideration predict several exotic decays of the SM Higgs boson, such as $H_1\to NN/Z_{BL}\gamma/\phi_D^\dagger\phi_D$,~\footnote{We do not consider the decay chain $H_1\to \phi_D^\dagger \phi_D(\to \chi N^{\star})$ as it is suppressed due to the coupling $Y_{D\chi}< 10^{-10}$.} among which $H_1\to Z_{BL}\gamma/\phi_D^\dagger\phi_D$ are closed kinematically and $H_1\to NN$ is open having BR$(H_1\to NN)=0.5\%$ for $\sin\theta = 0.3$ and $M_N=50$ GeV. For our benchmark point, decay length of RHN $c\tau_N\simeq40$ m~(for active-sterile mixing, $V\simeq 10^{-7}$) and its possible decay modes are $N\to ljj/\nu jj/ll\nu/3\nu$. Therefore, $N$ is a LLP undergoing displaced decays. The recent CMS search for displaced heavy neutral leptons limits the active-sterile mixing in the mass range $1-18$ GeV~\cite{CMS-PAS-EXO-20-009} with the most tight constraint appears $|V|^2 < 10^{-7}$ for $M_N \sim \mathcal{O}(10)$ GeV. Our choice of RHN mass $M_N=50$ GeV and active sterile mixing $V \sim 10^{-7}$ is beyond the range covered in this paper. There are other CMS and ATLAS searches for exotic decays of SM Higgs into two LLP states, which are instead applicable. The CMS and ATLAS have recently searched for exotic decays of the Higgs boson into LLP in the tracking system~\cite{CERN-EP-2021-106,CMS:2021uxj}. These searches are mainly sensitive to LLP with $c\tau=\mathcal{O}(1 ~\text{mm}-300~\text{mm})$. Other displaced vertex searches in the tracking system that are also sensitive to Higgs decays to LLP~\cite{CMS-PAS-EXO-18-003,CMS:2020iwv}. Our benchmark point is unconstrained from these searches owing to a very long lifetime of the RHN. The latest search for neutral LLP decaying into displaced jets in the ATLAS muon spectrometer~\cite{ATLAS-CONF-2021-032,ATLAS:2019jcm,ATLAS:2018tup} and in the CMS endcap
muon detectors~\cite{CMS:2021juv} are relevant for LLP with $c\tau\ge \mathcal{O}(1~\text{m})$. The RHN mostly decays in the muon spectrometer for our benchmark mass point. This is to note that our model prediction of BR$(H_1\to NN)=0.5\%$ for $M_N=50$ GeV is consistent with the observed bound on BR of Higgs to LLP decay. Note that this constraint is given for the Higgs decaying to scalar LLP and for the two-body decay of the LLP. In reinterpreting this analysis for our scenario, we assume similar signal selection efficiency as given in \cite{CMS:2021juv}.

\begin{figure}[]
	\centering
		\subfigure[]{\includegraphics[width=7cm,height=6.cm]{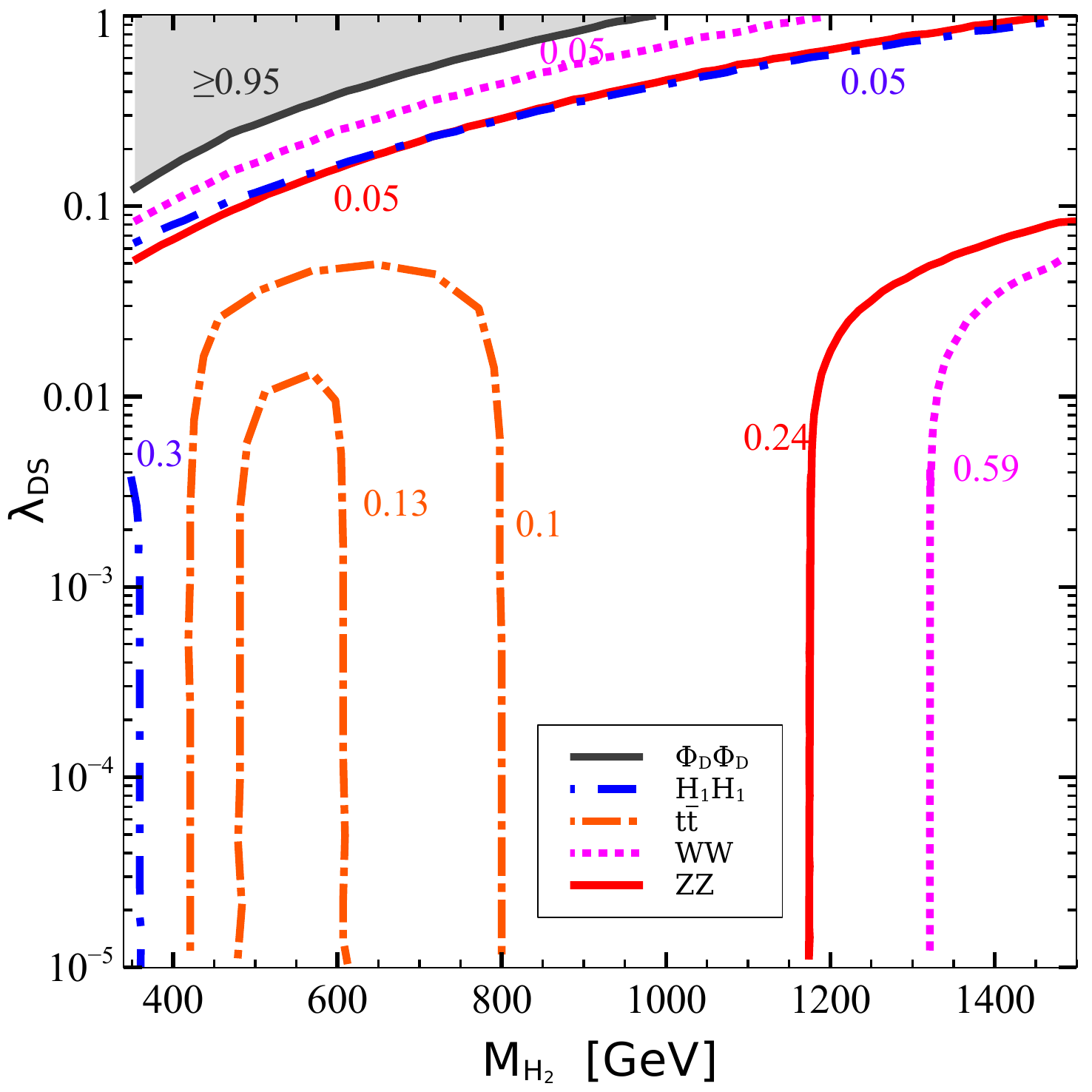}\label{fig:d1}}
\subfigure[]{\includegraphics[width=7cm,height=6.cm]{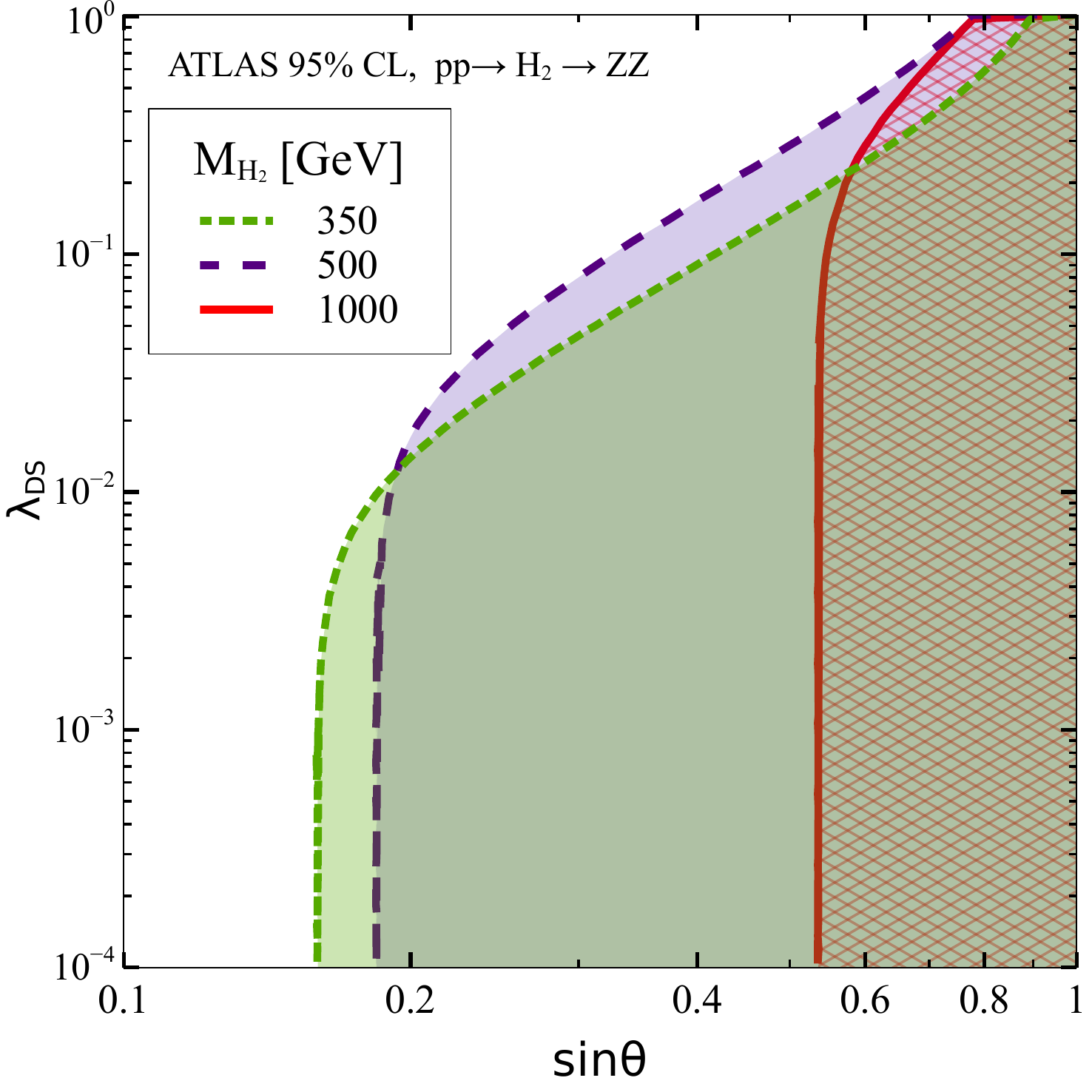}\label{fig:d2}}
	\caption{Fig.~\ref{fig:d1}: Contours of BR($H_2 \to \phi_D^\dagger\phi_D/ZZ/W^+W^-/H_1H_1/t\bar{t}$) in $M_{H_2}$ - $\lambda_{SD}$ plane for $\sin\theta= 0.3$. Fig.~\ref{fig:d2}: Constraints in $\sin\theta- \lambda_{SD}$ plane derived from the ATLAS search for heavy scalar resonance decaying to two $Z$ bosons, $pp\to H_2\to ZZ\to4l$~\cite{Aad:2020fpj}. 
	}  \label{fig:BR_lamDSsintheta}
\end{figure}

{\bf { Heavy Higgs searches:}} Other LHC searches~\cite{Aad:2020fpj,Aad:2020ddw,Aad:2019uzh,Sirunyan:2018zkk,Aad:2020kub} aimed at probing BSM Higgs via direct measurements can constrain  our model. These are  the searches to detect a   heavy scalar resonance~($H_2$) decaying into various final states, such as $ W^+W^-/ZZ//H_1H_1$.  Among them the strongest limits comes from the multi-lepton search in the channel $p p \to H_2 \to Z Z$~\cite{Aad:2020fpj}. In the model under consideration,  $H_2$ has an additional decay mode $H_2\to\phi_D^\dagger\phi_D$, which is governed by the coupling $\lambda_{SD}$. For large value of the coupling $\lambda_{SD}$ this decay mode can be dominant over $H_2\to H_1H_1/W^+W^-/ZZ$. Fig.~\ref{fig:d1} shows the contours of BR($H_2\to \phi_D^\dagger\phi_D/W^+W^-/ZZ//H_1H_1/t\bar{t}$) in the $M_{H_2}$ - $\lambda_{SD}$ plane. The expressions for the respective partial decay widths are given in the appendix. For this plot we fix the scalar mixing angle, $\sin\theta=0.3$. The gray shaded region represents $\text{BR}(H_2\to\phi_D^\dagger\phi_D)\ge0.95$. As $\text{BR}(H_2\to\phi_D^\dagger\phi_D)$ grows with $\lambda_{SD}$,  $\text{BR}(H_2\to ZZ)$ decreases and hence, the bound on scalar mixing angle $\sin\theta$ becomes weaker for a fixed mass of $H_2$. This is shown in Fig.~\ref{fig:d2} for three illustrative mass points  of $H_2$, $M_{H_2}=350,500,1000$ GeV. Here we translate the observed limit on $\sigma(pp\to H_2\to ZZ)$ from ATLAS search~\cite{Aad:2020fpj} into $\sin\theta- \lambda_{SD}$ plane. The shaded regions are disallowed for the respective values of $M_{H_2}$. For a smaller value of $\lambda_{SD} < 10^{-2}$, for which $H_2 \to ZZ$ branching is significantly larger, a very tight constraint $\sin \theta < 0.2$ appears for $M_{H_2}< 500$ GeV.

\begin{figure}[h]
	\centering
	\includegraphics[width=8.cm,height=6cm]{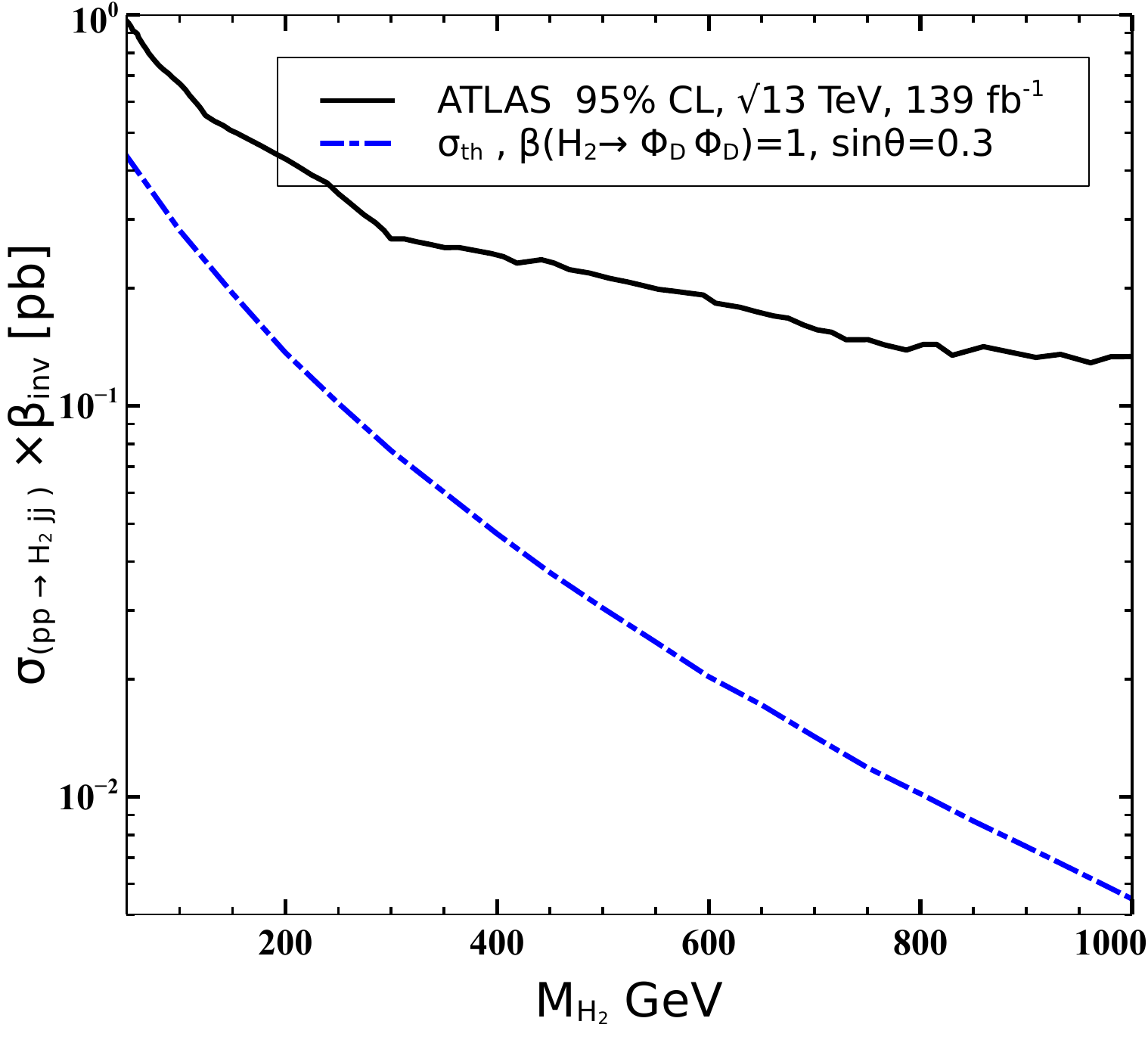}
	\caption{Upper limit on $\sigma(pp\to H_2 jj) \times \text{BR}(H_2\to \text{inv})$ as a function of $M_{H_2}$~\cite{ATLAS:2020cjb}, and its comparision with the theory prediction.}  \label{fig:limitinv}
\end{figure}

\begin{figure}[h]
	\centering
	\includegraphics[width=6.5cm,height=4cm]{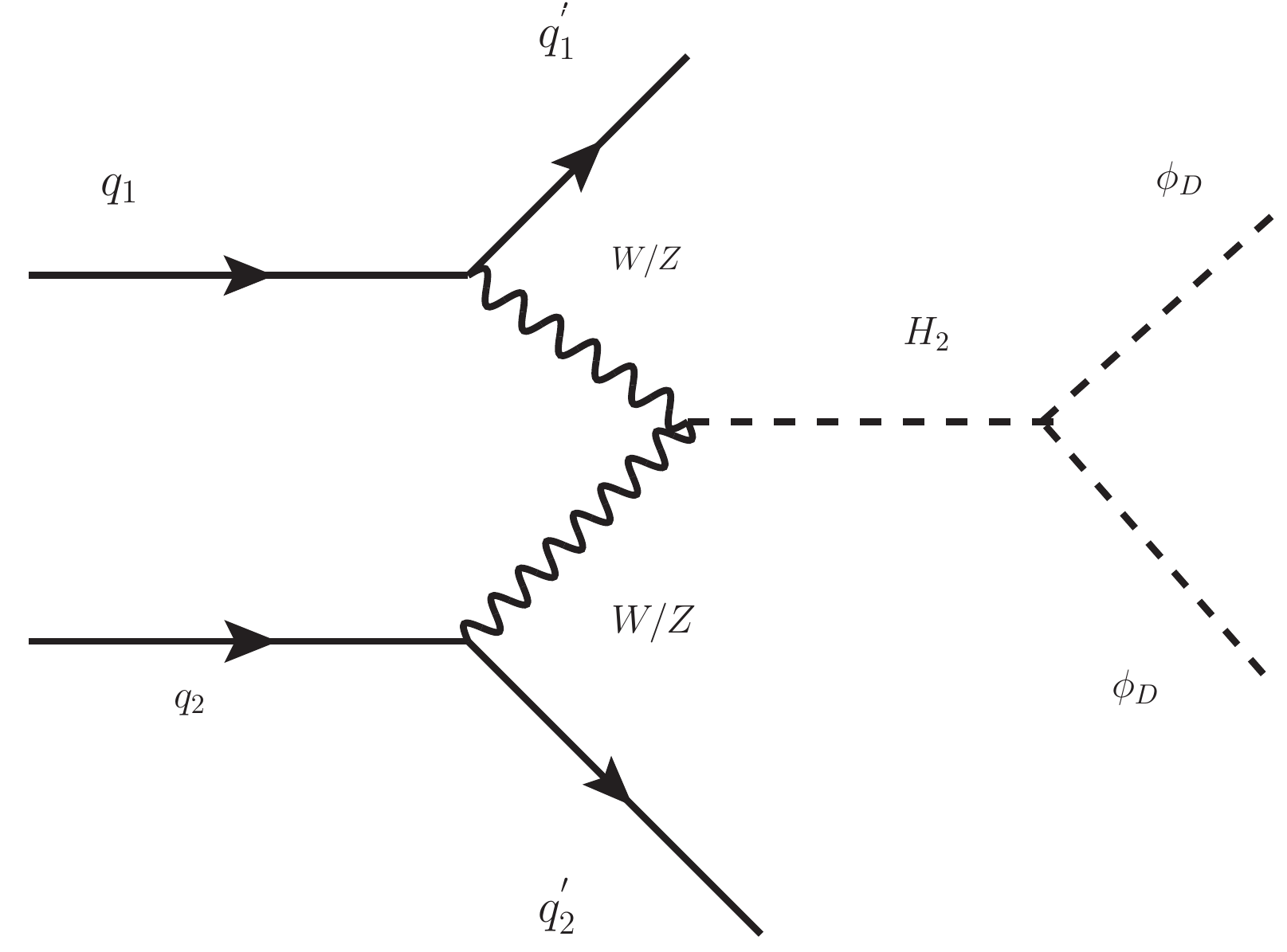}
	\caption{Feynman diagram for the VBF process producing a pair of $\phi_D$.}  \label{fig:vbf_fymn}
\end{figure}

The CMS and ATLAS collaborations have also performed searches for Higgs boson decaying invisibly. For our parameter choice $H_1 \to \phi_D^\dagger\phi_D$ is closed. Recently ATLAS has searched for such invisible decay of Higgs through vector boson fusion (VBF) production channel and interpreted the result for a heavy scalar particle~\cite{ATLAS:2020cjb}. In Fig.~\ref{fig:limitinv}, the black curve shows the observed bounds on the cross-section times branching ratio to invisible final states of the heavy Higgs from this ATLAS search. The blue-dashed curve represents the theory prediction for $\text{BR}(H_2\to \text{inv})=1$ and for the scalar mixing angle $\sin\theta=0.3$. In deriving this, we consider a simplistic parton-level analysis with MadGraph5 and do not consider any specific cut-efficiencies. Our theory cross-section agrees with the observed limit in the entire mass range, which is evident from this plot. In the upcoming section, we examined the reach of HL-LHC to search for $H_2$ decaying invisibly through VBF production mode. The Feynman diagram for this process is shown in Fig. \ref{fig:vbf_fymn}. 
\subsection{Search for $H_2\to\phi_D^{\dagger}\phi_D $ via VBF production mode}
The VBF process is one of the most promising channels to search for the invisible decay of Higgs boson~\cite{Eboli:2000ze}. Recently the ATLAS~\cite{ATLAS:2020cjb} and CMS~\cite{CMS:2018yfx} collaborations have studied the SM Higgs decay to invisible particles and constrained such production processes. Invisible decay of the SM Higgs boson through the VBF channel has been studied for the Higgs portal models ~\cite{Craig:2014lda,Heisig:2019vcj}, for Inert-doublet model~\cite{Dercks:2018wch}. Below, we investigate the production of the BSM scalar $H_2$ via the VBF process and its subsequent decay to the invisible state $\phi_D$. Note that, owing to a very tiny coupling $Y_{D\chi} \sim \mathcal{O}(10^{-12})$, $\phi_D$ state decays outside the detector.

 After gluon-fusion, VBF is the dominant channel for Higgs bosons production at the LHC, characterised by the two highly energetic forward jets~\cite{Kleiss:1987cj}. The two VBF jets are widely separated in pseudo-rapidity, lying in the opposite hemisphere of the detector. For the invisible decay $H_2\to\phi_D^{\dagger}\phi_D$, the signal is marked by a large transverse momentum imbalance. All these features allow us to discriminate between the signal and background. The dominant SM processes that mimic the signal are $pp\to Z(\to\nu\nu)jj$ and $pp\to W^{\pm}(\to\nu \ell^\pm)jj$. The latter process contributes when the charged lepton is
not detected. QCD multi-jet events with large missing transverse
momentum (MET), arising from the mismeasurement of jet energy, can also imitate the VBF signal. A suppressed central jet activity accompanies the VBF signal. On the contrary, QCD jets are more central in the detector. Therefore, the central jet veto and a strong cut on MET could reduce QCD multi-jet events. Another potential background $t\bar{t}$ can be suppressed by vetoing $b$ jets and leptons. 

\textbf{Event Simulation:-} 
We implement the Lagrangian of this model in FeynRules(v2.3)~\cite{Alloul:2013bka}. The generated UFO files are used in the MC event generator MADGRAPH5(v2.6)~\cite{Alwall:2014hca} to generate the signal events at the leading order. Partonic events are passed through PYTHIA8~\cite{Sjostrand:2014zea} to perform showering and hadronization.  We implement a cut-count analysis code in CheckMate~\cite{Drees:2013wra,Kim:2015wza}, to calculate the signal and background cut efficiencies. CheckMate makes use of Delphes~\cite{deFavereau:2013fsa} for the simulation of detector effect, and Fastjet~\cite{Cacciari:2011ma,Cacciari:2005hq} for jet clustering. We use anti-$k_t$ jet clustering algorithm~\cite{Cacciari:2008gp} with radius parameter, $R=0.4$. We estimate the sensitivity of the invisible signature of the $H_2$ produced via VBF process at the $pp$-collider, $pp\to H_2 jj \to \phi_D^{\dagger} \phi_Djj$. Here $\phi_D$ being a stable particle at the detector length scale gives rise to MET.  Thus, the process under consideration leads to $2j+MET$ signature at collider. Among the other SM processes that can fake the signal we simulate the two most dominant processes which are $pp\to Z jj \to \nu \nu jj$ and $pp\to W^{\pm} jj \to \nu \ell^{\pm} jj$. We consider the HL-LHC for this study, which is planned to operate with $\sqrt{s}=14$ TeV and $\mathcal{L}=3000/\text{fb}$. 

Although the signal consists of two jets at the parton level, additional jets can arise due to initial and final state radiation after the parton shower. Thus, we consider up to two extra jets in the final state to simulate backgrounds. 
During the generation of background events we demand transverse momentum ($p_T$) of the leading partons: $p_{T}(j_{1,2}) > 50\, \textrm{GeV}$, the pseudo-rapidity: $|\eta_{j}| < 5.0 $. We simulate merged sample with 2-4 jets for $W+jets$ and $Z+jets$ using the MLM matching scheme~\cite{MLM:2003ml}. We consider the parameter xqcut=55 GeV which is the minimum jet measure $(p_T/k_T)$ between partons. Partonic cross-sections for backgrounds $W+jets$ and $Z+jets$ are $528.755\times 10^3$ fb and $223.603\times 10^3$ fb, respectively. Next to leading order~(NLO) QCD corrections for $W+jets$ and $Z+jets$ are given in~\cite{Campbell:2003hd}, which are negative, and the corresponding K-factors are $0.87$ and $0.905$ , respectively. We perform the simulation with a harder $p_T$ cut on jets compared to that in the mentioned reference. As, K-factor depends on the kinematic cuts we do not normalise the cross-sections to NLO. For signal we also consider LO cross-section. For $M_{H_2}=(350-1000) $ GeV, K-factor varies in the range $1.018-0.979$~\cite{Bolzoni:2011cu}. The sensitivity can be improved for reduced background cross-sections and enhanced signal cross-section.

\begin{figure}
	\centering
    	\subfigure[]{\includegraphics[width=4.80cm,height=6.0cm]{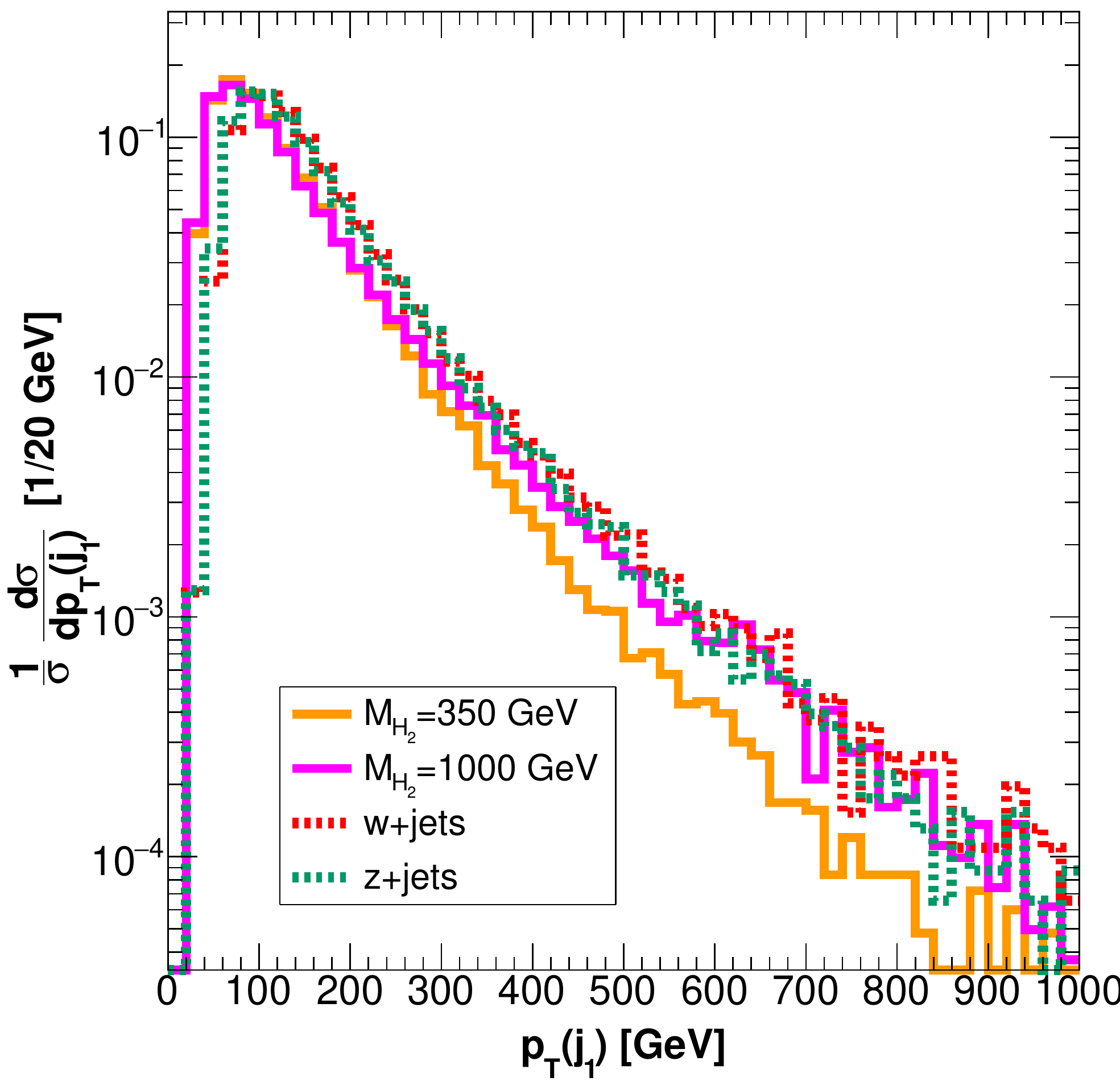}\label{fig:histo1}}
		\subfigure[]{\includegraphics[width=4.80cm,height=6.0cm]{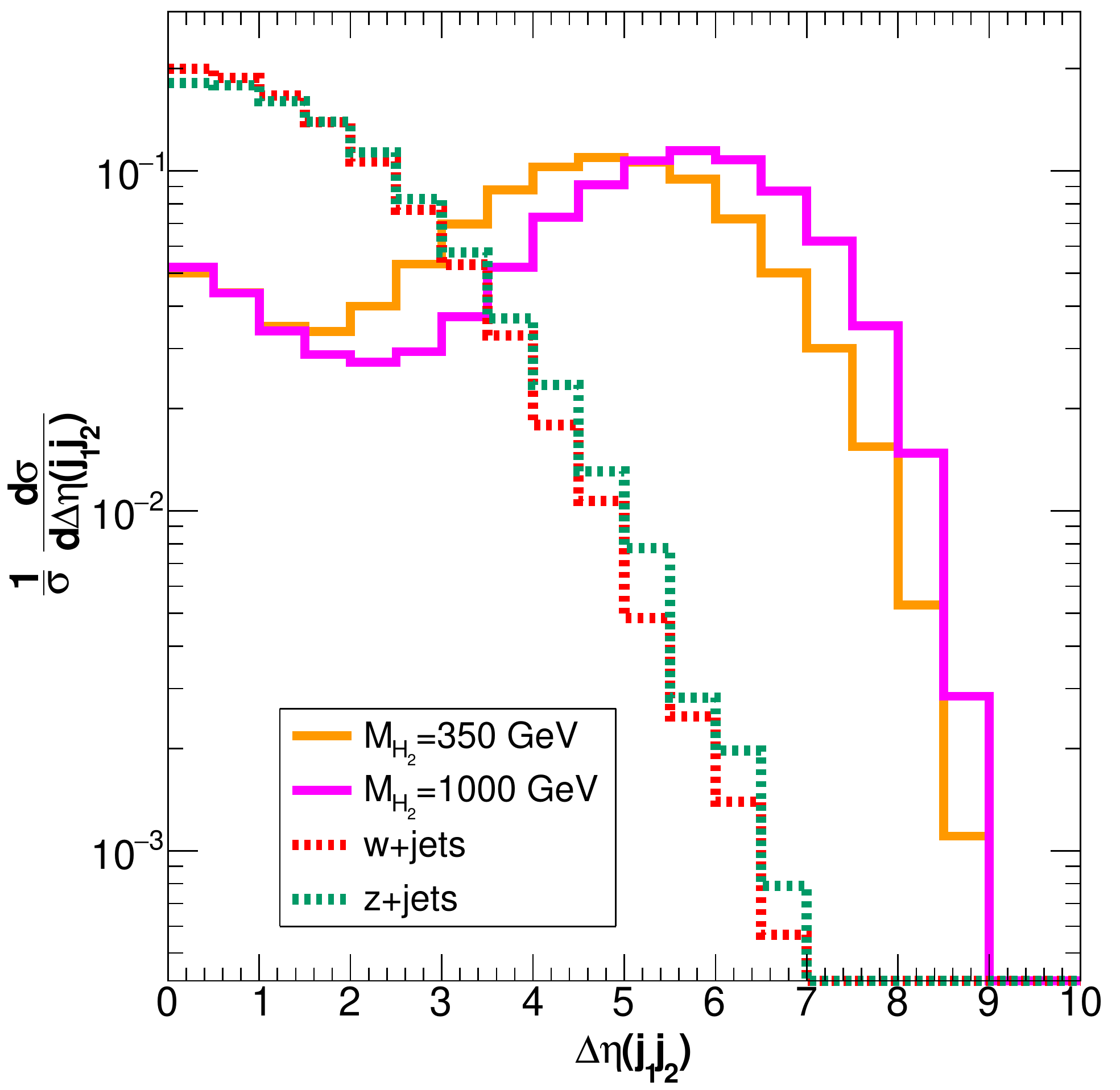}\label{fig:histo2}}
		\subfigure[]{\includegraphics[width=4.80cm,height=6.0cm]{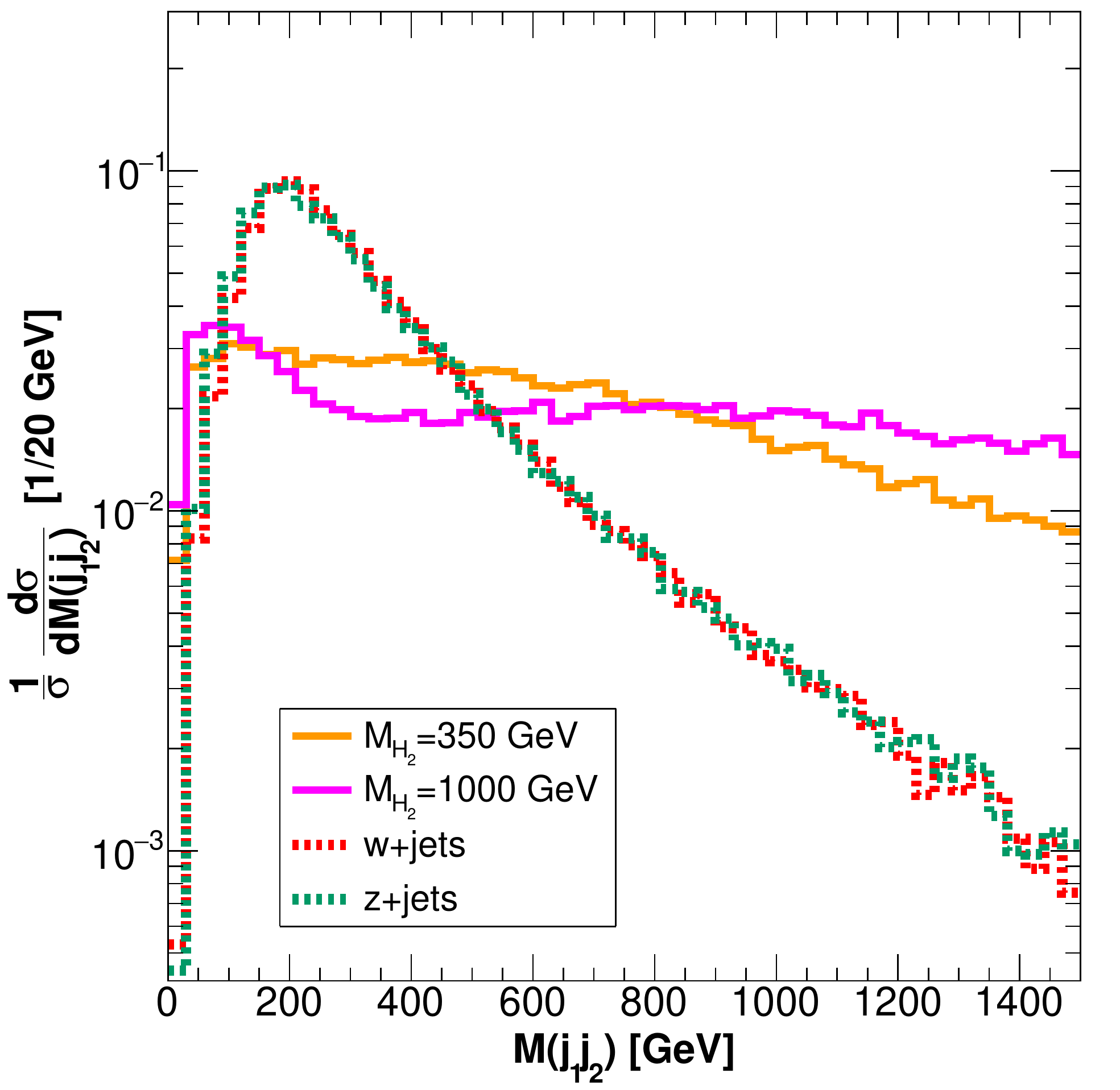}\label{fig:histo3}}
	\caption{Normalised Distribution of transverse momentum  of the leading jet~(Fig.~\ref{fig:histo1}), difference in pseudo-rapidity~(Fig.~\ref{fig:histo2}) and  invariant mass of the leading and sub-leading jets~(Fig.~\ref{fig:histo3}), for both signal and background events. The distributions are without any selection cuts.  }.\label{fig:histo}
\end{figure}

  Fig.~\ref{fig:histo} shows the normalised distributions of some of the kinematic variables for both signal and background which motivate to design the selection cuts. Fig.~\ref{fig:histo1} corresponds to the distributions for transverse momentum  of the leading jet~($p_T(j_{1})$) which shows that the background events possess comparatively harder jets than signal events due to the generation level $p_T$ cut. Fig.~\ref{fig:histo2} and Fig.~\ref{fig:histo3} represent the difference in pseudo-rapidity~($|\Delta\eta(j_1,j_2)|$) and  invariant mass~($M(j_1j_2)$) of the leading and sub-leading jets. Majority of the signal events hold higher $|\Delta\eta(j_1,j_2)|$ with respect to the background events and hence, larger  $M(j_1j_2)$ as both variables are connected with the equation $M(j_1j_2)\simeq \sqrt{p_T(j_1) \ p_T(j_2) \ e^{\Delta\eta(j_1,j_2)}}$~\footnote{This relation can be derived using the transformation of the four-momentum to $p_T,\ \eta\  \text{and}\  \phi$ variables.}.

\textbf{Event selection and Results:-}
We closely follow the cuts used in ATLAS search~\cite{ATLAS:2020cjb}, which are as follows
\begin{itemize}
	\item We select events with 2 to 4 jets with $p_T(j)\ge 25$ GeV and $|\eta(j)|<4.5$, $p_T(j_1)\ge 60$ GeV and $p_T(j_2)\ge 50$ GeV.
	\item We veto events with more than one b-tagged jets.
	\item We select the events with no lepton and photon candidates. $p_T$ and $\eta$ requirement for lepton and photon are: $p_T(e)\ge 4.5$ GeV, $p_T(\mu)\ge 4$ GeV, $p_T(\gamma)\ge 20$ GeV, $|\eta(e)|<2.47$, $|\eta(\mu)|< 2.7$, $|\eta(\gamma)|< 2.37$. 
	\item We demand missing transverse momentum $E_T^{\text{miss}}\ge150$ GeV.
	\item For the leading and sub-leading jet: $\Delta\phi(j_1,j_2)< 2.0$, $\eta_{j_1} \times\eta_{j_2}<0$ and $\Delta\eta(j_1,j_2)\ge3.8$
	\item The invariant mass of the leading and sub-leading jet: $M(j_1j_2)\ge 600$ GeV.
\end{itemize}
	\begin{table}
		\begin{tabular}{|l|l|l|l|l|l|}
			\hline
			& \multicolumn{3}{l|}{Signal efficiency $\epsilon_{\rm{s}}$ for $M_{H_2}$ } & \multicolumn{2}{l|}{Background efficiency $\epsilon_{\rm{b}}$}  \\ 
			\hline
			cuts& 350 GeV  &$500$ GeV &$1000$ GeV &$W^\pm +jets$& $Z+jets$   \\ \hline	   
			$p_T(j_{1,2})\ge (60,50)$ GeV&0.38722     &         0.39005  &   0.36685   &      0.77382    &     0.77113\\ \hline         
			$n_{b-jet}\le1$&0.38552     &         0.38827  &   0.36551   &      0.76276    &     0.76004\\ \hline         
			$n_{\ell^\pm,\gamma}=0$&0.3382      &         0.34376  &   0.32787   &      0.05173   &     0.62245\\ \hline         
			$E_T^{\text{miss}}\ge150$ GeV&0.14735     &         0.15873  &   0.15835   &   0.007814&     0.13348 \\ \hline        
			$\Delta\phi(j_1,j_2)<2.0$&0.1208      &         0.12989  &   0.12976   &      0.002453   &     0.05685\\ \hline         
			$\eta_{j_1}.\eta_{j_2}<0$, $\Delta\eta(j_1,j_2)\ge3.8$&0.07044     &         0.08143  &   0.08923   &   $7.9 \times 10^{-5}$      &     0.00293\\ \hline         
			$M(j_1j_2)\ge 600$ GeV&0.06922     &         0.08042  &   0.08863   &   $7.3 \times 10^{-5}$       &     0.00290\\ 
			\hline
			 \multicolumn{5}{|l}{\qquad signal cross-section $\sigma_{s}=2 \sqrt{\epsilon_{\rm{b}}\sigma_{\rm{b}}}/(\epsilon_s \sqrt{\mathcal{L}})$ for $2\sigma$ significance with $\mathcal{L}=3000/\text{fb}$ }&\\
			\hline
			&13.831 fb&11.905 fb&10.802 fb&--&--\\ \hline
		\end{tabular}
		\caption{Cumulative cut efficiencies for Signal:  $pp\to H_2 jj \to 2j+MET$ and the SM background: $W+jets$ and $Z+jets$. The numbers in last row are the required signal cross-sections to obtain $2\sigma$ significance using $3000/\text{fb}$ luminosity.
		}\label{tab:cuteff}
	\end{table}	
In Table.~\ref{tab:cuteff}, we present the cut efficiencies of the cuts mentioned above for signal and the SM background. We find the lepton veto to be most effective in reducing the $W^\pm +jets$ events. Finally, after the cuts on $\Delta\eta(j_1,j_2)$ and $M(j_1j_2)$ a significant fraction of both $W^\pm +jets$ and $Z +jets$ events are cut down. The effect of these two cuts is similar as both variables are related.

 In the last row, we write the required signal cross-section~(in fb) before applying selection cuts to obtain $2\sigma$ significance for $\mathcal{L}=3000/\text{fb}$ luminosity. This is calculated from the relation $\sigma_{s}= n_\sigma \sqrt{\epsilon_{\rm{b}}\sigma_{\rm{b}}}/(\epsilon_s \sqrt{\mathcal{L}})$, where $\sigma_{\rm{s}}$ ($\sigma_{\rm{b}}$) is the initial signal~(background) cross-section, $\epsilon_s$ and $\epsilon_b$ is the corresponding cut efficiency, and $n_\sigma$ is the significance.  

\begin{figure}[H]
	\centering
	\includegraphics[width=8cm,height=6.5cm]{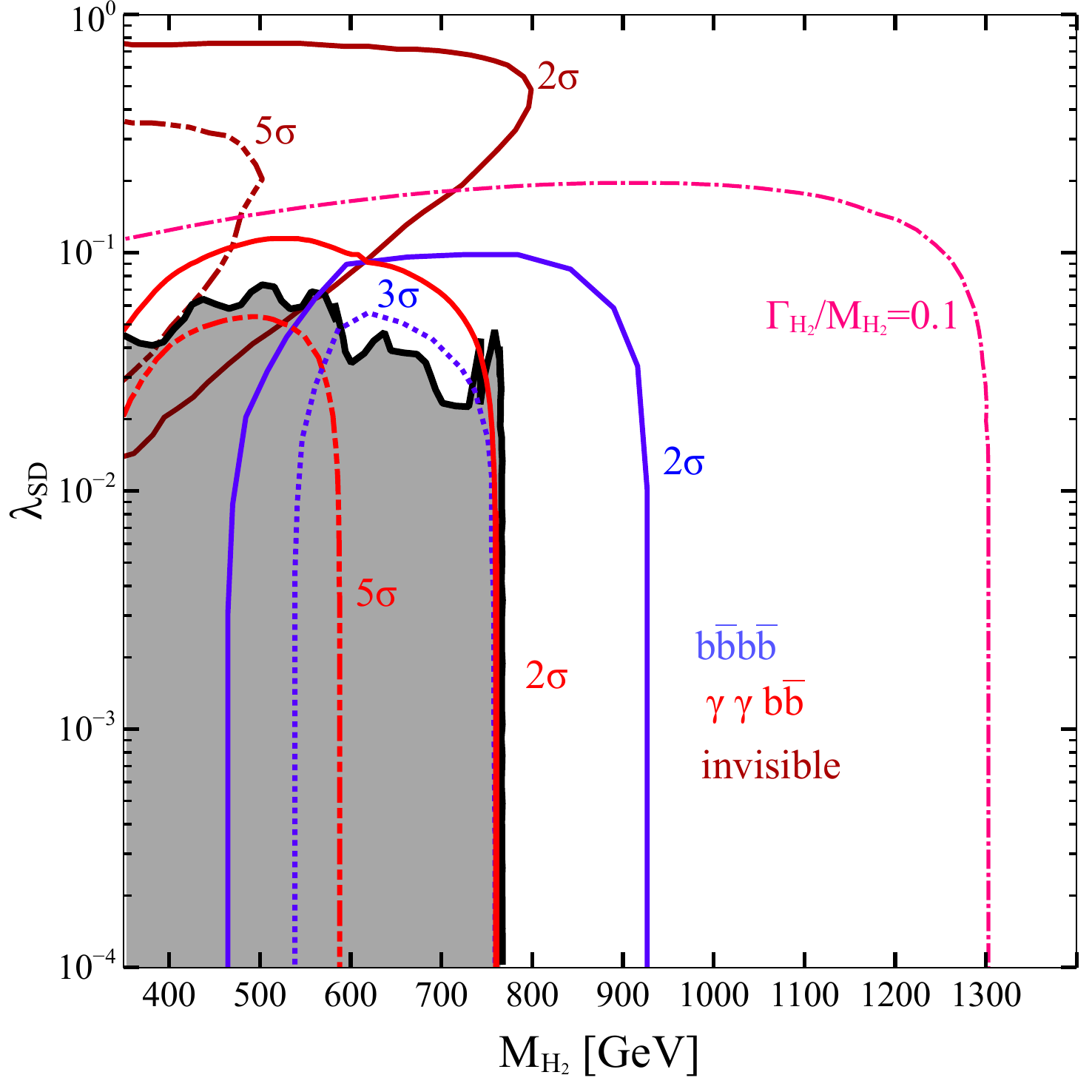}
	\caption{HL-LHC prediction for  $pp\to H_2jj\to jj +MET$, $pp\to H_2\to H_1 H_1\to 2 b +2 \gamma$  and $pp\to H_2\to H_1 H_1\to 4 b$ indicated by brown, red and blue color contours in $M_{H_2}-\lambda_{SD}$ plane. Black shaded region is ruled out from the ATLAS search for $H_2\to ZZ$~\cite{Aad:2020fpj}. See text for details. }.\label{fig:prediction}
\end{figure}

In Fig.~\ref{fig:prediction}, we show the HL-LHC prediction for invisible signature of $H_2$ in $M_{H_2}-\lambda_{SD}$ plane assuming $\sin\theta=0.3$. 
The signal significance is calculated for $3000/\text{fb}$ luminosity. The brown solid (brown dashed-dot) line indicates to $2\sigma$ ($5\sigma$) sensitivity. As per our results, $2\sigma$ significance can be achieved upto $M_{H_2}\simeq 800$ GeV for $\lambda_{SD}$ in between $10^{-2}$ and $1$. We also highlighted the region excluded from the ATLAS search for $H_2 \to ZZ$~\cite{Aad:2020fpj} by the black-shading. As these results are based on narrow width approximation, we show the $\Gamma/M_{H_2} =0.1$ contour by the pink line. The area enclosed by it corresponds to $\Gamma/M_{H_2} <0.1$. Note that to obtain the results for the invisible search we take into account the width effect.

 The sensitivity of the visible signature of $H_2$ has been analysed in ref.~\cite{Adhikary:2018ise}. Here the authors have considered the production of $H_2$ via gluon fusion process and subsequent decay to di-Higgs, $pp\to H_2\to H_1 H_1$. They have studied various final states depending on the decay of SM Higgs $H_1$ and have shown that $2 b +2 \gamma$ and $4 b$ signatures are more sensitive compared to others. In Fig.~\ref{fig:prediction}, we shows the HL-LHC predictions for di-Higgs channel in $2 b +2 \gamma$ and $4 b$ final state obtained from the ref.~\cite{Adhikary:2018ise}. Here we assume $\sin\theta=0.3$. For $2 b +2 \gamma$ signature, regions enclosed by red solid and dashed-dot curves represent $2\sigma$ and $5\sigma$ sensitivity, respectively. Similarly blue curves show $2\sigma$ and $3\sigma$ sensitivity for $4b$ final state.

\section{Conclusion}\label{conclusion}
We analyse the thermal freeze-in and non-thermal freeze-in production of DM in an extended, gauged $B-L$ model where dark sector fermion $\chi$ serves as the DM candidate. In this work, we have a secluded dark sector containing feeble interacting DM candidate $\chi$ and a complex scalar field $\phi_D$ charged under $B-L$ symmetry. The DM fails to thermalise with the surrounding plasma due to suppressed interaction with other bath particles owing to small coupling $Y_{D\chi}$. At an early epoch, it is primarily produced via the decay of $\phi_{D }$ via thermal freeze-in mechanism and through the late decay of $\phi_{D }$ via the non-thermal freeze-in mechanism. Contrary to the DM field $\chi$, the dark sector scalar field $\phi_{D }$ thermalises with the bath particles due to large gauge coupling as well as sizeable interactions with SM and BSM Higgs states. The correct abundance of $\phi_{D }$ at decoupling is therefore obtained via the freeze-out mechanism. The annihilation of $\phi_{D }$ mediated via $Z_{BL}$, i.e., $\phi_{D }^{\dagger}\phi_{D }\to Z_{BL}\to f\bar{f}$ etc decouples from the thermal bath at an earlier epoch compared to the processes mediated via SM and BSM Higgs. The abundance of $\phi_{D }$ is hence primarily governed by the interplay of scalar quartic couplings $\lambda_{ SD}$, $\lambda_{ Dh}$ and SM-BSM Higgs mixing angle $sin\ \theta$. For our choice of $\phi_D$ and $\chi$ masses, $\phi_{D }$ decaying to $\chi$ and RHN is kinematically open, and this is the primary production process for DM. We subdivide the discussion into two scenarios, {\it Scenario-I}, where the thermal freeze-in via $\phi_D \to \chi N$ contribution is significant, and {\it Scenario-II}, where the late decay of $\phi_D \to \chi N$ is primarily responsible in satisfying the DM relic abundance. This is to note that the late production of DM through out-of-equilibrium decay of $\phi_{D }$ will depend on the abundance of $\phi_{D }$ at decoupling obtained through the freeze-out mechanism. Therefore, for suppressed interaction of $\phi_{D }$ with the bath particles leading to high abundance of $\phi_{D }$ at decoupling significantly enhances the production of $\chi$. We also study mixed scenarios, where the correct relic abundance of $\chi$ is governed via both the thermal and non-thermal freeze-in mechanism. We majorly focus our discussion around two benchmark points, 1 and 2. We find that 
\begin{itemize}
\item
Due to relatively large SM-BSM Higgs mixing angle, as well as due to the choice of large quartic couplings $\lambda_{SD}$, $\lambda_{Dh}$, $\phi_D$ decouples much later in {\it Scenario I}, and hence its late decay contributes negligibly to the DM abundance. Here, the primary production of $\chi$ occurs due to thermal freeze-in. 
\item
For benchmark 2 ({\it Scenario II}), the small SM-BSM Higgs mixing and choice of $\lambda_{SD}$ and $\lambda_{Dh}$ enables $\phi_D$ to be out-of-equilibrium in an earlier epoch leading to a larger $\phi_D$ abundance. In this case, the $\chi$ abundance can be built-up primarily from the late decay of $\phi_D$. 
\item
We find that for {\it Scenario-I}, which corresponds to benchmark 1, the dark sector scalar $\phi_D$ can be produced at a $p p $ collider. This occurs because the SM-BSM Higgs mixing angle is relatively larger to realise this scenario. The $\phi_D$ is produced from BSM Higgs decay in the VBF channel.
\end{itemize}
In addition to the detailed DM analysis, we also evaluate the prospect of detection of this model at the HL-LHC. In particular, in Section.~\ref{sec:collider}, we investigate the possibility of probing the coupling $\lambda_{SD}$ of $\phi_D$ with the heavy scalar $H_2$ at the HL-LHC. This coupling enables an extra decay mode of $H_2$ to a pair of $\phi_D$. When this decay mode becomes dominant, the existing bounds on the mass of $H_2$ and the scalar mixing angle weaken. To study $H_2 \to \phi_D^\dagger \phi_D$, we consider the production of $H_2$ from the VBF process, characterised by two forward jets with a large pseudo-rapidity gap. In our case, $\phi_D$ is stable over the detector length scale resulting in an extra distinguishing feature, the missing transverse momentum. For a fix mass of $\phi_D$, we present the $5\sigma $ discovery and $2\sigma $ exclusion contours in the $M_{H_2}-\lambda_{SD}$ plane. Following a simple cut-count analysis we show that $5\sigma$ sensitivity can be obtained for $M_{H_2}\simeq(350- 500)$ GeV and $\lambda_{SD}$ in between $0.03$ and $0.35$. Similarly, $2\sigma$ exclusion limit can be placed for the mass range $\simeq(350- 800)$ GeV and for $\lambda_{SD}\simeq(10^{-2}-1) $. 
\section*{Acknowledgments}
MM acknowledges the DST-INSPIRE Research Grant  IFA14-PH-99   and research grant from CEFIPRA  (Grant no: 6304-2). PB acknowledges SERB CORE Grant CRG/2018/004971 and MATRICS Grant MTR/2020/000668. RP and AR acknowledges SAMKHYA: High-Performance Computing Facility provided by the Institute of Physics (IoP), Bhubaneswar. RP thanks Dr. Shankha Banerjee for useful discussion regarding the collider analysis.
\appendix
\section{Analytical Expressions of relevant cross-sections and decay widths}
We provide the expressions for the relevant cross-sections and decay widths, involved in the coupled Boltzmann equations.
\subsection{Decay width of $\phi_{D }$}\label{appen1}
\begin{itemize}
	\item $\Gamma(\phi_{D} \rightarrow \chi N ) =\frac{y_{D\chi}^2}{8\pi}\frac{m_{\phi_{D }}^2 -(m_{\chi}+m_{N})^2 }{m_{\phi_{D }}^3}\bar{\lambda}^{\frac{1}{2}} (m_{\phi_{D }^{2}},m_{\chi}^2,m_N^{2}),$
	
	\item $ \Gamma(\phi_{D} \rightarrow \chi \nu ) =\frac{1}{16\pi} \frac{y_{D\chi}^2 y_{N}^2}{m_N^{2}}m_{\phi_{D }}\Big(1-\frac{m_{\chi}^2}{m_{\phi_{D }}^2}\Big)^2,$\\
	where in the above, $\bar{\lambda}(x^2,y^2,z^2)=x^2+y^2+z^2-2xy-2yz-2zx$ is a K$\ddot{a}$len function.	
\end{itemize}
\subsection{Decay width of $N$}\label{appen2}
The two body decay width of $N_{R}$ when $m_N$ is larger than $m_W,m_Z$ and $m_{H_1}$ are as follows,  
\begin{itemize}
	\item $\Gamma(N \to \chi \phi) = \frac{y_{D\chi}^2}{16 \pi}
	\frac{(m_\chi + m_N)^2-m_{\phi_D}^2 }{m_N^3}\bar\lambda^{1/2}\left(m_{\phi_D}^2,m_\chi^2, m_N^2 \right ) ,$
	
	\item $\Gamma(N\to H_1 \nu)=\Gamma(N\to H_1 \bar{\nu}) = \frac{y_N^2 m_N}{64\pi} \left(1- \frac{m_{H_1}^2}{m_N^2}\right)^2,$ 
	
	\item $\Gamma(N\to \ell^- W^+)= \Gamma(N\to \ell^+ W^-)	= \frac{y_N^2 m_N}{32\pi} \left(1- \!\frac{m_W^2}{m_N^2}\right)^2\!\! \left(1+ 2 \frac{m_W^2}{m_N^2}\right),$ 
	
	\item $\Gamma(N\to Z \nu ) = \Gamma(N\to Z \bar{\nu})	=  \frac{y_N^2 m_N}{64\pi}  \left(1-\frac{m_Z^2}{m_N^2}\right)^2 \!\! \left(1+ 2 \frac{m_Z^2}{m_N^2}\right).$
\end{itemize}
For $M_{N} < m_{W^{\pm}} ,m_{Z} $, it decays to the three SM fermions through off-shell $W$, and $Z$ gauge bosons. The three body decay widths are as follows,

\begin{align}
\Gamma(N \to l_\alpha^-u \bar{d})=N_c |V^{CKM}_{ud}|^2 |U_\alpha|^2\frac{ G_F^2 M_{N}^5}{192 \pi^3} \mathcal{I}(x_{u},x_{d},x_{l})
\end{align}
Here, $\mathcal{I}(x_{u},x_{d},x_{l})= 12 \int_{(x_{d}+x_{l})^2}^{(1-x_{u})^2}\frac{dx}{x} (1+x_{u}^2-x) (x-x_{d}^2-x_{l}^2)\lambda^{\frac{1}{2}}(1,x,x_{u}^2)\lambda^{\frac{1}{2}}(x,x_{l}^2,x_{d}^2)$ and $x_{u/d/l}=\frac{m_{u/d/l}}{M_{N}}$, $\lambda(a,b,c)=a^2+b^2+c^2-2ab-2bc-2ca$, 
and  $N_c=3$ is the color factor.

\begin{align}
\Gamma(N \to l_\alpha^-\nu_\beta l^+_\beta)= |U_\alpha|^2\frac{ G_F^2 M_{N}^5}{192 \pi^3} \mathcal{I}(x_{l_\alpha},x_{l_\beta},x_{\nu_\beta})
\end{align}
\begin{align}
\Gamma(N \to \nu_\alpha f \bar{f})= N_c |U_\alpha|^2\frac{ G_F^2 M_{N}^5}{192 \pi^3}
\Big[ C_1^f\Big( (1-14x^2-2x^4-12x^6)\sqrt{1-4x^2}+12 x^4(x^4-1)L(x)\Big) + & \nonumber\\ 
4C_2^f\Big( x^2(2+10x^2-12x^4)\sqrt{1-4x^2}+6 x^4(1-2x^2+2x^4)L(x)\Big)\Big]
\end{align}
Here $x=\frac{m_{f}}{M_{N}}$, $L(x)=\log\Big[ \frac{1-3x^2-(1-x^2)\sqrt{1-4x^2}}{x^2(1+\sqrt{1-4x^2})}\Big]$. The values of $C_1^f$ and $C_2^f$ are given in \cite{Bondarenko:2018ptm}


\subsection{Cross-Sections for relevant processes}\label{appen3}
Here we give relevant annihilation cross-sections for $\phi_{D }$ depletion process are follows,
\begin{itemize}
	\item $\underline{\phi_{\mathrm{D}}^\dagger \phi_{\mathrm{D}}\rightarrow H_1 H_1 }$
	\begin{eqnarray}
	&&\lambda_{H_{1} H_{1} H_{1}}= -3\,[2\,v\lambda_{h}\cos^{3}\theta
	+ 2\,v_{BL}\,\lambda_{S}\sin^{3}\theta +
	\lambda_{Sh}\sin\theta\,\cos\theta\,
	(v\sin\theta + v_{BL}\cos\theta)],\nn\\
	&&\lambda_{H_1H_1H_2}= -\lambda_{ Sh}(v_{BL}\cos^{3} \theta +v \sin^{3} \theta)+2v_{BL}(-3\lambda_{ S}+\lambda_{ Sh})\cos\theta\sin^{2}\theta+2v(-3\lambda_{ h}+\lambda_{ Sh})\cos^{2}\theta\sin\theta,\nn \\
	&&\lambda_{H_1H_1\phi_{\mathrm{D}}^\dagger \phi_{\mathrm{D}}}=-(\lambda_{ Dh}\cos^{2} \theta+\lambda_{ SD}\sin^{2} \theta),\nn\\
	&&M_{H_{1} H_{1}} = \Big(\frac{\lambda_{H_1H_1H_1}\lambda_{H_1\phi_{\mathrm{D}}^\dagger \phi_{\mathrm{D}}}}{(s-m_{H_1}^{2}) +	i m_{H_1} \Gamma_{H_1}} +\frac{\lambda_{H_1H_1H_2}\lambda_{H_2\phi_{\mathrm{D}}^\dagger \phi_{\mathrm{D}}}}{(s-m_{H_2}^{2}) +
		i m_{H_2} \Gamma_{H_2}} \Big)-\lambda_{H_1H_1\phi_{\mathrm{D}}^\dagger \phi_{\mathrm{D}}}\ ,\nn\\
	&&\sigma_{\phi_{D}^{\dagger} \phi_{D}\rightarrow H_{1}H_{1}} =\frac{1}{16 \pi s}\sqrt{\frac{s-4m_{H_1}^2}{s-4m_{\phi_{D }}^2}}|M_{H_{1}H_{1}}|^{2}.
	\end{eqnarray}
	
	\item $\underline{\phi_{\mathrm{D}}^\dagger \phi_{\mathrm{D}}\rightarrow H_2 H_2 }$
	\begin{eqnarray}
	&&\lambda_{H_{2} H_{2} H_{2}}= 3\,[2\,v\lambda_{h}\sin^{3}
	\theta - 2\,v_{BL}\lambda_{S}\cos^{3}\theta +
	\lambda_{Sh}\sin\theta\cos\theta\,
	(v\cos\theta - v_{BL}\sin\theta)],,\nn\\
	&&\lambda_{H_2 H_2 H_1}=-[6\,v\lambda_{h}\sin^{2}
	\theta\cos\theta + 6\,v_{BL}\lambda_{S}\cos^{2}\theta\sin\theta 
	-(2-3\,\sin^{2}\theta)v_{BL}\lambda_{Sh}\sin\theta,\nn \\
	&&~~~~~~~~~+(1-3\sin^{2}\theta)v\lambda_{Sh}\cos\theta]\,, \nn \\
	&&\lambda_{H_2 H_2 \phi_{\mathrm{D}}^\dagger \phi_{\mathrm{D}}}=-(\lambda_{ Dh}\sin^{2} \theta+\lambda_{SD}\cos^{2} \theta),\nn\\
	&&M_{H_{2} H_{2}} = \Big(\frac{\lambda_{H_2 H_2 H_2}\lambda_{H_2\phi_{\mathrm{D}}^\dagger \phi_{\mathrm{D}}}}{(s-m_{H_2}^{2}) +i m_{H_2} \Gamma_{H_2}} +\frac{\lambda_{H_2 H_2 H_1}\lambda_{H_1\phi_{\mathrm{D}}^\dagger \phi_{\mathrm{D}}}}{(s-m_{H_1}^{2}) +i m_{H_1} \Gamma_{H_1}} \Big)-\lambda_{H_2 H_2\phi_{\mathrm{D}}^\dagger \phi_{\mathrm{D}}}\ ,\nn\\
	&&\sigma_{\phi_{D}^{\dagger} \phi_{D}\rightarrow H_{2}H_{2}} =\frac{1}{16 \pi s}\sqrt{\frac{s-4m_{H_2}^2}{s-4m_{\phi_{D }}^2}}|M_{H_{2}H_{2}}|^{2}.
	\end{eqnarray}
	
	\item $\underline{\phi_{\mathrm{D}}^\dagger \phi_{\mathrm{D}}\rightarrow W^{+}W^{-} }$
	\begin{eqnarray}
	g_{H_{1}WW} &=& \frac{2m_{W}^{2} \cos\theta}{v}\,,\nn \\
	g_{H_{2}WW} &=& \frac{2m_{W}^{2} \sin\theta}{v}, \nn \\
	M_{WW} &=&\frac{2}{9}\,\left(1 + \frac{(s - 2m_{W}^{2})^{2}}
	{8m_{W}^{4}}\right)\left(\frac{g_{H_{1}WW} \lambda_{H_{1}\phi_{D}^{\dagger}
			\phi_{D}}}{(s-m_{H_1}^{2}) + i m_{H_1} \Gamma_{H_1}}
	+ \frac{g_{H_{2}WW}\,\lambda_{H_{2}\phi_{D}^{\dagger} \phi_{D}}}
	{(s-m_{H_2}^{2}) + i m_{H_2} \Gamma_{H_2}}\right), \nn \\
	\sigma_{\phi_{DM}^{\dagger} \phi_{DM} \rightarrow WW} &=& \frac{1}{16 \pi s}\,\,
	\sqrt{\frac{s - 4m_{W}^{2}}{s - 4m_{\phi_{D }}^{2}}}|M_{WW}|^{2}\,.
	\end{eqnarray}
	
	\item $\underline{\phi_{\mathrm{D}}^\dagger \phi_{\mathrm{D}}\rightarrow ZZ }$
	\begin{eqnarray}
	g_{H_{1}ZZ} &=& \frac{2m_{Z}^{2} \cos\theta}{v}\,,\nn \\
	g_{H_{2}ZZ} &=& \frac{2m_{Z}^{2} \sin\theta}{v}, \nn \\
	M_{ZZ} &=&\frac{2}{9}\,\left(1 + \frac{(s - 2m_{Z}^{2})^{2}}
	{8m_{Z}^{4}}\right)\left(\frac{g_{H_{1}ZZ} \lambda_{H_{1}\phi_{D}^{\dagger}
			\phi_{D}}}{(s-m_{H_1}^{2}) + i m_{H_1} \Gamma_{H_1}}
	+ \frac{g_{H_{2}ZZ}\,\lambda_{H_{2}\phi_{D}^{\dagger} \phi_{D}}}
	{(s-m_{H_2}^{2}) + i m_{H_2} \Gamma_{H_2}}\right), \nn \\
	\sigma_{\phi_{DM}^{\dagger} \phi_{DM} \rightarrow ZZ} &=& \frac{1}{16 \pi s}\,\,
	\sqrt{\frac{s - 4m_{Z}^{2}}{s - 4m_{\phi_{D }}^{2}}}|M_{ZZ}|^{2}\,.
	\end{eqnarray}
	\item $\underline{\phi_{\mathrm{D}}^\dagger \phi_{\mathrm{D}}\rightarrow N N }$
	\begin{eqnarray}
	\lambda_{H_{1}N_{R}N_{R}} &=& \frac{y_{N}\sin \theta}{\sqrt{2}}  \nn \\
	\lambda_{H_{2}N_{R}N_{R}} &=& \frac{y_{N}\cos \theta}{\sqrt{2}}  \nn \\
	M_{N_{R}N_{R}} &=& \frac{\lambda_{H_{1}N_{R}N_{R}}\,
		\,\lambda_{H_{1}\phi_{D}^{\dagger} \phi_{D}}}{(s-m_{H_1}^{2}) +
		i m_{H_1} \Gamma_{H_1}} + \frac{\lambda_{H_{2}N_{R}N_{R}}
		\,\,\lambda_{H_{2}\phi_{D}^{\dagger} \phi_{D}}}{(s-m_{H_2}^{2}) +
		i m_{H_2} \Gamma_{H_2}}\,, \nn \\
	\sigma_{\phi_{D}^{\dagger} \phi_{D}\rightarrow N_{R}N_{R} } &=&
	\frac{(s - 4\,m_{N_R}^{2})}{32 \pi s}\,
	\sqrt{\frac{s-4 m_{N_R}^{2}}{s - 4m_{\phi_{D }}^{2}}}\,\,\,|M_{N_{R}N_{R}}|^{2}
	\end{eqnarray}
	\item $\underline{\phi_{\mathrm{D}}^\dagger \phi_{\mathrm{D}}\rightarrow f\bar{f} }$
	\begin{eqnarray}
	\lambda_{H_{1}ff} &=& \frac{m_{f}\cos \theta}{v}  \nn \\
	\lambda_{H_{2}ff} &=& \frac{m_{f}\sin \theta}{v}  \nn \\
	M_{ff} &=& \frac{\lambda_{H_{1}ff}\,
		\,\lambda_{H_{1}\phi_{D}^{\dagger} \phi_{D}}}{(s-m_{H_1}^{2}) +
		i m_{H_1} \Gamma_{H_1}} + \frac{\lambda_{H_{2}ff}
		\,\,\lambda_{H_{2}\phi_{D}^{\dagger} \phi_{D}}}{(s-m_{H_2}^{2}) +
		i m_{H_2} \Gamma_{H_2}}\,, \nn \\
	\sigma_{\phi_{D}^{\dagger} \phi_{D}\rightarrow f\bar{f} } &=&
	\frac{(s - 4\,m_{f}^{2})}{32 \pi s\  n_c}\,
	\sqrt{\frac{s-4 m_{f}^{2}}{s - 4m_{\phi_{D }}^{2}}}\,\,\,|M_{ff}|^{2}
	\end{eqnarray}
	In the above, $n_c$ is the color charge and is 1 for leptons and 3 for quarks. 
\end{itemize} 
Similarly, the expression of cross-section for the relevant processes in  $\chi$ production are as follows,
\begin{itemize}
	\item $\underline{\phi_{\mathrm{D}}^\dagger \phi_{\mathrm{D}}\rightarrow \chi\chi }$
	\begin{align}\sigma=\frac{y_{D\chi}^4}{16 \Pi  s \left(s-4 M_{\phi }^2\right)} \Big[-\left(s-2 \left(-M_N^2+M_{\phi }^2+M_{\chi }^2\right)\right) 
	\left(\log  \frac{t_{\text{max}}-M_N^2}{t_{\text{min}}-M_N^2}\right)
	+ & \nonumber\\ \frac{(t_{\text{min}}-t_{\text{max}}) \left(-M^2_N+(M_{\phi }-M_{\chi })^2\right)  \left(-M^2_N+(M_{\phi }+M_{\chi })^2\right)}{\left(M_N^2-t_{\text{min}}\right) \left(M_N^2-t_{\text{max}}\right)}+t_{\text{min}}-t_{\text{max}}\Big]
	\end{align}
	
	$t_{\text{min/max}}=\mp\frac{\sqrt{s-4 M_{\phi }^2} \sqrt{s-4  M_{\chi }^2}}{2 }+M_{\phi }^2+M_{\chi }^2-\frac{s}{2}$
	
	\item $\underline{N N\rightarrow \chi\chi }$
		\begin{align}\sigma=\frac{y_{D\chi}^4}{16 \Pi  s \left(s-4 M_N^2\right)}
	\Big[\frac{\left(t_{\max }-t_{\min }\right) \left(\left(M_N+M_{\chi }\right){}^2-M_{\phi }^2\right){}^2}{\left(M_{\phi }^2-t_{\max }\right) \left(M_{\phi }^2-t_{\min }\right)} & \nonumber\\-2 \left(\left(M_N+M_{\chi }\right){}^2-M_{\phi }^2\right) \log \left(\frac{M_{\phi }^2-t_{\max }}{M_{\phi }^2-t_{\min }}\right)+t_{\max }-t_{\min }\Big]	\end{align}
	
	$t_{\text{min/max}}=-\frac{\sqrt{s-4  M_N^2} \sqrt{s-4  M_{\chi }^2}}{2 }+M_N^2+M_{\chi }^2-\frac{s}{2}$
	\item $\underline{N H_i\rightarrow \chi \phi_{D } }$
\begin{align}\sigma=\frac{\lambda_{H_i\phi_D\phi_D}^2 y_{D\chi}^2}{32 \Pi  \left(s-\left(M_N-M_{H_i}\right){}^2\right) \left(s-\left(M_{H_i}+M_N\right){}^2\right)} \Big[\log\big(\frac{M_{\phi }^2-t_{\min }}{M_{\phi }^2-t_{\max }}\big)& \nonumber \\ -\frac{\left(t_{\min }-t_{\max }\right) \left(\left(M_N+M_{\chi }\right){}^2-M_{\phi }^2\right)}{\left(M_{\phi }^2-t_{\max }\right) \left(M_{\phi }^2-t_{\min }\right)}\Big]\end{align}
	
	\begin{align}t_{\text{min/max}}=\frac{\mp1}{2s}\sqrt{-2 s M_{\phi }^2-2 s M_{\chi }^2-2 M_{\phi }^2 M_{\chi }^2+M_{\phi }^4+M_{\chi }^4+s^2} \ \times & \nonumber\\ \sqrt{-2 M_N^2 M_{H_i}^2-2 s M_{H_i}^2+M_{H_i}^4-2 s M_N^2+M_N^4+s^2}& \nonumber\\
	\frac{-1}{2s}\left(-M_{\phi }^2+M_{\chi }^2+s\right) \left(-M_{H_i}^2+M_N^2+s\right)+M_N^2+M_{\chi }^2
	\end{align}
\end{itemize}
\bibliographystyle{utphys}
\bibliography{bibitem}  	
\end{document}